\documentclass[11pt, epsfig]{article}

\usepackage[hmargin=2.1cm,vmargin=2.5cm]{geometry}
\usepackage{graphicx}
\usepackage{enumerate}

\begin{document}

\title{The quantum measurement approach to particle oscillations}
\author {Charis Anastopoulos\footnote{anastop@physics.upatras.gr} \\ 
 {\small Department of Physics, University of Patras, 26500 Greece} \\ \\
 and \\ \\ Ntina Savvidou\footnote{ntina@imperial.ac.uk} \\ 
  {\small  Theoretical Physics Group, Imperial College, SW7 2BZ,
London, UK} }

\maketitle
\begin{abstract}
The LSND and  MiniBoone seeming anomalies in neutrino oscillations are usually attributed to physics beyond the Standard model. It is, however, possible that they may be an artefact of  the   theoretical treatment of particle oscillations that ignores fine points of quantum measurement theory relevant to the experiments. In this paper, we construct a rigorous measurement-theoretic framework for the description of particle oscillations, employing no assumptions extrinsic to quantum theory. The formalism leads to a non-standard oscillation formula; at low energy it predicts an `anomalous' oscillation wavelength, while at high energy it differs from the standard expression by a factor of 2. The key novelties in the formalism are the treatment of a particle's  time of arrival at the detector as a genuine quantum observable,  the theoretical precision in the definition of quantum probabilities, and the detailed modeling of the measurement process. The article also contains an extensive  critical review of existing theoretical treatments of particle oscillations, identifying key problems and showing that these are overcome by the proposed formalism.

\end{abstract}

\section{Introduction}

In this paper we  construct a {\em solid} measurement-theoretic
framework for the description of particle oscillations \cite{GePa, Ponte}, i.e., a framework that (i)  contains no {\em ad-hoc}
assumptions extrinsic to the general rules of quantum theory, (ii) allows for a precise correspondence between experimental operations and elements of the formalism,  and (iii) is sufficiently general to incorporate fine  details in the description of particle oscillations, when such are needed. The proposed formalism predicts an `anomalous' oscillation wavelength at the low end of the neutrino energy spectrum, and, as such, it might provide an  explanation for excess of  events at low energy in the MiniBooNE experiment and of the LSND anomaly \cite{LNSD, MiniB, PDG}.

Existing treatments of particle oscillations, as a rule, introduce assumptions extrinsic to quantum theory, hence important problems arise. In particular,  existing methods  contain ambiguities in the relation of elements of the formalism to the experimental setup  and in their treatment of time and quantum probabilities. These problems originate from the fact that particle oscillations touch upon fine points of quantum measurement theory (the quantum treatment of the time of arrival, the joint measurement of incompatible observables and the role of decoherence and coarse-graining in the measurement process), whose comprehensive treatment had been unavailable at the time the phenomenon was first predicted \cite{GePa, Ponte, Kay81}.

The original intent of this paper was to employ the techniques and ideas from quantum measurement theory, in order  to resolve the problems above
and to place the standard results on particle oscillations on a firm conceptual and mathematical footing. Impressively, we find that our method leads to a {\em non-standard} expression for the wavelength of particle oscillations, i.e., to an expression that disagrees with the result that has been obtained by the vast majority of existing approaches to the treatment of particle oscillations. We have verified our result to be robust: it is not an artefact of any calculational approximation, its derivation involves no {\em ad hoc} assumptions and it persists when more details are added in the models of the measurement process. The approach we propose is based on a novel treatment of the concept of time, as presented in the development of the History Projection Operator theory as a new approach to quantum gravity \cite{Sav, Sav2}. As far as the theory presented in this paper, the distinction between the kinematics and dynamics on the definition of time resulted to a new treatment of the time of arrival as a {\em genuine} quantum observable \cite{AnSav}.

Claims of  non-standard oscillation formulas
have appeared sporadically in the literature---for example, Refs. \cite{Sriv, Field}---and they have been
repeatedly refuted in favor of the standard result \cite{equaltime, Lip97, OST, BLSG, Lip05}.
We do agree with
the critique of such non-standard formulas, in the sense that the methodologies through which they are produced can be found wanting. However, we claim that the standard derivations are also wanting on the same grounds. To show this, we have undertaken an extensive critical review of existing theoretical treatments of particle oscillations (Sec. 4), identifying their shortcomings and showing that these shortcomings are, indeed, overcome with the method we develop here (Sec. 5.3).

One argument in favor of the non-standard result obtained here is that its derivation is not plagued by any mathematical ambiguities or conceptual inconsistencies. More importantly, our expression for the oscillation wavelength can be experimentally distinguished from the standard one, at least in principle. For neutrinos, the standard expression for the oscillation wavelength $\lambda$ at energy $E$ is
\begin{eqnarray}
\lambda = \frac{4 \pi E}{\Delta m^2}, \label{wlength}
\end{eqnarray}
where $\Delta m^2$ is the difference of squared neutrino masses. The non-standard oscillation formula that has been obtained before differ from Eq. (\ref{wlength}) by a factor of two, i.e.,
\begin{eqnarray}
\lambda = \frac{2 \pi E}{\Delta m^2}. \label{wlength2}
\end{eqnarray}
In absence of an independent precise measurement of neutrino masses Eqs. (\ref{wlength}) and (\ref{wlength2}) cannot be experimentally distinguished. In contrast, here we find
\begin{eqnarray}
\lambda = \frac{2 \pi E}{\Delta m^2( 1 - \frac{\epsilon_{th}}{2 E})}, \label{wlength3}
\end{eqnarray}
where $\epsilon_{th}$ is the energy threshold for the process through which the neutrino is detected. While for $E >> \epsilon_{th}$, Eq. (\ref{wlength3}) coincides with Eq. (\ref{wlength2}), for $E$ near  $\epsilon_{th}$, Eq. (\ref{wlength3}) predicts that  the ratio $\lambda/E$ is not a constant but increases with energy.

Hence,  the method we develop here leads to the following  predictions.
\begin{itemize}{}
\item At low energies, the oscillation wavelength depends strongly on the value of the threshold energy $\epsilon_{th}$, hence, it may vary according to the process through which neutrinos are detected\footnote{To be precise, Eq. (\ref{wlength3}) is obtained from the study of  reactions in which the neutrino is annihilated, i.e., charged current interactions---see, Sec. 5.2. Neutrino detection through scattering processes can also be studied in the present framework and it will be treated in a different publication.}. In other words, data subsets for different reaction channels may exhibit different oscillation patterns.
\item For a given reaction channel, the expected number of events as a function of energy will differ strongly from what is predicted by the standard oscillation formula at low energies. This difference  can be seen heuristically from the plots in Fig. 2  (Sec. 5.3).
\end{itemize}

 The prediction of  `anomalous' neutrino oscillations at low energies might provide an explanation for the excess of neutrino detection events  in the LSND \cite{LNSD} and MiniBooNE \cite{MiniB} experiments.  In particular, our approach makes no recourse to physics beyond the Standard Model and introduces no additional undetermined parameters. Hence, it can be directly tested against the standard expression for particle oscillations in terms of better fit to existing data.

\bigskip

 We will refer to the formalism developed here as the {\em quantum measurement approach to particle oscillations}. Its main points are the following.
\begin{itemize}

\item Particle oscillation experiments do not correspond to measurements taking place at a predetermined moment of time. The time of arrival $t$ of the particle at the detector is a genuine physical observable that should be described in terms of quantum theory. In particular, the time of arrival is to be distinguished, both physically and mathematically, from the time parameter $t$ of Schr\"odinger's equation \cite{Sav}.

\item For a quantum measurement the most general consistent assignment of probabilities  is through the use of Positive-Operator-Valued-Measures (POVMs)---see Sec. 3.1.3 for definition. The probability densities for particle oscillation are represented as $p_{\alpha}(t, L) = Tr \left(\hat{\rho}_0 \hat{\Pi}_{\alpha}(t, L)\right)$, in terms of a family of POVMs $\hat{\Pi}_{\alpha}(t, L)$, where $t$ is the  time of arrival, defined in terms of the detection of a  particle produced through a channel of flavor $\alpha$, and $L$ is the distance from the production region. Hence, we describe particle oscillation experiments as
      {\em joint} measurements of the time of arrival $t$ and the  position $L$. This is a key difference of our approach from existing ones, and it leads to a non-standard expression for the oscillation wavelength.

\item The family $\hat{\Pi}_{\alpha}(t, L)$ is constructed in terms of the detection of product particles, i.e., in terms of directly observable quantities.  Moreover, the method for constructing $\hat{\Pi}_{\alpha}(t, L)$  involves only the Hamiltonian of the system and the specification of the measured variables. In particular, it does not depend on properties of the initial state and contains no assumptions extrinsic to quantum theory.

\end{itemize}

 \bigskip
 The paper is structured as follows. Sec. 2 describes our motivation in undertaking this work, in order to correct certain conceptual and mathematical inadequacies in existing treatments of particle oscillations.
We identify theoretical assumptions that enter  into existing derivations that are not justified by the rules of quantum theory and, sometimes,  they are  erroneous. In that section we also present the main ideas of our approach.  The presentation is largely non-technical, and it requires little background to the issue.

Sec. 3 provides background material necessary for the development of our approach. In particular, we  focus on quantum measurement theory providing a review of main results that are relevant to our critique of existing approaches to particle oscillations.

Sec. 4 presents a specific and more technical critique of existing approaches to particle oscillations, in light of the material presented in Sec.3.

The development of a new framework for particle oscillations is contained in Sec. 5. In Sec. 5.1 we construct the necessary tools in the context of an abstract Hilbert space formalism. In particular, we develop a methodology for defining probabilities for general measurement schemes, where the time of detection is treated as a physical observable. Particle oscillation experiments constitute   a special case of such schemes, and they are treated in Sec. 5.2, where  we derive the formula for the oscillation wavelength. In Sec. 5.3 we undertake a critical analysis of our method, demonstrating that it resolves the problems we identified in existing approaches to the issue. In Sec. 5.4 we further develop the formalism, applying it to more general settings, and we demonstrate that our results are robust.

\section{Motivation for a quantum measurement approach to particle oscillations}

The problems in the existing treatments of particle oscillations arise from their identification of probabilities as the modulus squared of matrix elements of the evolution operator $e^{-i\hat{H}t}$. While this is the most common probability assignment in  quantum theory, it is not universal. In particular, it does not apply to particle oscillations: the resulting probabilities depend on the time $t$ and there is no rigorous way to relate them to what is actually being measured in an experiment. When particle oscillations were first predicted, the researchers did not have the benefit of a mature  quantum theory of measurement that could provide a solution in terms of first principles. As a result, the predictions relied on heuristic arguments and assumptions that are not warranted by  quantum theory. These created ambiguities in the relation of the formalism to experimental procedures.

Later work on the quantum theory of measurement  provided a precise language for the treatment of the subtle issues that enter the theoretical description of particle oscillations. As it turns out, the more precise treatment yields information inaccessible from the  heuristic ones, namely, the prediction of `anomalous' oscillation wavelength at low energies that could be relevant for the explanation for the LSND and MiniBooNE experiments.

\medskip

In section 2.1, we provide an extensive discussion of  problems in existing treatments of particle oscillations, explaining why they arise and how they affect the existing treatments. In section 2.2, we describe the main ideas of our approach, and how they resolve these problems.

\subsection{Main problems in particle-oscillations theory}

We proceed to an analysis of the following issues: (i)
the improper treatment of the time of arrival, (ii) the lack of a first principles derivation of the relation between amplitudes and probabilities, and (iii) the improper elimination of the time variable in the probability assignment. In our opinion, these  issues constitute the most important problems  in existing treatments of particle oscillations.

\subsubsection{The time-of-arrival issue}
Current methods for
the treatment of particle oscillations do not establish a sharp
correspondence between mathematical formalism and experimental
operations. The  reason  is that these methods
 conflate two  {\em distinct} notions of time as it appears in quantum
theory: time as a parameter of evolution in Schr\"odinger equation
and the time of arrival (or time of event, or time of transition) \cite{Sav}.

Time as a parameter of evolution, like in Schr\"odinger equation, is an essentially classical parameter \cite{Ish}, i.e., it does not correspond to a  Hilbert space operator;
hence,   it is not an `observable'. On the other hand, the time of arrival is a variable defined in terms of coincidences of {\em events}: we observe an event (e.g., the creation of a lepton in a detector) and
we correlate its occurrence with the reading of an external clock. The time of arrival is, therefore, a physical observable.

In existing treatments of particle oscillation the distinction above is ignored. This results into the definition of probabilities in terms of the Schr\"odinger time parameter which, unlike time of arrival, is not a physical observable. To obtain physical predictions from these probabilities it is necessary to remove this time dependence; this is achieved by procedures that are not justified from first principles---see Sec. 2.1.3.

The time of arrival  is
defined in terms of the physical event that a particle is detected by a measuring device. Since particle detection is a quantum process, time of arrival is a  genuine quantum observable
 associated to physical processes on the measured
system. (See, Sec. 3.3 for details and references.)
 As such, time of arrival may exhibit interferences. In a
hypothetical experiment where a source prepares particles in a
superposition of states with different mean momentum, the times of
arrival recorded at a detector, at distance $L$ from the source, will
be distributed according to an oscillatory pattern (i.e.,
interference) around  a mean value of arrival time.

\subsubsection{The  physical interpretation of amplitudes}

The vast majority of the theoretical studies of particle
oscillations concentrate on the evaluation of  amplitudes between certain vector states on a Hilbert space that describes the physical system. For example, these states may be
 `eigenstates' for flavors $\alpha$ and $\beta$, leading to amplitudes of the form
\begin{eqnarray}
{\cal A}_{\alpha \rightarrow \beta}(t, {\bf L}) = \langle \alpha,  {\bf x} = 0|
e^{- i \hat{H}t} |\beta, {\bf x} = {\bf L} \rangle. \label{ampl0}
\end{eqnarray}
  In Eq. (\ref{ampl0}) the states $|{\bf x}, \alpha \rangle$ refer to particles localized around the point ${\bf x}$ and they correspond to detection of particles characterized by flavor $\alpha$. In other approaches that employ quantum  field theory, analogous amplitudes are computed between states of incoming and product particles,
rather than neutrinos or neutral bosons, as in Eq.\ (\ref{ampl0}). The dependence of the amplitude on the flavor indices arises through the flavor dependence of an interaction Hamiltonian. The critique in this section applies to those approaches, as well. The critique refers to a structural point of the quantum formalism (i.e.,  the relation of matrix elements of  $e^{-i\hat{H}t}$ to the probabilities of measurement outcomes), and not to properties of specific physical systems or calculational techniques. It also applies to cases that the amplitude is defined in terms of elements of the $S$-matrix rather than matrix elements of the finite-time propagator $e^{-i \hat{H}t}$.

Having computed an amplitude of the form Eq. (\ref{ampl0}), the modulus square $|{\cal A}_{\alpha \rightarrow \beta}(t, {\bf x})|^2$ is then interpreted as a measure of
probability, from which  physical
predictions about particle oscillation experiments are extracted. In Sec. 4, we will examine the various methods employed to extract physical predictions from such amplitudes.

A key question then  arises: in what sense are amplitudes of the form Eq.\ (\ref{ampl0}) relevant to the description of particle oscillation experiments? In particular, Eq. (\ref{ampl0})
defines the matrix elements of the unitary evolution operator between states of the form $|\alpha, {\bf x} \rangle$. How
does this relate to probabilities relevant to particle oscillation experiments?

 The squared amplitude
$|{\cal A}_{\alpha \rightarrow \beta}(t, {\bf L})|^2$ defines  probabilities for the measurement of a particle's position, at a specific, predetermined and  sharply defined moment of time $t$. Besides the idealization involved in talking about measurements taking place at a sharply defined moment of time, the main problem in the use of the squared amplitude $|{\cal A}_{\alpha \rightarrow \beta}(t, {\bf L})|^2$ is that in particle oscillation experiments the timing of the detection event  is not predetermined by the experimentalist. What is controlled is the {\em location} of the detector. In other words, particle oscillation experiments do not fall in the domain of validity of the expression $|{\cal A}_{\alpha \rightarrow \beta}(t, {\bf L})|^2$ for probabilities of measurement outcomes. We shall elaborate on this point in Sec. 3.1.

In order to construct a proper probability assignment that fully reflects the setup of particle oscillation experiment, it is necessary to work within the broader context of quantum measurement theory \cite{Dav, WZ83, BLM}. The latter allows for probability assignments more general than squared amplitudes (for example, probabilities for joint measurements \cite{Dav}, continuous-time measurements \cite{ctm}, weak measurements \cite{AAV}, time-of-arrival measurements \cite{Kij74, toarev, AnSav}, and so on). In particular, it is necessary to  derive the relevant probability from {\em first principles}, taking into account the physics of the detection scheme; we shall provide such a derivation in Sec. 5.1.

In most physical applications, a first-principles derivation of the probability assignment through quantum measurement theory makes little difference to the theory's predictions,
 hence, it might be dismissed as too fine  to be relevant.
However, this is not the case for particle oscillation experiments, in which
 the measured quantities are directly correlated
with tiny microscopic variations (oscillations) in the
time of arrival of the particles. A proper treatment of the probabilities does affect the physical predictions of the theory.

\subsubsection{Elimination of the time variable}

The problems in the use of the squared amplitude
$|{\cal A}_{\alpha \rightarrow \beta}(t, {\bf L})|^2$ for the description of
 particle oscillations have been
noticed early on in the theory, as the presence of the time $t$ in addition to the position ${\bf L}$ appears superfluous when  relating the
theoretical expression to experimental results. However, this issue has not led
to a reconsideration of the starting point,  that is, to the derivation of the probabilities for particle oscillations from first
principles. Instead, the relevance of the squared amplitudes has been taken
for granted, and various procedures have been invoked in order to
eliminate time from the probabilities. Clearly, since the starting
point is accepted without justification, any
remedy to the  consequent problems can be introduced only {\em ad hoc},
i.e., without justification on the basic principles of quantum
theory.

Here we  describe the two most common such procedures, for eliminating time from squared amplitudes.

\paragraph{Use of the classical arrival time.} The oldest approach
consists in using $|{\cal A}_{\alpha \rightarrow \beta}(t, {\bf L})|^2$ as a
probability density for the detection, by substituting $t$ with a
classical time of arrival $t_{cl} = L / \bar{v}$; $L = |{\bf
L}|$ is the distance from the detector to the source and $\bar{v}$
the mean velocity of the wave-packet.

The obvious problem with this
method is that it is mathematically inconsistent: the time $t$, that a measurement takes
place, is substituted by an `average' time-of-arrival, which depends on the
system's initial state, through $\bar{v}$. The resulting
probabilities are then not linear functionals of the initial state
of the system $\hat{\rho}$, as required by quantum theory. Hence, this solution can only be  a heuristic device, perhaps valid
for a specific class of initial states.

The method above
could  be viewed as a kind of semi-classical approximation. As such, it can only be justified by demonstrating that it is a good approximation, i.e., by comparing with a {\em full quantum mechanical treatment} of the problem. Heuristic arguments based on estimations of uncertainties have pitfalls, and they are no substitute for a full calculation.

\paragraph{Integration of probabilities over time.} A second
commonly employed procedure for the elimination of time $t$ from the
squared amplitude $|{\cal A}_{\alpha \rightarrow \beta}(t, {\bf L})|^2$ is
integration. The related argument is that, since time is not a
directly measured quantity in the experiment, we integrate over the
single-time probabilities $|{\cal A}_{\alpha \rightarrow \beta}(t, {\bf L})|^2$. We then
 obtain a probability density for position alone, i.e.,
\begin{eqnarray}
p_{\alpha \rightarrow \beta} ({\bf L}) \sim \int_0^T dt |{\cal A}_{\alpha \rightarrow \beta}(t, {\bf L})|^2, \label{sumprob}
\end{eqnarray}
where $T$ is an integration time for the experiment.

The argument above is invalid for the following reasons.
First, the squared amplitudes $|{\cal A}_{\alpha \rightarrow \beta}(t, {\bf
L})|^2$ are probability densities with respect to position  {\em
and not\/} with respect to time, and thus time-integration  does not lead to meaningful probabilities. This can be seen  even on dimensional
grounds: a density with respect to time as well as space ought to
have dimensions $[time]^{-1} \times [length]^{-3}$, while the squared
modulus $|{\cal A}_{\alpha \rightarrow \beta}(t, {\bf L})|^2$ has dimensions
$[length]^{-3}$. Moreover, it does not transform as a density under a time-rescaling $ t \rightarrow a t$.

This obvious problem is sidestepped by a procedure invoked for the solution of another problem, namely, the fact that $\int_0^T dt |{\cal
A}_{\alpha \rightarrow \beta}(t, {\bf L})|^2$ is not normalized to unity. One
 is then forced to normalize it by dividing by a constant $ C$, for example,  $C =
\sum_{\beta} \int d^3L \int_0^T |{\cal A}_{\alpha \rightarrow \beta}(t, {\bf L})|^2$. However, $C$
 depends on the initial state
$\hat{\rho}_0$, hence, the probability assignment
 is not a linear functional of  $\hat{\rho}_0$, thus violating a fundamental property of quantum mechanics. For
an elaboration of these points, see, Sec. 4.2.3.

The mathematical and conceptual consistency of this procedure apart,
there are also physical problems with the integration over time in
Eq.\ (\ref{sumprob}). Suppose we were able to  properly define
 amplitudes ${\cal B}_{\alpha \rightarrow \beta}(t, {\bf L})$,
different from Eq.\ (\ref{ampl0}), in which the parameter $t$ stands for the time of arrival which is a genuine observable (we shall see how this is done in
Sec. 5.1). Would we be justified then in defining a probability density for position
\begin{eqnarray}
p_{\alpha \rightarrow \beta} ({\bf L}) \sim \int_0^T dt |{\cal B}_{\alpha \rightarrow \beta}(t,
{\bf L})|^2, \label{sumprob2}
\end{eqnarray}
if we assume that the time of arrival is not directly observable?
In classical probability, the answer would be affirmative. However,
the defining feature of quantum probability, clearly manifested in
the two-slit experiment, is that alternatives are summed over at the
level of amplitudes. Therefore  interference terms appear. Only if a
macroscopic distinction of alternatives is possible
  are the interference terms suppressed.  Hence, Eq. (\ref{sumprob2}) for the probability
  would  be justified only if time of arrival is measured very finely. In particular, Eq. (\ref{sumprob2}) is justified only
  if the resolution of the time measurements captures the full
  dependence of
 ${\cal B}_{\alpha \rightarrow \beta}(t, {\bf L})$ on time.  Unless there is no time measurement----or if the time measurements
 are too coarse to distinguish oscillating terms in the distribution
 of the time of arrival---then one should expect that  the relevant probabilities are of the form
 \begin{eqnarray}
p_{\alpha \rightarrow \beta} ({\bf L}) \sim |\int_0^T dt {\cal B}_{\alpha \rightarrow \beta}(t,
{\bf L})|^2,
 \end{eqnarray}
where interference terms will be in general present.

In other words, even if all the mathematical problems associated to
integration over time at the level of squared amplitudes were resolved,
Eq. (\ref{sumprob2}) does not follow from the basic principles of
quantum theory, unless one postulates a decoherence mechanism  that destroys all interferences for the time of arrival
in the amplitudes ${\cal B}_{\alpha \rightarrow \beta}(t, {\bf L})$.

\subsection{The main ideas of the quantum measurement approach to particle oscillations}

In Sec. 2.1, we argued that the existing methods for particle oscillations suffer from  problems of consistency with the basic principles of quantum theory. These problems can be concisely summarized in the following statements.

\begin{enumerate}{}

\item A key issue in the theoretical study of
particle oscillations is the definition of the time of arrival as a genuine quantum variable. In existing
treatments, time of arrival is either conflated with the
time-parameter of Schr\"odinger's equation, or it is treated
classically. The former case contains a conceptual error that leads to a
discrepancy between the formalism and the experiment.  The
latter case involves  an approximation, whose validity
can only be ascertained by comparison with the results of a full quantum
description.

\item There is no first-principles derivation of the
probabilities that pertain to particle oscillation experiments. The
modulus square of the amplitude Eq.\ (\ref{ampl0}) is introduced without
justification as a probability measure relevant to particle oscillation experiments.  {\em Ad-hoc} procedures
are employed in order to extract physical predictions.
\end{enumerate}

These issues are resolved in the present paper: the quantum measurement approach to particle oscillations is based on two key ideas which can be concisely summarized as follows.

\begin{enumerate}
\item The distinction between the time of arrival and the time parameter of Schr\"odinger's equation is the groundwork of our approach. The importance of this distinction originates from previous work on the concept of time in quantum theory \cite{Sav, Sav2}, as we stated in the Introduction.
    The formalism presented here strongly separates between those two different notions of time, by representing them by different mathematical objects.
      In particular, the formalism treats the time of arrival as a physical observable, directly relevant to  particle oscillation experiments and, it provides a probabilistically consistent treatment of its quantum fluctuations.

\item Treating the time of arrival as a quantum observable, implies that the probabilities relevant to particle oscillations correspond to the {\em joint} measurement of two observables: the time of arrival $t$ and the location of a detection event ${\bf L}$. In quantum theory, probabilities for joint measurements cannot, in general, be obtained by squared amplitudes. However, they are described, with full consistency and rigor, in terms of Positive Operator Valued Measures (POVMs) \cite{Dav, BLM}. Hence,  we obtain the  probabilities relevant to particle oscillation experiments through the construction of
 a POVM  $\hat{\Pi}_{\alpha}(t, {\bf L})$  that describes the probabilities for the  detection of $\alpha$-flavor events at location ${\bf L}$ and time $t$. The corresponding probabilities are then of the form
\begin{eqnarray}
p_{\alpha}({\bf L}) = Tr_H \left[ \hat{\rho}_0 \hat{\Pi}_{\alpha}(t, {\bf L}) \right], \label{probbas}
\end{eqnarray}
where $\hat{\rho}_0$ is the initial state of the system. Since in Eq. (\ref{probbas}) the time of arrival $t$ is an observable, it can be {\em consistently} integrated out, leading to a probability density that is a function of ${\bf L}$ alone. The oscillation wavelength Eq. (\ref{wlength3}) is then obtained.
\end{enumerate}

The implementation of the ideas above involves the construction of amplitudes ${\cal A}_{\alpha}(t, {\bf L})$ that depend on the time of arrival rather than the time parameter of Schr\"odinger's equation. These amplitudes  are then employed for the construction of the POVM $\hat{\Pi}_{\alpha}(t, {\bf L})$ using a method that was developed in Ref. \cite{AnSav}.
In fact, we do not  construct a single POVM for the task at hand but
we provide a general methodology for their construction. The results, therefore, do not
depend on simple idealizations;  there exists a systematic procedure that allows us to incorporate as many details in their derivation as needed.

The measurement process is described in terms of the detection of particles produced by reactions of the oscillating particles at the detector, hence the use of quantum field theory is necessary.  We also take into account the fact that the detector is a macroscopic system with a large number of degrees of freedom.  In spite of the complexity involved in a realistic description of the measurement process, our main result Eq. (\ref{wlength3}) for the oscillation wavelength is robust, and it does not depend on variations in the modeling of the measurement. Moreover,
it does not require any fine-tuning in the properties of the initial state.


\section{Background}

This section consists of two parts that address  issues important  for the critique of  existing approaches to particle oscillation.  It also serves as a theoretical background for the formalism we develop in Sec. 5. In particular, in Sec. 3.1 we review the main ideas of quantum measurement theory, placing special emphasis on measurements of von Neumann type, explaining why they are not suitable for the description of particle oscillations and discussing the properties and wide applicability of POVMs. In Sec. 3.2 we briefly review the quantum treatment of time of arrival in the language of POVMs, showing that, like any other quantum observable, the time of arrival may be characterized by interferences.

\subsection{Quantum measurement theory}

\subsubsection{von Neumann  measurements}

First, we study the simple model about the measurement of a
variable corresponding to the self-adjoint operator $\hat{A}$, with generalized (continuous) eigenvalues $a$. This class of models originates
with von Neumann \cite{vN, Lud}---see also \cite{BLM}---the variation we present here is from Ref. \cite{AnSav07}.

To this end, let ${\cal H}_s$ be the Hilbert space
of the microscopic system and ${\cal H}_{app}$ the Hilbert space
corresponding to the degrees of freedom of a macroscopic apparatus.
The values of the observable $\hat{A}$ of the quantum system are
correlated to the values of an operator $\hat{X}$ on ${\cal H}_{app}$.
$ \hat{X}$ is usually called the `pointer variable', its values
are  accessible to macroscopic observation. Hence, its spectrum is highly degenerate. We denote the
generalized eigenstates of $\hat{X}$ as $|X, \lambda \rangle$;
$\lambda$ labels orthonormal  bases in the degeneracy subspaces of
$\hat{X}$.

 $\hat{H}_s$ is the Hamiltonian operator describing the
self-dynamics of the quantum system.
 The interaction between the
measured system and the apparatus is described by  a Hamiltonian of
the form
$\hat{H}_{int} =  f(t)  \hat{A} \otimes \hat{K}$.
$ \hat{K} = \sum_{\lambda} \int dk \,k \,|k, \lambda \rangle
\langle k, \lambda|$  is the `conjugate momentum' of $\hat{X}$, i.e., the generator of translations for the pointer variable
$\hat{X}$,  where $\langle X, \lambda| k, \lambda' \rangle = \frac{1}{\sqrt{2\pi}} e^{ikX} \delta_{\lambda \lambda'}$.
The function $f(t)$ in $\hat{H}_{int}$ describes the `switching-on' of the interaction between system and measurement device.

For simplicity, we assume that the self-dynamics of the apparatus during the interaction are negligible, so that the total Hamiltonian is $\hat{H}_{tot} = \hat{H}_s + \hat{H}_{int}$. The corresponding evolution operator is $\hat{U}(t) = {\cal T} e^{- i \int_0^t ds \hat{H}_{tot}(s)}$, where ${\cal T}$ denotes the time-ordered exponential.

The microscopic system is prepared in a state $\hat{\rho}_0$ and the initial state of the apparatus is $|\Psi_0 \rangle$. The initial state of the combined system is then $\hat{\rho}_0 \times |\Psi_0\rangle \langle \Psi_0|$, and it evolves into $\hat{\rho}_{tot}(t) = \hat{U}(t) \hat{\rho}_0 \hat{U}^{\dagger}(t)$, under the action of the unitary operator $\hat{U}(t)$.

With the above assumptions, the probability distribution $p_t(X)$ over the
value of the pointer variable at time $t$ is
\begin{eqnarray}
p_t (X) = \sum_{\lambda} \langle X, \lambda|Tr_{H_s}
\hat{\rho}_{tot}(t)|X, \lambda \rangle
= Tr_{H_S} \left[\hat{\rho_0} \hat{\Pi}(X)\right],
\end{eqnarray}

where $\hat{\Pi}(X) = \sum_{\lambda} \hat{C}(X, \lambda) \hat{C}^{\dagger}(X,
\lambda)$ is a family of positive operators on $H_S$, and where

\begin{eqnarray}
\hat{C}(X, \lambda) =   \int \frac{dk}{\sqrt{2 \pi}} e^{i k X}
\langle k, \lambda|\Psi_0 \rangle
 \left[{\cal T} e^{-i \int_0^t ds (H
+ k f(s) \hat{A})} \right].
\end{eqnarray}

A common approximation is to assume that the function $f(t)$  is
narrowly concentrated around a value $t = \bar{t}$, i.e.,  $f(t) \simeq
\delta (t - \bar{t})$. In that case,

\begin{eqnarray}
p_{\bar{t}}(X) = Tr_{H_S} \left(e^{- i \hat{H} \bar{t}} \hat{\rho_0} e^{i
\hat{H} \bar{t}}  w( X- \hat{A})\right), \label{pr1}
\end{eqnarray}
where
\begin{eqnarray}
w(X) = \sum_{\lambda} |\Psi_0(X, \lambda)|^2 \label{wx},
\end{eqnarray}
and $w(X-\hat{A})$ stands for the positive operator $
\int da \; w(x - a) |a\rangle \langle a|$. The function $w(X)$
determines the correlation between the pointer variable $X$ and
the values of the operator $\hat{A}$.

Eq. (\ref{pr1}) corresponds to the probability for a measurement
of the observable $\hat{A}$, at a single moment of time
$\bar{t}$ \cite{inter}.
While  the degrees of freedom of the
apparatus have been included in the description, the end result for the probabilities is expressed solely
in terms of operators on the Hilbert space ${\cal H}_s$ of the microscopic
system.

\subsubsection{Coarse-graining and measurement uncertainties}

Next we elaborate on the physical origin of uncertainties and the role of coarse-graining in quantum measurements. These quantum measurement theory issues are crucial for the theoretical treatment of particle oscillations.

\medskip

 The accuracy by which the pointer variable $\hat{X}$ determines the
value of $\hat{A}$ is given by the weight function $w(X)$ in Eq.\ (\ref{wx});  $w(X)$ depends on the properties of the
measurement device. The resolution in the measurement of
$\hat{A}$  is at least of order $\delta X$, where
$\delta X$ is the mean deviation of $\hat{X}$ for the state $|\Psi_0\rangle$. In general, $\delta X$ is of macroscopic magnitude, since the apparatus has a large number of degrees of freedom. For example, $\delta X$ cannot be smaller than the thermal fluctuations of $\hat{X}$.

A measurement is a macroscopic amplification of a microscopic event. The inability to distinguish between values of $\hat{A}$ at a scale smaller than $\delta X$ is an objective fact of the system. There is no way in standard quantum theory---without introducing hidden variables---to infer properties of a microscopic system at a scale smaller than the resolution $\delta X$  \cite{Omn1, Omn2, Gri, GeHa}. In quantum theory, microscopic system and macroscopic apparatus become entangled during measurement and the association of definite properties requires a
sufficient degree of coarse-graining.

 Eq. (\ref{pr1}) reduces to the usual probability assignment, in terms of squared amplitudes only {\em when sufficiently coarse alternatives are considered}, i.e., at scales much larger than $\delta X$.  In that case, the weight $w(X - \hat{A})$, in Eq. (\ref{pr1}), can be substituted by a delta function and
 Eq. (\ref{pr1}) corresponds to a squared amplitude $|\langle \psi_0|e^{- i \hat{H} \bar{t}}|a\rangle|^2$.

\bigskip

 Measurements of von Neumann type are not universally applicable because they involve special assumptions and idealizations. In particular, the description of time in von Neumann measurements is artificial: we assumed an interaction between microscopic system and apparatus determined by a specific profile function $f(t)$.  The timing of the interaction between microscopic system and apparatus is not controlled by the experimentalist and it is not independent of the initial state. For example, the time of arrival of particle at a detector, located at distance $L$ from the particle source is not predetermined by the experimentalist and it varies according to  the energy distribution of the incoming particles. In this setup, the
   parameter controlled by the experimentalist is the distance $L$. Hence, an accurate modeling of such measurements should involve a profile function for position rather than time.

  The idealization in the treatment of time in von Neumann measurements  is a good approximation when the precise timing of the  event is not significant to the measurement outcomes. However, for  measurements where the time of detection is a physical variable (particle oscillations),
 a description in terms of von Neumann measurements is  insufficient.

\bigskip

 A realistic measurement scheme has a finite duration that corresponds to the time it takes for the pointer variable to settle to a definite  value, correlated to a property  of the microscopic system. This time-scale is determined by the physics of the apparatus, it typically lies in the macroscopic domain, and it sets a lower limit to the  temporal resolution of the apparatus.

     In order to provide an estimation for the duration of the measurement we generalize the model presented in Sec. 3.1.1, by including
      a Hamiltonian term $\hat{H}_{app}$ for the apparatus's degrees of freedom, so that the total Hamiltonian is $\hat{H} = \hat{H}_s \otimes 1 + 1 \otimes \hat{H}_{app} + \hat{H}_{int}$.
A good pointer variable $\hat{X}$ must be stable with respect to the self-dynamics of the apparatus, so that evolution under $\hat{H}_{app}$ does not significantly affect the recorded values after the measurement's completion.
  This means that $e^{i \hat{H}_{app} t } \hat{X} e^{- i \hat{H}_{app}t } |\Psi_0 \rangle \simeq \hat{X} |\Psi_0 \rangle$, or more precisely
\begin{eqnarray}
\xi(t) := | \langle \Psi_0 | e^{i \hat{H}_{app} t } \hat{X} e^{- i \hat{H}_{app}t } - \hat{X} | \Psi_0 \rangle |
<< |\langle \Psi_0| \hat{X} | \Psi_0\rangle| \label{xi}
\end{eqnarray}
and a similar condition must hold for the operator $\hat{K}$ that generates translations in $\hat{X}$. The condition above can be satisfied if $\hat{X}$ and $\hat{K}$  are coarse-grained quasi-classical variables, i.e., if they have  large degeneracy eigenspaces that remain approximately invariant under time
evolution. In general, the evolution of quasi-classical variables involves noise  \cite{GeHa, hartlelo, HaDo, Brun, PZ93, An01}, so $\xi(t)$ in Eq. (\ref{xi}) can be modeled by a classical stochastic process. When the effects of the noise can be ignored in the evolution of the system, the previous analysis passes unchanged and Eq. (\ref{pr1}) is obtained.  However, this involves a specific scale of coarse-graining both for the uncertainties $\delta X$ and for the duration $\tau$ of the measurement.

The effect of noise, such as $\xi(t)$ above, in the evolution  on the quasi-classical variables is often studied within the quantum theory of open systems \cite{Dav, BP}. The ignored degrees
of freedom of the apparatus are then treated as  an environment and they are modeled by a thermal reservoir---see,  Refs. \cite{Omn1,  Zur1, Schlos, GKZK, Omn3}. It is natural to expect  that the characteristic timescale $\tau$
of such measurement corresponds to the time that it takes
a variable $\hat{X}$ to settle to an  asymptotic value, characterized by purely thermal fluctuations. Therefore,  $\tau$  is  of the order of the relaxation time $\tau_{rel}$ of the reservoir.

The limit above is essentially classical, in the sense that it arises from the statistical fluctuations of a system with a large number of degrees of freedom. There is a more stringent, unavoidable
  lower limit for the resolution time scale $\tau$ arising from the quantum nature of the pointer variable. This limit is the {\em localization
 time}  $\tau_{loc}$, i.e., the time it takes  the pointer variable to establish classical behavior due to its interaction with the remaining degrees of freedom \cite{PZ93, PHZ93, AnHa}. Modeling the pointer variable by an oscillator undergoing quantum Brownian motion in a thermal reservoir of temperature $T$ \cite{BP, CaLe, HPZ}, one finds a localization time of order $\sqrt{\frac{\hbar \tau_{rel}} {T}}$, where $\tau_{rel}$ is the relaxation time \cite{PHZ93, AnHa, HaZo}. Such systems are characterized by a time scale separation  $T^{-1} << \tau_{loc} << \tau_{rel}$. For $T = 300^o K$, this implies that $\tau_{loc} >> 10^{-13}s$. We note  that the temporal resolution in a solid state detector is much coarser than these lower bounds: it is at best of the order of $10^{-9}s$.

Models such as the  above are  simplistic in comparison to the physics of actual particle detectors. However, they serve to identify the irreducibly {\em quantum} origin of the temporal resolution in a measurement scheme, namely, that the pointer variables can be assigned definite classical properties  only if a substantial degree of coarse-graining is involved.

\subsubsection{Positive-Operator-Valued Measures}

In Sec. 3.1.1, we constructed the positive operators $\hat{\Pi}(X)$ for a measurement
device by considering a single pointer variable $\hat{X}$, whose values
are correlated with one quantum variable $\hat{A}$. In general, a
measurement may involve many pointer variables $\hat{X}^i$ each
correlated with a different property $\hat{A}^i$ of the quantum
system. Again, the probabilities for the measurement outcomes can be
expressed in terms of a POVM. Here, we briefly describe the basic properties and uses of POVMs in quantum measurement theory.

Let ${\cal H}$ be the Hilbert space of a quantum system and let $a^i$ be
the values of observables correlated to pointer variables of a
measurement device. The possible outcomes of the measurements span a
space $\Omega$. A POVM is a map that assigns a positive operator $\hat{Pi}_U$ on
${\cal H}$ to each (measurable)
subset $U$ of $\Omega$, such that (i) $\hat{\Pi}_{U_1 \cup U_2} = \hat{\Pi}_{U_1} + \hat{\Pi}_{U_2}$,  for
$U_1, U_2 \subset \Omega $, and (ii)  $\hat{\Pi}_{\Omega} = \hat{1}$.
Usually, the POVM $\hat{\Pi}_U$ can be expressed as an integral over
$U$ in terms of operators $\hat{\Pi}(a), a \in \Omega$, such that
$\hat{\Pi}_U = \int_U da \hat{\Pi}(a)$.

The most general probability assignment for measurement outcomes
in $\Omega$ ,compatible with the rules of quantum mechanics is
\begin{eqnarray}
p(U) = Tr(\hat{\rho}_0 \hat{\Pi}_U), \label{POVM}
\end{eqnarray}
in terms of a density matrix $\hat{\rho}_0$ that incorporates the
initial state and preparation of the system and some POVM
$\hat{\Pi}_U$. If a procedure provides probabilities for a
measurement outcome that cannot be brought into the form Eq.
(\ref{POVM}), then these probabilities are {\em incompatible} with quantum theory. They may approximate a valid quantum mechanical expression (for a special class of initial states), but they are
not themselves fundamental and cannot be taken as a standard for the
theoretical description.

\subsection{The quantum description of the time of arrival}

This section contains a brief review on the quantum treatment of the time of arrival, with an emphasis on its description in terms of POVMs. The key point relevant to particle oscillations is to demonstrate that the probabilities associated to the time of arrival, like those of any other      quantum observable, may  exhibit quantum interferences.

\subsubsection{Key issues in the quantum description of the time of arrival}
 An idealized description of a time-of-arrival measurement is the following. At time $t = 0$ a source emits
particles prepared at an initial state $\hat{\rho}_0$. At a
macroscopic distance $L$ from the source a particle detector is
located. An external clock determines the time $t$ coincident with
each detection event. Given a large number of events, a probability
distribution for the time of arrival is constructed. The issue is to determine this
probability distribution from first principles in quantum theory.

The difficulty in describing the time of arrival in a quantum system
stems from the fact that given a solution to Schr\"odinger's
equation $\psi(x,t)$, its modulus square $|\psi(x,t)|^2$ is a
probability density with respect to $x$ at each  time $t$, and {\em
not} a probability density with respect to $t$.  Moreover,
there is no operator
 representing time
in the system's Hilbert space.  In that case one would use the Born
rule to determine a probability density for the time of arrival. The existence of a time operator $\hat{T}$, conjugate to the
Hamiltonian (so that the Hamiltonian generates time-translations:
$e^{i\hat{H}s} \hat{T} e^{-i \hat{H}s} = \hat{T} + s$), is ruled out
by the requirement that the Hamiltonian is bounded from below
\cite{PaEn, UnWa89}. For some systems, one may still define a
quantum time variable, by choosing some degrees of freedom of the
system as defining an internal
`clock'.  However,  clock times fail to be conjugate to the
Hamiltonian of the system, with the result that they do not forward
under Hamiltonian evolution. In other words, quantum fluctuations
invariably force clocks to move occasionally `backwards in time'.

There are various approaches for the description of the time of arrival \cite{toarev, time_operators, MSP98}, and related problems, e.g., transversal time
in tunneling \cite{tunnel}. Some of the approaches go beyond the standard quantum mechanical
formalism by employing properties of `paths' identified
either through path integrals, or the Wigner function, or Bohmian
mechanics. In this
 paper we are interested in well-defined probabilities for the time of arrival in
 concrete measurement situations, which means the identification of a POVM $\hat{\Pi}(t)$ for the time of arrival \cite{AnSav, Kij74, BGL, BSPME00, HSMN}. We also note that the method we adopt in this paper for the construction of a POVM for the time of arrival shares many common features with the studies based on the decoherent histories approach
\cite{Haltoa1, HalTOA, HalYea}.

\subsubsection{Time-of-arrival interferences}

 Next, we present  Kijowski's POVM \cite{Kij74} for the time of arrival of a
non-relativistic particle of mass $M$ in one spatial dimension.
Given an initial state $|\psi_0\rangle$ at time $t = 0$,  Kijowski's
POVM assigns the following probabilities for the time of arrival of
the particle at a detector located at $x = 0$

\begin{eqnarray}
p(t) = |\int_0^{\infty} dp \left(\frac{p}{ 2\pi m}
\right)^{1/2} e^{-ip^2t/2M} \psi(p)|^2
+ |\int^0_{-\infty} dp \left(\frac{-p}{ 2\pi m} \right)^{1/2}
e^{-ip^2t/2M} \psi(p)|^2 \label{Kij}.
\end{eqnarray}

The POVM Eq. (\ref{Kij}) is uniquely constructed by the requirements of
Galilean invariance, of invariance under parity and time-reversal transformations, and of having the correct classical limit. It is
defined on wave functions with support on either positive or
negative values of momentum, and the values of $t$ extend to the
whole real axis. It can be extended to all states if we also include
 the alternative ${\cal N}$ of no detection \cite{AnSav},
as $p({\cal N}) = 1 - \int_{-\infty}^{\infty} p(t)$.

In order to demonstrate that the POVM Eq. (\ref{Kij}) may lead to interferences in the time of arrival, we evaluate Eq. (\ref{Kij}) for an initial superposition of two
Gaussian wave packets with the same mean position $-L$, but different mean
momentum $\bar{p}_1$ and $\bar{p}_2$
\begin{eqnarray}
\tilde{\psi}_0(p) = \left(\frac{a^2}{8\pi} \right)^{1/4} \left[ e^{-
\frac{a^2}{4} (p - \bar{p}_1)^2 + i p L} + e^{- \frac{a^2}{4} (p -
\bar{p}_2)^2 + i p L } \right]. \label{gsup}
\end{eqnarray}
We assume that $a |\bar{p}_1 - \bar{p}_2| >> 1$ so that the two wave
packets do not significantly overlap.

For $a \bar{p}_1 >> 1$, $a \bar{p}_2 >> 1$, and ignoring dispersion,
the leading contribution to the probability distribution obtained by
the POVM Eq. (\ref{Kij}) is

\begin{eqnarray}
p(t) \simeq \frac{1}{\sqrt{2 \pi} M a} \left[ \frac{\bar{p}_1}{2}
e^{ - \frac{2}{a^2} (L - \bar{p}_1 t/M)^2} + \frac{\bar{p}_2}{2} e^{
- \frac{2}{a^2}(L - \bar{p}_1 t/M)^2} \right. \nonumber
\\ \left. + \sqrt{\bar{p}_1 \bar{p}_2} \; e^{ - \frac{1}{a^2}[(L -
\bar{p}_1t/M)^2 + (L - \bar{p}_2 t/M)^2]} \cos\left((\bar{p}_1 -
\bar{p}_2) (L - \frac{\bar{p}_1 + \bar{p}_2}{2m}t)\right) \right],
\label{222}
\end{eqnarray}
i.e., there are oscillations of frequency $\omega = \frac{1}{2M}
|\bar{p}_1^2 - \bar{p}_2^2|$ extending for a time interval with
width of order $ \Delta t = L M |\hat{p}_1 - \hat{p}_2|/(\bar{p}_1
\bar{p}_2)$ around the mean value of arrival time $t_{cl} =
\frac{2LM}{\bar{p}_1 + \bar{p}_2}$. If the momentum difference $\Delta p =
|\bar{p}_1 - \bar{p}_2|$  is much
smaller than the mean momentum $\bar{p} = \frac{1}{2} (\bar{p}_1 +
\bar{p}_2)$, then $\omega = \bar{p}
\Delta p /M$ and $\Delta t = ML\Delta p/\bar{p}^2$ . Clearly $\Delta
t/t_{cl} = \Delta p /\bar{p} << 1 $, but this has nothing to do with
the measurability of the interferences in the time of arrival. The interference are observable if $\Delta t >> \tau$, where $\tau$ is the temporal
resolution of the detector. For  measurements of the time of arrival, see Ref.  \cite{toarev} and references therein.


The discussion above demonstrates that there exist regimes in which
the semiclassical description of the time of arrival is inadequate. A common theme in the
study of particle oscillations is to identify the relevant
probabilities by substituting into the square amplitude Eq.
(\ref{ampl0}) the classical value of time of arrival at the detector.
The meaning of such an expression apart, this procedure
may gravely misrepresent the probabilities of the system. Indeed,
for the initial superposition state Eq. (\ref{gsup})
substituting, for example, $ t = t_{cl}$ into the square amplitude $|\psi(t, x)|^2$, leads
invariably to an expression that does not exhibit an interference
pattern.


\section{Critique of existing theoretical treatments of particle oscillations}

In this section, we undertake a more detailed critique of existing approaches to particle oscillations.
  First, we present the existing approaches for the treatment of particle oscillations, separated into three broad categories as in the reviews \cite{Zra, Beuth, Giunti07, Akhm}, namely, plane wave methods, wave packet methods and quantum field theoretic methods. We then proceed to a critique of their assumptions and formalism,  elaborating on the issues  sketched in  Sec. 2.

\subsection{A brief presentation of existing formalisms for the description of particle oscillations}

\subsubsection{The plane wave description of particle oscillations}
We present the elementary treatment of particle oscillations based on plane waves, which is the most well-known and widely used due to its simplicity. This approach is known to be fundamentally inconsistent \cite{Kay81, Beuth}, as it relies on infinitely extended wave packets with zero momentum spread in each mass eigenspace. The wave packet methods, presented in Sec. 4.1.2, and the quantum field theory methods, presented in Sec. 4.1.3, were developed partly in order to resolve outstanding issues in the plane wave treatment and to place the derivation of the oscillation formulas on a firmer mathematical ground.

By definition, an oscillating particle is described in terms of a Hilbert space
 ${\cal H} = \oplus_i{\cal H}_i$, where the subspaces ${\cal H}_i$ carry irreducible unitary representation of the Poincar\'e group with the same spin but different masses $m_i$. Ignoring for simplicity the spin degrees of freedom and particle decay, each mass eigenspace ${\cal H}_i$ is spanned by the generalized eigenstates $|{\bf p}, i \rangle$ of the Hamiltonian $ \hat{H}_i = \sqrt{{\bf \hat{p}}^2 + m_i^2}$ for a free particle of mass $m_i$. One defines `flavor eigenstates'
\begin{eqnarray}
|{\bf p}, \alpha \rangle = \sum_i U^*_{\alpha i} | {\bf p}, i \rangle, \label{flavoreig}
\end{eqnarray}
where $U_{\alpha i}$ is the mixing matrix appearing in the interaction Hamiltonian for the oscillating particles. The transition amplitude between a generalized eigenstate of flavor $\alpha$ ,at time $t = 0$ and location ${\bf x} = 0 $, and a generalized  eigenstate of flavor $\beta$, at time $t$ and location ${\bf x} = {\bf L}$, is
\begin{eqnarray}
{\cal A}_{\alpha \rightarrow \beta} (t, {\bf L}) = \langle {\bf p}, \beta|e^{- i \hat{H} t + i \hat{p}\cdot {\bf L}}|{\bf p}, \alpha \rangle = e^{i {\bf p} \cdot L} \sum_i e^{ - i E_i t} U^*_{\alpha i} U_{\beta i} \langle {\bf p}, i |{\bf p}, i \rangle,
\end{eqnarray}
where $E_i = \sqrt{{\bf p}^2 + m_i^2}$. The term $\langle {\bf p}, i |{\bf p}, i \rangle$ is really infinite, but in the plane wave method we are interested primarily in the relative phases between the mass eigenstates, so this issue can be ignored.

The squared amplitude
\begin{eqnarray}
|{\cal A}_{\alpha \rightarrow \beta} (t, {\bf L}) |^2 = \sum_{ij} U^*_{\alpha i} U_{\beta i} U_{\alpha j} U^*_{\beta j}  \langle {\bf p}, i |{\bf p}, i \rangle \langle {\bf p}, j |{\bf p}, j \rangle e^{ - i E_i t + iE_j t} \label{sqamp}
\end{eqnarray}
depends explicitly only on $t$. It would also carry a dependence on ${\bf L}$ if we had assumed different values of ${\bf p}_i$ in the definition of flavor eigenstates, Eq. (\ref{flavoreig}), but this is not necessary for the present demonstration.  In order to relate the square amplitude Eq. (\ref{sqamp}) with a measurable quantity, it is necessary to convert its time dependence into  spatial dependence. There is no rule of quantum mechanics that allows one to do so, so one has to rely on {\em ad hoc} prescriptions. To this end, we note that  each subspace is characterized by a velocity $v_i = p/E_i$, where $p = |{\bf p}|$. At the classical level, this corresponds to the  time of arrival $t_i = L/v_i$, for a particle  starting at ${\bf x} = 0$ to a detector at ${\bf x} = {\bf L}$, where $L = |{\bf L}|$.

The standard oscillation phase is obtained by substituting the time parameter $t$ with a mean  time of arrival $\bar{t}$ at the detector, e.g., for the time of arrival calculated for a mean energy $\bar{E}$,---this procedure is usually called an `equal times prescription'. We then obtain a phase difference
\begin{eqnarray}
\phi_i - \phi_j = (E_i  - E_j) \bar{t} = (E_i - E_j) L \bar{E}/p,
\end{eqnarray}
which renders the square amplitude Eq. (\ref{sqamp}) an oscillating function of $L$, from which  the standard oscillation wave-number
$k_{ij} = \frac{\bar{E}}{p} |E_i - E_j|$ follows.
In the ultra-relativistic regime, relevant to neutrinos, the oscillation wave-number is
\begin{eqnarray}
k_{ij} = \frac{|m_i^2 - m_j^2|}{2 p}. \label{standardur}
\end{eqnarray}

On the other hand, if we substitute the different values $t_i = L/E_i$ for arrival time in each mass eigenspace (unequal-times prescription), one obtains
$
\phi_i - \phi_j = E_i t_i - E_j t_j = (E_i^2 - E_j^2) L/p = (m_i^2 - m_j^2) L/p$,
corresponding to an oscillation wavenumber
\begin{eqnarray}
k_{ij} = \frac{|m_i^2 - m_j^2|}{ p}, \label{nonstandard}
\end{eqnarray}
which is twice the standard expression Eq. (\ref{standardur}) in the ultra-relativistic limit.

We note that the non-standard result Eq. (\ref{nonstandard}) includes  the contribution from interferences in the time of arrival corresponding to the different values $t_i$ in the different mass eigenstates. The derivation of the standard result Eq. (\ref{standardur}) involves an  assumption that such interferences are somehow suppressed.

\subsubsection{Wave-packet methods}

The natural generalization of the plane wave method is the consideration of amplitudes between properly square-integrable wave-packets \cite{Kay81, Kim1, KG98,  Nau, Zra}. That is, one defines Hilbert space vectors
\begin{eqnarray}
|\Psi^P(0), \alpha \rangle = \sum_i U^*_{\alpha i} \Psi^P_i({\bf x}) |{\bf x}, i \rangle, \hspace{2cm}
|\Psi^D({\bf L}) \beta \rangle = \sum_i U^*_{\alpha i} \Psi^D_i({\bf x}) |{\bf x}, i \rangle,
\end{eqnarray}
in terms of profile functions $\Psi^P$ and $\Psi^D$ centered around the  production ( ${\bf x} = 0$) and detection (${\bf x} = {\bf L}$) region respectively; $|{\bf x}, i \rangle$ stand for the generalized eigenstates of position in the $i$-th mass eigenspace. One then computes the amplitude
\begin{eqnarray}
{\cal A}_{\alpha \rightarrow \beta}(t, {\bf L}) = \langle \Psi^D({\bf L}), \beta|e^{-i \hat{H}t} | \Psi^P(0), \alpha \rangle. \label{amplwp}
\end{eqnarray}

The simplest case that allows for analytic calculations corresponds to isotropic Gaussian wave-packets
\begin{eqnarray}
\Psi_i^P ({\bf x}) = \frac{1}{(2 \pi \sigma_P)^{\frac{1}{4}}} e^{ - \frac{|{\bf x}|^2}{4 \sigma_P^2} + i p_i x}, \hspace{1.5cm}
\Psi_i^D ({\bf x}) = \frac{1}{(2 \pi \sigma_D)^{\frac{1}{4}}} e^{ - \frac{|{\bf x} - {\bf L}|^2}{4 \sigma_D^2} + i p_i x},
\end{eqnarray}
where one assumes that the production region is centered around ${\bf x} = 0$ with spread equal to $\sigma_P$ and the detection region is centered around ${\bf x} = {\bf L}$ with spread equal to $\sigma_D$.

The amplitude Eq. (\ref{amplwp}) can be computed in the no-dispersion regime, defined by the approximation $E_i (p) \simeq E_i({\bf p}_i) + {\bf v}_i \cdot ({\bf p} - {\bf p}_i) $. Restricting for simplicity to one spatial dimension, one finds
\begin{eqnarray}
{\cal A}_{\alpha \rightarrow \beta}(t,  L) = \sqrt{\frac{2 \sigma_P \sigma_D}{\sigma^2}} \sum_i U_{\beta i}U^*_{\alpha i} e^{- i E_i t + i p_i L - \frac{( L -  v_i t)^2}{4 \sigma^2}}, \label{amplwp2}
\end{eqnarray}
where $\sigma^2 = \sigma_D^2 + \sigma_P^2$.

In order to express the probabilities in terms of observable quantities alone, one needs to remove the time-dependence from the amplitude in Eq. (\ref{amplwp2}). This can be done in two ways.
\begin{enumerate}{}
\item The interference phases in the modulus square of the amplitude Eq. (\ref{amplwp2}) 
\begin{eqnarray}
\phi_i - \phi_j =  (E_i  - E_j )t -  L(p_i - p_j). \label{phiij}
\end{eqnarray}
are the same with the ones obtained from the plane wave methods. They can, therefore, be evaluated with the same methods, i.e., by making a  specific choice of time $t$ corresponding to a `mean' time of arrival.
The standard expression for the oscillation wavelength necessitates the postulate of an additional  prescription for the terms in Eq. (\ref{phiij}).
The most usual form of such prescription is the assumption of equal-times $t_i = t_j$, i.e. an assumption that interference takes place only for equal propagation times \cite{equaltime}. Alternatively, the standard oscillation formula may be obtained by an equal energy assumption $E_i = E_j$ \cite{ Lip97, GL97},  by an equal velocity assumption $v_i = v_j$ \cite{equalvel}, or by a specific choice of `mean velocity' for the wave packet \cite{BLSG, Lip05,  Levy09}.

\item The introduction of {\em ad hoc} prescriptions is avoided in the wave-packet method by integrating the squared amplitude $|{\cal A}_{\alpha \rightarrow \beta}(t,  L) |^2$ over time \cite{Kim1}, i.e., by defining a probability for particle detection
\begin{eqnarray}
p_{\alpha \rightarrow \beta}(L) = C \int_0^T dt |{\cal A}_{\alpha \rightarrow \beta}(t,  L) |^2, \label{ta0}
\end{eqnarray}
in terms of some long integration time $T$. The constant $C$ is introduced for purposes of normalization of the probability to unity.
\end{enumerate}

\subsubsection{Quantum field theoretic methods}

The oscillating particles are not directly observable, therefore, the natural framework for the  treatment of particle oscillations is quantum field theory \cite{GKLL93,  GriSto, GKL98, KW, IP99, Sachs, Rich, GS99,  Stan98,  Gi02, Cardall, Beuth2,  Beuth, Akhm}. In the latter, oscillating particles are described to internal lines of Feynman diagrams,  whose external lines  correspond to incoming particles in the production region and outcoming particles at the detector.

In this approach, one computes amplitudes of the form
\begin{eqnarray}
\langle \Psi^D, t_D| {\cal T}e^{- i \int d^4 x {\cal H}_I}    | \Psi^P, t_P \rangle, \label{Smat}
\end{eqnarray}
where $|\Psi^P, t_P\rangle$ describes the state of the incoming particles in the production region, $|\Psi^D, t_D \rangle$ describes the state of the particles produced in the detection process, ${\cal T}$ denotes time-ordering and ${\cal H}_I$ is the interaction Hamiltonian for the total process. Typically, $|\Psi^P, t_P \rangle$ correspond to wave-packets localized at the production region at a specific time $t_P$ and $|\Psi_D, t_D \rangle$ corresponds to wave-packets localized at the detector region at a time $t_D$.
The calculation of the amplitude Eq. (\ref{Smat}) to leading order in perturbation theory introduces  the characteristic dependence  of the amplitude on the mixing matrix $U_{\alpha i}$ in Eq. (\ref{amplwp2}). Usually, one constructs the relevant probabilities from an integration of the amplitudes Eq. (\ref{Smat}) squared over time $T = t_D - t_P$.

 Ref. \cite{GriSto} avoids the use of integration over time in squared amplitudes by following a different procedure. One computes the $S$-matrix elements for the weak process $n \rightarrow p + e^- + \bar{\nu}$, for the production of the antineutrino localized around ${\bf x} = 0$ and the subsequent scattering process $\bar{\nu} + e^- \rightarrow \bar{\nu} + e^-$ localized around ${\bf x} = {\bf L}$ for the antineutrino detection without reference to an initial and final time. The $S$-matrix formalism in this case implies that the initial and final states are `in' and `out' respectively, i.e., they are defined at $t \rightarrow - \infty$ and $t \rightarrow + \infty$. The authors then construct the scattering cross-section at the limit $L \rightarrow \infty$ recovering the standard result. Ref. \cite{Rich} avoids the $S$-matrix formalism, relying instead on  time-dependent perturbation theory.

The quantum field theoretic method provides a substantial improvement in the theoretical treatment of particle oscillation, because one needs not introduce ill-defined flavor eigenstates and also because it allows for a more elaborate treatment of the production and detection of oscillating particles---for a comparison of the methods, see Ref. \cite{Akhm}. A quantum field theoretic treatment is essential to a first-principles treatment of the problem and we shall employ it in Sec. 5.2.

However, there is little difference at the abstract level of formalism between the quantum field theoretic and the wave packet methods. Both make the same use of vector states and evolution operators in the definition of probabilities, essentially presupposing a description in terms of von Neumann measurements. This elementary point may be obscured by the $S$ matrix language commonly employed in quantum field theory,
so we elaborate here on the meaning of the amplitude Eq. (\ref{Smat}). In quantum theory, the dependence of a vector state on time arises from unitary time evolution. The vector states $|\Psi_{D,P}, t_{D,P}\rangle$ are then states of the form $e^{-i \hat{h}t} |\Psi'_{D, P} \rangle$ for some vector states $|\Psi'_{D, P} \rangle$ and a Hamiltonian $\hat{h}$. The choices for  $|\Psi_{D,P}, t_{D,P}\rangle$ in the literature imply an identification of $\hat{h}$ with the Hamiltonian $\hat{H}_0$ of free particles. Then $|\Psi'_{D, P} \rangle$ corresponds to single-particle states for outcoming and incoming particles, centered in the detection and production region respectively. The amplitude Eq. (\ref{Smat}) is then equivalent to
\begin{eqnarray}
\langle \Psi'_D|e^{ -i \hat{H}_0 t_D} {\cal T}e^{- i \int dt   H_I} e^{i \hat{H}_0t_P}|\Psi'_P \rangle, \label{Sm2}
\end{eqnarray}
where $H_I = \int d^3x {\cal H}_I$.

 Eq. (\ref{Sm2}) is a meaningful quantum expression only if the integration in the time-ordered exponential is for the time interval $[t_P, t_D]$, in which case, Eq. (\ref{Sm2}) equals $ \langle \Psi'_D| e^{-i (\hat{H}_0 + \hat{H}_I)(t_D - t_P)}| \Psi'_P \rangle$. The squared amplitude then provides the usual expression for probabilities in  von Neumann measurements. Hence, as far as the description of measurements is concerned, both wave packet and quantum field theoretic methods employ the same assumptions.

\subsection{Critical analysis of the existing formalisms for particle oscillations}

\subsubsection{Justification in terms of first principles.}

The first point of critique towards the particle-oscillations methods described above is that the construction of probabilities does not follow from the first principles of quantum theory. The vast majority of these methods identifies the relevant probabilities with the modulus square of amplitudes, such as Eq. (\ref{amplwp}). However, this expression is relevant only for measurements that take place at a pre-specified  moment of time (or time-interval)---see, Sec. 3.1. In particle oscillations,   the experimentally fixed quantity is the location of the detector, {\em not the time} that a particle is detected. The time of detection is a variable rather than an external parameter. This inconsistency in the definition of probabilities is the reason why, in order to obtain a connection with measurable quantities, one needs to remove the time dependence from
the square amplitudes---either by introducing additional prescriptions or by time integration.

The construction of probabilities using squared amplitudes is the most common practice in applications of quantum theory, but it is neither necessary nor unique. The most general probability assignments in quantum theory correspond to POVMs (see, Sec. 3.1.3), and these can be applied to cases where the simple definition in terms of squared amplitudes fails. We maintain that the correct probability assignment for each particular case should be constructed from first principles, in order to reflect precisely the physics of the experiment.
The use of a probability assignment without prior justification  may lead to inconsistencies in the correspondence of elements of the formalism to experimentally determined variables.
Hence, the modeling of particle oscillation experiments by von Neumann measurements is problematic, because the setup of the former violates a main assumption in the latter, namely, that measurements take place at a pre-determined instant of time.


\medskip

 Even should one attempt to model particle oscillations by von Neumann measurements, amplitudes such as  Eq. (\ref{amplwp}) or Eq. (\ref{Smat}) are insufficient: they ignore the effect  of the detector, which is a macroscopic system with a large number of degrees of freedom. To see this, we note that the squared amplitude $|{\cal A}_{\alpha \rightarrow \beta}(t, L)|^2$ can be  brought in the standard form $Tr(\hat{\rho} \hat{\Pi}_L)$ by identifying the state $\hat{\rho} = e^{-i \hat{H}t} |\Psi^P(0)\rangle \langle \Psi^P(0)| e^{i \hat{H}t}$ and the projector $\hat{\Pi}_L = | \Psi^D(L)\rangle \langle \Psi^D(L)|$, that describes the measurement scheme. The projector $\hat{\Pi}_L$ is one-dimensional ($Tr \hat{\Pi}_L = 1 $), i.e., it corresponds to an extremely fine measurement that determines a phase space volume for the particle of size $\hbar$ \cite{Dav, Omn1}. One-dimensional projectors may be relevant for sharp measurements (observables with discrete spectrum), but they are unsuitable for measurements of continuous variables. As explained in Sec. 3.1,  the interaction with the measuring apparatus necessitates the use of much coarser projectors ($Tr \hat{\Pi} >> 1$) for continuous variables. Hence, the use of the squared amplitude $|{\cal A}_{\alpha \rightarrow \beta}(t, L)|^2$ for the probabilities misrepresents the coarse-graining inherent in the detection process.

 For example, a projector describing an unsharp measurement of position is
\begin{eqnarray}
\hat{\Pi}_L = \int dy f(y - L) | \Psi^D(y)\rangle \langle \Psi^D(y)|,
\end{eqnarray}
where $f(y)$ is a positive function centered around $y = 0$ with a spread of order $\delta L$ \cite{comment}. The corresponding probabilities  are
\begin{eqnarray}
Tr(\hat{\rho} \hat{\Pi}_L)  = \int dy f(y - L) |{\cal A}_{\alpha \rightarrow \beta}(t, y)|^2. \label{trrp}
\end{eqnarray}
Such modifications in the derivation of probabilities would not  affect the phases relevant to the oscillations (as long as $\delta L << L$), but it would make a difference in the expression for  the so called `coherence length' \cite{Kay81,   Kim1, Nus76,  Cardall, KNW} i.e., a scale $L_{coh}$ such that, if $k_{ij} L_{coh} >> 1$ no oscillations are observable.

\subsubsection{Times of arrival and the different prescriptions}

In Sec. 4.2.1, we argued that the definition of probabilities from the squared amplitude $|{\cal A}_{\alpha \rightarrow  \beta}(t, L)|^2$ is not suitable for particle oscillation experiments. Its use presents an immediate problem of reconciling the theoretical description with the experimental procedure. In particular, it is necessary to remove the time dependence from the squared amplitudes  $|{\cal A}_{\alpha \rightarrow  \beta}(t, L)|^2$.
  In this section, we examine the most common procedure employed for this purpose, namely, the substitution of the time parameter $t$ in the squared amplitude by the `mean' or classical time of arrival for a particle propagating from the source to the detector. The second common procedure of eliminating the time variable, by time-integration, is considered in Sec. 4.2.3.


The  substitution of a mean time of arrival in the squared amplitudes $|{\cal A}_{\alpha \rightarrow  \beta}(t, L)|^2$ has no fundamental justification. In fact, this procedure is inconsistent with quantum theory. The time of arrival depends on the initial state
(for example, its mean energy), and its substitution into the squared amplitudes $|{\cal A}_{\alpha \rightarrow  \beta}(t, L)|^2$ leads to a probability assignment that is not a linear functional of the system's density matrix, thus contradicting a fundamental requirement  for any probability assignment in quantum theory. In particular, as we explain in more detail in Sec. 4.2.3, a probability assignment that does not respect linearity does not have a consistent statistical interpretation.

  The lack of a solid mathematical justification for the aforementioned procedure implies that there is no independent theoretical criterion to settle any ambiguities that may arise in its application. This is a problem, because, as we saw in Sec. 4.1.1, a variation of this procedure that is equally plausible {\em a priori}---the so-called   unequal-times prescription---leads to a prediction for the particle oscillation wavelength that differs  from the standard one by a factor of two. Hence, a significant part of the related bibliography is devoted in showing that the unequal-times prescription cannot be valid, aiming to strengthen the standard result by the exclusion of its alternative. 
  
   In what follows, we will show that  (i) that the physical arguments developed against the non-standard oscillation formulas do not apply to our result which is obtained by a different method, (ii) the standard derivations ignore the probabilistic and quantum aspects of the time of arrival, and (iii)   the timescales involved in particle oscillation experiments suggest that interferences in the time of arrival contribute to the oscillation wavelength thus leading to a non-standard particle oscillation formula.


\medskip

  A strong argument against the unequal-times prescription is that it leads to an unphysical  prediction of oscillations for recoil particles; for instance, the muon in the reaction $\pi \rightarrow \mu \nu$ or the $\Lambda$ in the reaction $\pi^-p \rightarrow \Lambda K^0$. This is highly counterintuitive, and it is generally considered as a severe deficiency of the method. The formalism we develop here leads to a non-standard oscillation formula, but {\em it does not  predict oscillations for recoil particles}---see Sec. 5.3. A careful description of the measurement process, in terms of particles that are actually being detected, shows that oscillations are only possible for particles with different mass eigenstates.  Therefore, we maintain that the oscillations for recoil particles is the artefact of an improper mathematical framework for the evaluation of quantum probabilities. It is not a consequence of the general statement  that  different values of the time of arrival contribute coherently into the particle oscillation probability.

\medskip

Common arguments against the unequal prescription involve statements  that quantum interference only occurs between states defined at the same spacetime point  \cite{equaltime, Lip97, Lip2}, or that measurements take place in space and not in time \cite{GL97}. However, the restrictions implied by such statements do not follow from the basic principles of quantum theory. Any observable is recorded by pointer variables on a measuring apparatus, and interferences are observed through the distribution of values of these variables in a large number of runs. Nothing in quantum theory forbids a pointer variable from becoming correlated with properties of the measured system that correspond to different instants of time. As shown in Sec. 3.2.2, the time of arrival may exhibit interferences, just as any other quantum observable. In general, we maintain that any statements about what cannot be measured in quantum theory can be made only on the basis of a precise modeling of the measurement process.

\medskip

In Sec. 2.1, we stated that a key problem in existing approaches to particle oscillations is that they conflate the time-parameter of Schr\"odinger's equation with the time of arrival, which is a physical observable. In the present context, this problem is manifested by the {\em ad hoc} substitution of the time parameter $t$---arising from the solution of Schr\"odinger's equation---in the squared amplitudes $|{\cal A}_{\alpha \rightarrow  \beta}(t, L)|^2$ by a `mean' time of arrival. This procedure ignores the fact that the time of arrival is a
 genuine quantum observable. Indeed, the time of arrival is determined by the  coincidence of a microscopic {\em quantum} event (for example, a particle reaction in the detector)
   with the reading of an external clock. While the {\em value} of the time of arrival is a classical variable, it is correlated to a microscopic process that is governed by the rules of quantum theory. Consequently, the values of the time of arrival exhibit fluctuations, and these must be described by quantum probabilities.
 The substitution of a mean value for the time of arrival in the squared amplitudes $|{\cal A}_{\alpha \rightarrow  \beta}(t, L)|^2$  ignores
all effects pertaining to quantum fluctuations; in particular, it excludes {\em a priori}  the possibility of quantum interference in the values of the time of arrival.

Acknowledging the time of arrival as a quantum observable, brings new light in the distinction between the standard oscillation formula and the non-standard one following from the unequal-times prescription. The former essentially assumes that quantum interferences in the time of arrival are either suppressed or negligible; the latter implies that such interferences persist and they contribute significantly to the particle oscillation probability. As with any other quantum variable, the issue whether such interferences persist or not can be resolved only through a careful study of the physical situation at hand, namely, of the quantum system and the measuring apparatus. To this end, a {\em fully quantum description} of the time of arrival is necessary. This, we undertake in Sec. 5.

\medskip

An additional problem in the use of a `mean'  time of arrival in the squared amplitudes $|{\cal A}_{\alpha \rightarrow  \beta}(t, L)|^2$ is that the very notion of a `mean' time of arrival is ambiguous in absence of a full quantum  treatment. For example, in Refs. \cite{BLSG, Lip05, Levy09} the standard expression for particle oscillation wavelength is obtained by substituting a mean
 time of arrival $\bar{t} = L/\bar{v}$ in Eq. (\ref{sqamp}), where $L$ is the source-detector distance, and
 the mean velocity $\bar{v}$ of the wave-packet is defined as
\begin{eqnarray}
\bar{v} = \frac{p_1 + p_2}{E_1 + E_2}, \label{meanv}
 \end{eqnarray}
where the indices $1$ and $2$  refer to two mass eigenspaces.
Eq. (\ref{meanv}) has no fundamental justification in quantum theory, where the only notion of a mean velocity is the expectation value of the operator $\hat{p}\hat{H}^{-1}$. There is no reason why the mean velocity should not  be defined by any other expression, such as the
 the arithmetic or the  geometric mean of the velocities $v_i = p_i/E_i$.  A justification for Eq. (\ref{meanv}) is that it guarantees that no oscillations for recoil particles are predicted, without making an equal-times assumption. If, however, oscillations of recoil particles is an artefact of a mathematically ambiguous  formalism, as our results in Sec. 5 indicate, this argument loses much of its power.


\medskip

    In our opinion, one of the most important arguments   in support of prescriptions for the standard oscillation formula (it is also quoted as definitive in the Review of Particle Physics \cite{PDG})
 is essentially the following \cite{Lip05}.
 Consider Eq. (\ref{phiij}) for the oscillation phases for two mass eigenspaces, $i,j = 1, 2$. Substitute for $\bar{t}$ the expression
 \begin{eqnarray}
 \bar{t} = \frac{L}{\bar{v} } + \delta t, \label{lip0}
 \end{eqnarray}
  where $\bar{v}$ is given by Eq. (\ref{meanv}) and $\delta t << L/\bar{v}$ is ``a correction to describe any discrepancies, including quantum fluctuations". Then,
  \begin{eqnarray}
  \phi_1 - \phi_2 = \frac{m_2^2-m_1^2}{p_1 + p_2}L - \frac{E_1^2 - E_2^2}{E_1 + E_2} \delta t. \label{lip}
  \end{eqnarray}
  The first term in the right-hand-side of Eq. (\ref{lip}) is the standard oscillation phase, while the second term is a correction much smaller than the first term. The correction $\delta t$ is deemed to be also much larger than the time-scale $\delta t_{12} = |L/v_1 - L/v_2|$, that corresponds to the different centers
 of the wave-packets in the different mass eigenspaces. The term $\delta t_{12}$ is responsible for the different prediction of the unequal-times prescription. One then  concludes that the effect of $\delta t_{12}$ in the oscillation phase is negligible and thus the standard oscillation formula holds.

  We note that the introduction of the correction $\delta t$   in the phase difference is essential for the argument to work. However, if $\delta t$ corresponds to the fluctuations of the time of arrival, its introduction into the phase difference is not justified. For the argument's sake, let us assume  that the parameter
   $t$ in the amplitude  ${\cal A}_{a \rightarrow \beta}(t, L)$ can indeed be substituted by the time of arrival. Let $p(t)$ be the probability distribution describing any fluctuations of the time of arrival---$\delta t$ corresponds to the mean deviation of this distribution. According to standard probability theory, the incorporation of these fluctuations involves weighting the squared amplitudes $|{\cal A}_{a \rightarrow \beta}(t, L)|^2$ with $p(t)$, i.e., one should consider the integral
   \begin{eqnarray}
   \int dt p(t) |{\cal A}_{a \rightarrow \beta}(t, L)|^2. \label{intampl}
   \end{eqnarray}
   Then, the integration Eq. (\ref{intampl}) does not lead to the introduction of the mean deviation $\delta t$ into the
   phase difference $\phi_1 - \phi_2$. For example, if $p(t)$ is a Gaussian $\sim exp[-(t - L/\bar {v})^2/(2 \delta t^2)]$, the integration in Eq. (\ref{intampl}) would only lead to a suppression of the oscillation's amplitude. Hence, the fact that $\delta t << L/\bar{v}$ or that $\delta t >> \delta t_{12}$ has nothing to do with the relative size of terms in the oscillation phase, and, hence, with the oscillation wavelength.

   Moreover, the specification of a single parameter $\delta t$ for the description of the fluctuations in the time of arrival is insufficient if these fluctuations are  quantum. A single parameter  does not allow for distinction between incoherent and coherent fluctuations because these have very different physical behavior\footnote{For example, a Gaussian wave function $exp(-x^2/L^2)$ has approximately the same spread $L$ with the superposition $exp(-x^2/\sigma^2) + exp(-(x - L)^2/\sigma^2)$, if $\sigma << L$, but their physical properties are very different.}. The former tend to {\em suppress}  oscillations while the latter would contribute additional terms to the oscillation phase.  Ref. \cite{Lip05} emphasizes that the understanding of  time fluctuations is fundamental for the proper theoretical description  of particle oscillations. We strongly agree on this point. Furthermore, we maintain that such a treatment should follow precise probabilistic reasoning and it should take into account the quantum nature of the time of arrival.

\medskip

The treatment of the time of arrival as a physical observable
has important implications for the description of particle oscillations. A particle-detection event takes place at a specific locus and a specific moment of time. In the context of particle oscillations the detection time corresponds to the time of arrival. Since the latter is a quantum observable, the detection process corresponds to a   {\em joint} measurement of time of arrival and position, which can be described in terms of a suitable POVM---see, Sec. 3.1.3. For a {\em joint} measurement, the separation of the timescales $\delta t$ and $\delta t_{12}$, noted in Ref. \cite{Lip05}, leads to the opposite conclusion from the one reached in that reference.
If the sampling of one of the two variables in a joint measurement is much coarser than the relevant interference scales in the quantum state, then the measurement of this variable does not affect the probability distribution for the other variable. In other words, the fact that the timescale $\delta t$
and $\delta t_{12}$, defined earlier, satisfy $\delta t >> \delta t_{12}$, implies that the measurement of the time of arrival does not disturb the contribution of the time-of-arrival interferences in the oscillation probability.

 To see this, one may consider the two-slit experiment.
  Let $\delta L$ be the distance between the two slits. Consider now a special case of a joint measurement: first, a  recording of the position of a particle as it passes through the slits and its
later position on a screen.
 If the   accuracy $d$  of the first measurement satisfies $d >> \delta L$, the slit through which the particle crossed is undetermined, so the interference pattern in the second measurement will persist. Only for a finer-grained  first measurement $d \sim \delta L$ are the  interferences suppressed\footnote{This property is seen most clearly in the so-called ``which-way" experiments \cite{whichway}, where a microwave cavity is used in order to perform a measurement of position as the particle crosses through the slits: the state of the electromagnetic field in the cavity determines the accuracy of this measurement.}. The key point  in this version of the two-slit experiment, is that we also have a {\em joint measurement of two observables}. A coarse measurement of one observable does not affect the coherence in the measurement of the second observable.

\subsubsection{Integration of probabilities over time}

In Sec. 4.2.2, we examined the procedure of removing the time-dependence from the squared amplitude $|{\cal A}_{\alpha \rightarrow \beta}(t, L)|^2$ by substitution with a `mean' value for the time of arrival. In this section we consider the alternative procedure of integrating the squared amplitude $|{\cal A}_{\alpha \rightarrow \beta}(t, L)|^2$ over time, as in Eq.
 (\ref{ta0}). This procedure is more common in the recent works on particle oscillations, and it is often applied in a field theory setting. The arguments presented here do not depend on the way such amplitudes are calculated, but follow from basic properties of the formalism of quantum theory.

  From a mathematical point of view,  $|{\cal A}_{\alpha \rightarrow \beta}(t, L)|^2$ is not a density with respect to $t$, so integration with respect to $t$ is ill-defined. This problem arises because $t$ in $|{\cal A}_{\alpha \rightarrow \beta}(t, L)|^2$ is an {\em external parameter} and not a random variable (i.e, an observable). The integration over parameters (rather than over random variables) does not define proper probability densities, neither in classical nor in quantum probability theory.
 The reason for this is that the probabilities for different alternatives can be added {\em only if the alternatives  are disjoint}. If
the alternatives are not disjoint, the joint probability is not obtained by the sum of the individual probabilities\footnote{For  two overlapping alternatives $U_1$ and $U_2$ and  probability distribution $p(\cdot)$, the joint probability $p(U_1 \cup U_2) = p(U_1) + p(U_2) - p(U_1 \cap U_2)$.}.

An elementary example from ordinary probability theory serves to illustrate that events distinguished only by the value of the time parameter $t$ fail to be independent. Let us consider a variable $X$ taking values in some set $\Gamma$ and consider probabilities defined on paths $X(\cdot)$ of such variables labeled by the time parameter $t$.
  The event (alternative) ``$X = x$ at time $t_1$" is not disjoint to the
  event that ``$X = x$ at time $t_2$". There are paths that satisfy {\em both} propositions above, for example a
path that satisfies $X = x$ at {\em all} times $t$. One cannot
therefore obtain the joint probability for the event ``$X = x$ at $t_1$ {\em or} $X = x$ at time $t_2$" by summing the
individual probabilities.

The same also holds in quantum theory \cite{sumampl}. As a result, the integral in Eq. (\ref{ta0}) does not lead to well-defined probabilities. In particular, it does not preserve normalization: to normalize one has to divide by a constant $C$. In general, the normalization constant $C$ depends on the initial state of the system, so the probability assignment is a non-linear functional of the initial density matrix $\hat{\rho}$, a property that is incompatible with the rules of standard quantum theory\footnote{One might respond to our argument by stating that the normalized probabilities Eq. (\ref{ta0}) are in fact conditional probabilities at time $t$, provided that detection has taken place. But in this case one should prove that the normalization constant $C$ is proportional to the total detection probability, as predicted by quantum theory. This is problematic because $C$  has no analogue in standard expressions for detection probabilities. Moreover, the recourse to conditional probabilities does not remove the problem of defining unconditioned probabilities. In particular,  a proper normalization  of unconditioned probabilities is necessary for the description of experiments at which neutrino detection takes place at two different locations---see, for example, the proposal in Ref. \cite{Rub}.}.

Linearity with respect to the initial state is an essential property for any probability assignment. It guarantees that one can combine sub-ensembles through statistical mixing. This means the following: let  $\hat{\rho}_{\xi}$ be the density matrix corresponding to a sub-ensemble of events distinguished by the macroscopically determined quantities $\xi$, and let $p_{\hat{\rho}_{\xi}}(A)$ be the probability distribution corresponding to an observable $A$---see the Appendix B for details. The ensemble constructed by weighting each sub-ensemble with a factor $w_{\xi}$ is described by the convex combination $\sum_{\xi} w _{\xi} \hat{\rho}_{\xi}$. Linearity guarantees that the probability for the variable $A$ in the total ensemble will also be a convex combination $\sum_{\xi} w_{\xi} p_{\hat{\rho}_{\xi}}(A)$. The lack of linearity in the probability assignment Eq.  (\ref{ta0}) implies the lack of a consistency in the combination of different subsets of data \cite{comment2}.

The requirement of linearity and positivity in the probability assignment is very restrictive. It implies that any probability assignment {\em must} correspond to a POVM. In the present context, the POVM should describe a joint measurement of position $L$ and arrival time $t$. Given such a POVM $\hat{\Pi}(t, L)$ we can integrate consistently over time of arrival in the probabilities $Tr[\hat{\rho} \hat{\Pi}(t, L)]$ obtaining a probability density that depends only on position
\begin{eqnarray}
p(L) = \int_0^T dt \; Tr[\hat{\rho} \hat{\Pi}(t, L)]. \label{povmpart}
\end{eqnarray}
It is only in the context of equations of the form Eq. (\ref{povmpart}) that integration of probabilities over time can be justified in accordance with the rules of quantum theory.

\medskip

The problems above would be resolved if one constructed amplitudes that depend on the time of arrival rather than the parameter $t$ of Schr\"odinger equation. However, neither in this case would Eq. (\ref{ta0}) be justified.
In quantum theory a summation of different alternatives at the level of probabilities is justified only when the alternatives are macroscopically distinct (for example, if they are distinguished by a measurement scheme). In absence of such distinction, the summation ought to take place at the level of amplitudes. This property is clearly manifested in the two-slit experiment. If we can determine macroscopically through which slit the particle passed, we sum over the corresponding probabilities. If we cannot, then we sum over amplitudes. In the latter case, we observe an interference pattern, in the former we do not. Hence, the assumption of an integral over time, at the level of probabilities, presupposes that we can distinguish temporal alternatives at a scale smaller than any interference terms in the system's state. This is not the case in particle oscillation experiments. As also discussed in Sec. 4.2.2, the temporal resolution of any particle detection corresponds to a time-scale $\tau_{res}$ which is at best of the order of $10^{-9}s$ for a solid state detector. On the other hand, the characteristic timescale of interferences $\delta t_{ij}$ is of the order of $10^{-20}s$, i.e., many order of magnitudes smaller than $\tau$.

Since the temporal resolution does not suffice to decohere the interferences, quantum theory would suggest an integration over different temporal alternatives at the level of amplitudes. Indeed, in this paper we describe the probabilities relevant to particle oscillations through a POVM $\hat{\Pi}(t, L)$, and we show (see Sec.5.2) that in the physically relevant regime, the resulting probabilities are proportional to
 $|\int_0^T dt {\cal A}_{\alpha \rightarrow \beta}(t, L)|^2$. A non-standard expression for the particle oscillation wavelength then follows.

Finally, we comment on a recent work Ref. \cite{Akhm}, which contains a very thorough treatment of the normalization in the time-integrated probabilities. In that work, a normalized probability distribution is obtained by integrating squared S-matrix  amplitudes with respect to both production time $t_P$ and detection time $t_D$. Here, one assumes that the particle flux is constant at both the production and the detection region. This methodology applies in the ultra-relativistic regime and for quasi-degenerate neutrino masses; in the general case, the oscillation probability is undefined.

 In our opinion, the assumption of a constant particle flux at detection is too strong. Such an assumption is reasonable at a highly coarse-grained, macroscopic time-scale  but there is no justification in its extrapolation to small time scales. In particular, this assumption excludes {\em a priori} possible variations of the flux resulting from
  different `propagation velocities' in different mass eigenspaces, i.e., the potential contribution of interferences in the time of arrival.

\subsubsection{The inapplicability of the S-matrix formalism}

In quantum field theoretic treatments of particle oscillation, one often applies the $S$-matrix formalism. The S-matrix  is well-suited for the construction of probabilities in scattering experiments. Its domain of validity, however, is not universal: it has been constructed for the description of ordinary scattering experiments in which the detector is far from the scattering region, the measured observables are the outcoming-particles' momenta, and one is not interested in the location and timing of the scattering event. In contrast, in particle oscillation experiments the main observable is the location and timing of the reaction through which the particle is detected.

To this end, we consider a simple model for the detection of particles produced from the scattering of a beam  into a target, using the formalism  of Sec. 3.1. In this case, the particle detector is
located at a macroscopic distance from the scattering region. Let
$t_0$ be an arbitrary time after the scattering has taken place, so
that after $t_0$ evolution is governed by the Hamiltonian
$\hat{H}_0$ of free particles.

 We assume that the measured observable $\hat{A}$ is
momentum (which commutes with the free particle Hamiltonian), or
indeed any other quantity such that $[\hat{A}, \hat{H}_0] = 0$.
Since the time-of-arrival is not a measurable quantity, the description
of measurement in term of a profile function $f(t)$, as in Sec. 3.1, suffices. We assume that the function $f(t)$ takes non-zero values only for $t
> t_0$, and that $t = T$ is an upper
limit to the time the particles spend in the detection region. It is
easy to show that, since $[\hat{A}, \hat{H}] = 0$,  Eq. (\ref{pr1}) becomes
\begin{eqnarray}
p(X) = Tr_{H_S}   \left( \hat{\rho}_{t_0}    w( X-  b
\hat{A})\right), \label{pr3}
\end{eqnarray}
i.e.,  the values of $\hat{A}$ are correlated with the values of the
rescaled pointer observable $X/b$, where $b = \int dt f(t)$.
 The density matrix $\hat{\rho}_{t_0}$ in Eq. (\ref{pr3})
arises from the evolution of the initially prepared state under the
full Hamiltonian $\hat{H}$ including interaction terms:
$\hat{\rho}_{t_0} = e^{- i \hat{H}t_0} \hat{\rho}_0 e^{i \hat{H}
t_0}$. The key point in Eq. (\ref{pr3}) is
that there is no dependence of the probabilities on an instant of measurement or on the
duration of the measurement. Since the observable $\hat{A}$ commutes
with the Hamiltonian, the measurement records the distribution of
$\hat{A}$, contained in the state $\hat{\rho}_{t_0}$, after
scattering. Since $t_0$ is itself arbitrary, we can make it
arbitrarily large, so that $\hat{\rho}_t = S \hat{\rho}_0
S^{\dagger}$, where $\hat{S}$ is the $S$-matrix.

This argument perhaps belabors the obvious, but it demonstrates that  $S$-matrix formalism provides accurate
probabilities without any reference to an instant of measurement,
only for observables that commute with the free-particle
Hamiltonian. For any other observables (in particular, position), the probabilities remain dependent on the time the measurement takes place and  the limit $t \rightarrow \infty$ is unwarranted. Therefore, the application of the $S$ matrix formalism in the context of particle oscillations, where the relevant observable is the location of the scattering event,  takes it outside its domain of validity. For another critique of the use of the S-matrix, see Ref. \cite{Boy}.

Another way to see the unsuitability of the $S$-matrix for
the description of particle oscillations follows from
Eq. (\ref{Sm2}). This equation is  quantum mechanically meaningful only if the domain of integration in the exponential is $[t_P, t_D]$. However, the time integral in the field theory treatments extends along the full real axis, i.e., the $S$ matrix formalism is employed. In the perturbation theory treatment of scattering experiments, the integral $\int_{t_i}^{t_f} dt$ in the time-ordered exponential is substituted for $\int_{-\infty}^{\infty} dt$ because one assumes that $t_i$ and $t_f$ are much earlier and much later respectively than the time of the scattering event, so that the interaction terms are effectively ``switched-off'. In this case, substituting  ${\cal T} e^{- i \int_{t_i}^{t_f} \hat{H}_Idt}$ with the $S$-matrix is an excellent approximation.
However, in a particle oscillation experiment $t_P$ and $t_D$ are the times at which scattering events are assumed to take place. There is, therefore, no justification in exchanging ${\cal T} e^{- i \int_{t_P}^{t_D} \hat{H}_Idt}$ with the $S$-matrix.

The issue of the applicability of the $S$ matrix formalism to particle oscillation experiments is a key problem in quantum field theory treatments that do not employ integration over time in their derivation \cite{GriSto, IP99}. In particular, the interpretation of the $S$-matrix elements on  spatially localized states as corresponding to position measurements takes the $S$-matrix formalism outside its domain of applicability.

\subsubsection{Special initial states}
A class of arguments in favor of an equal-energy prescription \cite{Lip97, Stod} states that the time of emission of a neutrino is unknown, so one should average the initial state over time. This leads to an initial state that is a sum over incoherent energy eigenstates, so that the energies of the initial state are the same in all mass eigenspaces: $E_i = E_j$. We find the following problems with this argument.

First, in any particle  production process we always obtain some information, such as macroscopic records of position and momentum of other particles produced in the reaction. In the Appendix B, where we discuss in detail the initial state issue,
we  denote these parameters collectively as $\xi$. Then, for each value of $\xi$ there is a different density matrix $\hat{\rho}_{\xi}$ corresponding to the sub-ensemble of neutrino creation events characterized by macroscopic records $\xi$. The total ensemble is then a convex combination $\sum_{\xi} p_{\xi} \hat{\rho}_{\xi}, 1 \geq p_{\xi} \geq 0$.  There is no general argument why the expression $\sum_{\xi} p_{\xi} \hat{\rho}_{\xi}$, referring in general to a large number of observed values $\xi$, and corresponding to different particle production processes,
 would always be proportional to
  an integral $\int dt  \hat{\rho}(t)$ over the initial state of the particles.

Moreover, averaging a density matrix over the time of particle production suffers from the same mathematical and conceptual problems as averaging the density matrix over time of arrival; the discussion in Sec. 4.2.3 applies to this case as well.
 Like the time of arrival, the time of production is a physical observable and it does not coincide with the time parameter of Schr\"odinger's equation. A proper description of the dependence of probabilities on the time of particle production $t_0$ should proceed from the inclusion of $t_0$ into the variables $\xi$ of the POVM $\hat{\Pi}_{\xi}$, that corresponds to the variables determined during the production process. Only with such a treatment would it make sense to average the state over the production time $t_0$, but in this case, integration over $t_0$ would be a part of the convex combination defining the neutrino initial state.

In general, the recourse to a fine specification of the initial state is undesirable in any theoretical prediction or explanation, because it involves strong assumptions that cannot be independently verified. It is highly unlikely that such fine-tuning could persist in different experiments characterized by different preparations of the system. Even the justification of the properties of the neutrino wave-packet by the physics of the production process \cite{Gi02},  requires the introduction of a number of unobserable parameters and a choice of specific regimes.
In the method we develop in this paper, the results do not require any fine-tuning in the parameters of the initial state. We obtain the non-standard oscillation formula even if we assume an initial state for the neutrinos with equal energies $E_i$ in different mass-eigenspaces. In fact, our results are valid for a  large class of initial states that only share some broad qualitative characteristics.

\subsubsection{Time integration in probability currents.}

A different class of theoretical approaches for the   description of particle oscillations considers the particle flux at the detector, defined by time-integration of  a probability current $J_{\alpha}^i({\bf, x}, t)$ for flavor $\alpha$---see, Ref. \cite{ABMN96}.  The  current $J_{\alpha}^i({\bf, x}, t)$ for flavor $\alpha$ is constructed as to  (approximately) satisfy a continuity equation
\begin{eqnarray}
    \frac{\partial \rho_{\alpha}(t, x)}{\partial t} + \nabla_i J^i_{\alpha}({\bf x}, t) = 0, \label{concur}
    \end{eqnarray}
where $\hat{\rho}_{\alpha}({\bf x}, t)$ is the quantum probability distribution at time $t$.  $J^i_{\alpha}$ is interpreted as a probability current. One then defines a flux for the particles of flavor $\alpha$, detected in a region $A$,  at distance $L$ from a source, as
\begin{eqnarray}
\Phi_{\alpha}(\partial A) = \int_0^T dt \int_{\partial A} d^2\sigma_i J^i_{\alpha}(x, t), \label{concur2}
\end{eqnarray}
where $T$ is an integration time for the experiment and $\partial A$ the boundary of the detector region $A$.

There are open theoretical issues in this approach: (i) how to properly define flavor currents in the relativistic regime, relevant to neutrinos \cite{Blas2}, and (ii) how to interpret the fact that the conservation equation is not exact. The latter issue, in particular, creates problems in the normalization, since the total flux along a closed surface $S$ around the production region is not conserved.

However, even if such problems were resolved, a more fundamental problem in
the consistency of the integration over time in Eq. (\ref{concur2}) would have to be addressed.
  Probability currents in quantum theory may also take  negative values \cite{negat}, so  the  flux in Eq. (\ref{concur2}) does not, in general, correspond to  the total number of detected particles.  This issue has been discussed in relation to the definition of probabilities for the time of arrival \cite{MJ,BrMe, negat2}. In that context, it has been argued that in regimes at which probability currents are approximately positive, they do yield physically acceptable distributions for the time of arrival  \cite{curr}.
  This procedure, however, cannot provide a rule for computing the expected   particle number for a {\em general} initial state. Moreover,  the probability current  defines positive probabilities only in the quasi-classical regime for the time of arrival \cite{HalTOA}. In contrast, particle oscillations are a consequence of quantum interference, and one should expect that they cannot be adequately described in the quasi-classical regime.

\subsubsection{Quantum models of the production and detection process}

A variation of the quantum field theory approach to particle oscillation involves the description of the production and detection events in terms of quantum systems playing the role of `sources' and `sinks'. These can be  modeled, for example, by a pair of families of harmonic oscillators, labeled by flavor, and localized in the production and detection region. The interactions of the oscillators   with the neutrino field is switched on by profile functions dependent on time \cite{KW}. The coupling of the oscillators to the neutrino field is constructed so that the desired  dependence of the probabilities on the mixing matrices $U_{\alpha i}$ is obtained. We denote by $|n_{\alpha}, m_{\beta}, 0 \rangle$ the state for the source oscillator for the $\alpha$ flavor  in the n-th energy eigenstate, for the sink oscillator for flavor $\beta$ in the m-th energy eigenstate and for the neutrino vacuum, one constructs the amplitudes
\begin{eqnarray}
{\cal A}_{\alpha \rightarrow \beta} = \langle 1_{\alpha}, 0_{\beta}, 0| {\cal T}e^{- i \int d^4 x {\cal H}_i} |0_{\alpha}, 1_{\beta}, 0 \rangle. \label{ampl12}
\end{eqnarray}
The precise form of the amplitudes depends on the choice of the profile functions for the time duration of the  interaction in the detector. In Ref. \cite{KW} two cases have been considered, (i) a profile function for the detection narrowly centered around a detection time $t_D$ and (ii) a profile function with support on the time interval $[t_1, t_2]$, where $t_1$ is later than the time of emission, and $t_2$ is eventually taken to infinity. In case (i) the final expression for the probability is computed by integrating the amplitude Eq. (\ref{ampl12}) squared, with respect to $t_D$ and by introducing a suitable normalization factor. In case (ii) the final expression for the probability is obtained by evaluating the amplitude Eq. (\ref{ampl12}) squared at the limit $t_2 \rightarrow \infty$.
The results in both cases recover the standard oscillation formula.


Regarding   case (i) above, we note that the detection time $t_D$ is also introduced as an external parameter to the system and not as a variable, so our previous critique about the problems in integrating the squared amplitudes over time stands as in Sec. 4.2.3. Case (ii) is more unphysical, as the authors of Ref.  \cite{KW} state, because it involves the limit $t \rightarrow \infty$. This case also corresponds to the treatment of neutrino production and detection in Ref. \cite{Rich}, even though in the latter reference the initial and final states are not discrete.
The reason is that these works employ  the first-order  term in time-dependent perturbation theory, while at the long-time limit there exist significant higher-orders contributions  to the total probability.
 The leading-order terms dominate only at early times, when time-of-arrival interferences have not been built up. To see this, we note that this description of production and detection---i.e., treating the neutrino as an intermediate state---is
 formally  similar to atom-field interactions in quantum optics, where photons cause transitions between energy levels in pairs of atoms.

 In the quantum optics context, the validity of such approximations has been studied in detail \cite{qop}, and it is known that higher-order terms are necessary for the description of the long-time limit. The leading-order terms in the perturbative expansion of the {\em finite-time} propagator dominate only at times much smaller than emission time and their behavior cannot be extrapolated to the $t \rightarrow \infty$ limit. For example, the Wigner-Weisskopf description of atomic decay \cite{WW}, which is valid for long times, involves the solution of an integro-differential equation \cite{qop}, and it is equivalent to a resummation of an infinite number of Feynman diagrams from time-dependent perturbation theory \cite{KA71}.

\section{The quantum measurement approach to particle oscillations}

In Sec. 4, we presented our review of the main problems in the existing theoretical approaches to particle oscillations. These problems originate from the lack of a first-principles derivation of the probability assignment relevant to the particle oscillation experiments. The probabilities defined in terms of squared amplitudes carry a time-dependence, which must be removed in order to obtain experimental predictions. The procedures employed so far for the removal of this time-dependence, such as the substitution of a `mean' time of arrival in the squared amplitude, or the integration of the squared amplitude over time, are mathematically unjustified. Moreover, they do not treat the time of arrival as a quantum variable.

In this section, we present the quantum measurement approach to particle oscillations that aims to resolve the problems of the existing methods. The key point in this method is the construction of the  probabilities relevant to particle oscillation experiments from first principles in quantum theory. To this end, we develop a general framework for measurements (Sec. 5.1) in which the time of detection (coinciding in the particle-oscillations context with the time of arrival) is a quantum variable, rather than a pre-determined parameter as in the common case of von Neumann measurements. This is a proper framework for the description of  particle oscillation experiments, because  the experimentally fixed quantity in the latter is not the time of detection but the location of the detector. The result is a non-standard expression for the particle-oscillation wavelength (Sec. 5.2).

\subsection{A general framework for quantum measurements}

In Sec. 5.1, we develop a general formalism for quantum measurements (to be applied in particle oscillations in Sec. 5.2) that treats the time of detection as a quantum variable.  To this end,  we generalize a method developed in Ref. \cite{AnSav},  for the
construction of probabilities starting with amplitudes that are defined at
{\em different} moments of time. This method was applied to various
problems, such as probabilities for the time of arrival, tunneling time \cite{AnSav2}
and for the study of non-exponential decays \cite{An08}. It contains ideas from the
decoherent histories approach to quantum mechanics \cite{Omn1, Omn2, Gri, GeHa, hartlelo, I, IL} and it
has many similarities to the Srinivas-Davies photo-detection theory
\cite{SD81}.

\subsubsection{Detection time as a variable in probability amplitudes}
The first step in our study is the derivation of a general formula
for the probability of detection outcomes that does not treat detection time
as a predetermined parameter.

 Let ${\cal H}$ be the
Hilbert space of the quantum system under study; ${\cal H}$ may
include the degrees of freedom of the apparatus in addition to the ones of the microscopic system. In the most general case, a quantum event, such as a particle detection, can be described as a transition of the system from one subspace of ${\cal H}$ to another subspace orthogonal to it. For example,
\begin{itemize}{}
 \item the emission of a photon from an atom corresponds to a
transition from the one-dimensional subspace, defined by the
electromagnetic field vacuum, to the subspace of single-photon
states \cite{SD81};
\item a von Neumann measurement corresponds to  a transition from the subspace, in which the pointer variable $\hat{X}$ takes its pre-measurement values, to a subspace  corresponding to possible measurement outcomes;

\item a time-of-arrival measurement through a detector, located at $x = 0$, may be idealized as a transition from the subspace corresponding to $x \in (-\infty, 0]$ to the subspace corresponding to $ x \in [0, \infty)$ \cite{AnSav};

\item any particle reaction  can be described as a transition from the subspace of states for the initial particles to the subspace of the product particles. In particular, a reaction in which a neutrino is annihilated can be described as
     a transition from the subspace of neutrino single-particle states to (other) lepton single-particle states.

\end{itemize}

Hence, we assume that ${\cal H}$ splits into two subspaces ${\cal
H}_+ \oplus {\cal H}_-$, where ${\cal H}_+$ describes the accessible
states of the system, given that a specific event is realized, and ${\cal H}_-$ is the complement of ${\cal H}_+$. We denote by $\hat{P}$ the
projection operator onto ${\cal H}_+$ and by $\hat{Q}$
the projector onto ${\cal H}_-$.

Once the transition has taken place, it is possible to
measure the values of various observables through their correlation to a pointer variable. Let us denote by $\hat{P}_\lambda$   projection operators corresponding to different values $\lambda$ of an observable that can be measured only if a detection event has occurred. For example, in neutrino oscillations, $\hat{P}_\lambda$ may correspond to a coarse-grained position variable for one of the product particles. The set of projectors $\hat{P}_\lambda$ is exclusive ($\hat{P}_{\lambda} \hat{P}_{\lambda'} = 0, $ if $\lambda \neq \lambda'$) and exhaustive, provided a detection has occurred; i.e., $\sum_\lambda \hat{P}_\lambda = \hat{P}$.

In general, it suffices that the operators $\hat{P}_{\lambda}$  be  positive rather than projectors;  they still have to be
exhaustive on ${\cal H}_+$, but they do not have to be exclusive.

We assume that  the system is initially prepared at a
state $|\psi_0 \rangle \in {\cal H}_+$, and that the dynamics is
governed by the self-adjoint Hamiltonian operator $\hat{H}$.
Our aim is to construct the amplitude $|\psi_0; \lambda, [t_1, t_2]
\rangle$ that, given the initial state $|\psi_0 \rangle$, at some
instant in the time interval $[t_1, t_2]$, a transition occurs with a
recorded value $\lambda$ for the measured observable. We assume that $[t_1, t_2] \subset [0, T]$, where $T$ is a final moment of time,
corresponding to a long integration time of an experiment.

 We first consider the case that the relevant time
interval is small, i.e., we set $t_1 = t$ and $t_2 = t + \delta t$, and we  keep only terms of leading  to $\delta t$. The
construction proceeds as follows.

Since the transition took place within the interval $[t, t + \delta
t]$, at times less than $t$ the state lay within ${\cal H}_-$. This is taken into account by evolving  $|\psi_0 \rangle$ with the
restricted propagator \cite{repr} into ${\cal H}_-$, i.e., with the one-parameter family of operators
\begin{eqnarray}
\hat{S}_t =  \lim_{N \rightarrow \infty}
(\hat{Q}e^{-i\hat{H} t/N} \hat{Q})^N.
\end{eqnarray}

By assumption, the transition took place some time within the time interval $[t, t+\delta t]$ and at the end of this interval the outcome $\lambda$ has been recorded. This means that in the time-interval $[t, t+\delta t]$ the amplitude transforms under
full unitary operator for time evolution  $e^{-i \hat{H} \delta t}
\simeq 1 - i \delta t \hat{H}$. At time $t + \delta t$ the event
corresponding to $\hat{P}_{\lambda}$ is recorded, so the amplitude
is transformed by the action of $\hat{P}_{\lambda}$, or of $\sqrt{\hat{P}_{\lambda}}$, if $\hat{P}_{\lambda}$ is not a
projector. Then, since there is no constraint, so the amplitude evolves
unitarily until time $T$.

At the limit of small $\delta t$, the
successive operations above yield
\begin{eqnarray}
|\psi_0; \lambda, [t, t+ \delta t] \rangle =  - i \, \delta t \,
\,e^{-i\hat{H}(T - t)} \hat{P}_{\lambda} \hat{H} \hat{S}_t |\psi_0
\rangle. \label{amp4}
\end{eqnarray}

We must emphasize here the important physical distinction on the role of time that is manifested in the derivation of the amplitude Eq.
(\ref{amp4}). The time of detection $t$ is {\em not} identical to
the evolution parameter of Schr\"odinger' s equation. Instead, it is
a {\em dynamical variable} that determines the moment that a
physical {\em event} has taken place \cite{Sav}. The construction of
the amplitude $|\psi; \lambda, [t, t + \delta t] \rangle$ above
highlights this distinction: the detection time $t$ is {\em distinct} from
the  time $T$ at which the amplitude is evaluated.

\medskip

 The amplitude $|\psi_0; \lambda, [t, t + \delta t] \rangle$ is proportional to $\delta t$, hence it defines  a {\em density} with respect to time $|\psi_0;  \lambda, t \rangle := \lim_{\delta t \rightarrow 0}
\frac{1}{\delta t} | \psi_0; \lambda, [t, t + \delta t] \rangle$, which by Eq. (\ref{amp4}) equals
\begin{eqnarray}
|\psi_0;  \lambda, t \rangle = - i   \,e^{-i\hat{H}(T - t)}
\hat{P}_{\lambda} \hat{H} \hat{S}_t |\psi_0 \rangle = - i e^{- i
\hat{H} T} \hat{C}_{\lambda, t} |\psi_0 \rangle, \label{ampl5a}
\end{eqnarray}
in terms of the class operator
\begin{eqnarray}
\hat{C}_{\lambda, t} = e^{i \hat{H}t} \hat{P}_{\lambda} \hat{H}
\hat{S}_t.
\end{eqnarray}

 Since the amplitude $|\psi_0;  \lambda, t \rangle $ is a density
with respect to the time of detection $t$, integration over time is well-defined. In particular, this means that alternatives labeled by different values of $t$ are disjoint  \cite{indep}. Hence, we can integrate  over $t$ \cite{probcom},  in order to obtain the total amplitude for detection happening at {\em any time} in the interval $[t_1,
t_2]$ as

\begin{eqnarray}
| \psi; \lambda, [t_1, t_2] \rangle = - i e^{- i \hat{H}T}
\int_{t_1}^{t_2} d t \hat{C}_{\lambda, t} |\psi_0 \rangle.
\label{ampl5}
\end{eqnarray}

We note that if $[\hat{P}, \hat{H}] = 0$, i.e., if the Hamiltonian
evolution preserves the subspaces ${\cal H}_{\pm}$, then $|\psi_0;
\lambda, t \rangle = 0$. When the Hamiltonian   is of the form $\hat{H} = \hat{H_0} + \hat{H_I}$, where $[\hat{H}_0, \hat{P}] = 0$, and $H_I$ a small interaction, then to leading order in the perturbation
\begin{eqnarray}
\hat{C}_{\lambda, t} = e^{i \hat{H}_0t} \hat{P}_{\lambda} \hat{H}_I
e^{-i \hat{H}_0t}. \label{perturbed}
\end{eqnarray}

Finally, we note that the  amplitude Eq. (\ref{amp4}) has a more  intuitive interpretation in the sum-over-histories formulation of quantum theory.  If the operators $\hat{P}, \hat{Q}$ and
$\hat{P}_{\lambda}$ are spectral projectors of configuration space
variables, the amplitude Eq.  (\ref{amp4}) corresponds to a path integral with a domain of integration restricted over a suitably defined class of paths \cite{hartlelo, scg, YaTa, BoHa}.  Such path-integration domains are said to define {\em spacetime coarse-grainings}.

\subsubsection{Probabilities for detection time}

From Eq. (\ref{ampl5}), we find that the probability $p (\lambda)\/$
that at some time in the interval $[t_1, t_2]$ a detection with
outcome $\lambda$ occurred equals
\begin{eqnarray}
p(\lambda, [t_1, t_2]) = \langle \psi; \lambda, [t_1, t_2] | \psi;
\lambda, [t_1, t_2] \rangle =   \int_{t_1}^{t_2} \,  dt
\, \int_{t_1}^{t_2} dt' \; Tr (e^{i\hat{H}( t - t')}
\hat{P}_{\lambda} \hat{H} \hat{S}^{\dagger}_t \hat{\rho}_0
\hat{S}_{t'} \hat{H} \hat{P}_{\lambda} ), \label{prob1}
\end{eqnarray}
where $\hat{\rho}_0 = |\psi_0\rangle \langle \psi_0|$.

Since the probabilities Eq. (\ref{prob1}) are obtained by the square of
an amplitude, they are in general characterized by interference, i.e., the
probability for an interval $[t_1, t_3] = [t_1, t_2] \cup [t_2,
t_3]$ equals
\begin{eqnarray}
p(\lambda, [t_1, t_3]) = p(\lambda, [t_1, t_2]) + p(\lambda, [t_2,
t_3])
+ 2 Re \left[ \int_{t_1}^{t_2} \,  dt \, \int_{t_2}^{t_3} dt'
Tr\left(\hat{C}_{\lambda, t} \hat{\rho_0}\hat{C}^{\dagger}_{\lambda,
t'}\right)\right]. \label{add}
\end{eqnarray}

Probabilities are a measure of event counts; hence,  the probabilities Eq.
(\ref{prob1}) cannot  describe the outcomes an
experiment, unless the terms
\begin{eqnarray}
2 Re \left[
\int_{t_1}^{t_2} \,  dt \, \int_{t_2}^{t_3} dt'
Tr\left(\hat{C}_{\lambda, t} \hat{\rho_0}\hat{C}^{\dagger}_{\lambda,
t'}\right)\right] \nonumber
 \end{eqnarray}
 vanish. A measurement apparatus is a macroscopic system with a large number of degrees of freedom, therefore  one expects that consistent probabilities can be defined given a substantial degree of coarse-graining \cite{Gri, GeHa}. In particular,  there is
 a coarse-graining time-scale $\tau$, such that the non-additive terms in Eq. (\ref{add}) are strongly suppressed if $ |t_2 - t_1| >> \tau$ and $|t_3 - t_2| >> \tau$. Then, Eq. (\ref{prob1}) does define a probability measure when restricted to intervals of size  larger than $\tau$. We explain how this coarse-graining scale arises in the Appendix A, where we consider a model measurement scheme.

Note that we are interested in constructing a
POVM that describes measurement outcomes, i.e., the physical system
under consideration includes the measuring apparatus, which is a
macroscopic body with a large number of degrees of freedom. The
pointer variables are essentially classical and classical
fluctuations (thermal or statistical) impose a lower limit $\tau$ to
the temporal resolution for the system. Hence,  there is no meaning
in discussing about the probabilities associated to time-intervals
of  width smaller than $\tau$ (see the discussion in Sec. 3.2).

\paragraph{Temporal smearing.}
Assuming a finite resolution scale $\tau$, one may use the
amplitudes Eq. (\ref{ampl5}) to define a POVM. It is more convenient to smear the
amplitudes at a time-scale defined by $\tau$ rather than
considering sharply defined intervals as in Eq. (\ref{prob1}).

To this end, we introduce a family of functions $f_{\tau}(s)$,  localized around $s = 0$ with width $\tau$, normalized so that
$\lim_{\tau \rightarrow 0} f_{\tau}(s) = 2\delta(s)$ \cite{del}. For example, we may employ the Gaussians
\begin{eqnarray}
f_{\tau}(s) = 2\frac{1}{\sqrt{2 \pi \tau^2}}
e^{-\frac{s^2}{2\tau^2}}. \label{gauss}
\end{eqnarray}
We then define the amplitude $|\psi_0; \lambda, t\rangle_{\tau}$,
localized around the value $t$ with width $\tau$, as
\begin{eqnarray}
|\psi_0; \lambda, t\rangle_{\tau} = \int ds \sqrt{f_{\tau}(s -t)}
|\psi_0; \lambda, s \rangle = \hat{C}_{t, \lambda, \tau} |\psi_0
\rangle, \label{smearing}
\end{eqnarray}
where we wrote
\begin{eqnarray}
\hat{C}_{t, \lambda, \tau} = \int ds \sqrt{f_{\tau}(s - t)}
C_{\lambda, s}.
\end{eqnarray}
The square amplitudes
\begin{eqnarray}
p_{\tau}(\lambda, t) = {}_{\tau}\langle \psi_0; \lambda, t|\psi_0;
\lambda, t\rangle_{\tau} = Tr \left(\hat{C}^{\dagger}_{\lambda, t,
\tau} \hat{\rho}_0 \hat{C}_{\lambda, t, \tau}\right) \label{ampl6}
\end{eqnarray}
are of the form $Tr[\hat{\rho}_0 \hat{\Pi}_{\tau}(\lambda, t)]$,
where
\begin{eqnarray}
\hat{\Pi}_{\tau}(\lambda, t) = \hat{C}_{\lambda, t, \tau}
\hat{C}^{\dagger}_{\lambda, t, \tau}. \label{povm2}
\end{eqnarray}

The positive
operator
\begin{eqnarray}
\hat{\Pi}_{\tau}(N) = 1 - \int_0^{\infty} dt \int d \lambda
\hat{\Pi}_{\tau}(\lambda, t), \label{nodet}
\end{eqnarray}
 corresponds to the alternative ${\cal N}$ that no detection took place
in the time interval $[0, \infty)$. $\hat{\Pi}_{\tau}(N)$ together with the positive operators Eq. (\ref{povm2})
 define a POVM on $ \left([0, \infty) \times
\Omega \right) \cup \{{\cal N} \}$, where $\Omega$ is the space of possible values of
$\lambda$. The POVM Eq. (\ref{povm2}) determines the probability {\em
densities} that a transition took place at time $t$, and that the outcome
$\lambda$ for the value of an observable has been recorded.

In Eq. (\ref{povm2}) the time of arrival is a genuine observable (a random variable) and not a parameter, so integration over time in Eq. (\ref{povm2}) is well defined and it leads to a partial POVM
$\hat{\Pi}_{\tau}(\lambda) = \int_0^{\infty} dt  \; \hat{\Pi}_{\tau}(\lambda, t)$. This equals
\begin{eqnarray}
\hat{\Pi}_{\tau}(\lambda) = \int_0^{\infty} ds \int_0^{\infty} ds G_1(s',s) \hat{C}_{\lambda, s} \hat{C}_{\lambda, s'}^{\dagger}, \label{partialpovm}
\end{eqnarray}
in terms of the kernel
\begin{eqnarray}
G_1(s', s) = \int_0^{\infty} dt \sqrt{f_{\tau}(s-t) f_{\tau}(s'-t)}. \label{kernel1}
\end{eqnarray}
If the POVM acts on states with support on values for the time-of-arrival much larger than $\tau$, the Gaussian smearing functions Eq. (\ref{gauss}) can be used. In this case,
\begin{eqnarray}
G_1(s, s') =  e^{-\frac{(s-s')^2}{8 \tau^2}}. \label{kernel1b}
\end{eqnarray}

Finally, we note that if the temporal resolution $\tau$ of the detection is much larger than any time-scales in the amplitudes $\hat{C}_{\lambda,t}|\psi_0\rangle$, then smearing is not necessary.  Then, simply consider the probabilities Eq. (\ref{prob1}) for the interval $[0, \infty)$, to obtain
\begin{eqnarray}
p(\lambda) = \int_0^{\infty} dt \int_0^{\infty}dt' \; Tr \left(\hat{C}^{\dagger}_{\lambda, t} \hat{\rho}_0 \hat{C}_{\lambda, t'}\right). \label{probdd}
\end{eqnarray}


 Appendix A contains an application of the above formalism to a system consisting of a microscopic particle and a measurement
apparatus. Then, we show that the apparatus defines a time-scale $\tau_{dec}$, such that a choice of $\tau$ in the smearing functions of the order of $\tau_{dec}$,  guarantees that the probabilities in Eqs. (\ref{povm2}) and (\ref{partialpovm}) are well defined.  Hence, $\tau_{dec}$ sets a scale of minimal temporal coarse-graining necessary for the quasi-classical attribution of definite macroscopic properties to the pointer variables. Thus it corresponds to the maximal temporal resolution of the detector. In the Appendix A, we also show that von Neumann measurements are  a special case of the framework we developed in Sec. 5.1.

\subsection{Particle oscillations}
\subsubsection{The construction of a POVM for particle oscillations}
We now apply the formalism developed in Sec. 5.1 to  the
treatment of particle oscillations. The oscillating particles are
detected by means of their reactions, hence, it is necessary to include into
the description the degrees of freedom of all particles
participating in such processes and to employ quantum field theory.

In what follows, we will not specify the type of oscillating
particles, so our results are valid for both neutrino and
neutral boson oscillations. To this end, let us denote the
oscillating particles as $A$ and assume that they are detected by
means of the process
\begin{eqnarray}
A + B_1 + \ldots B_M \rightarrow D_1  \ldots + D_N, \label{proc}
\end{eqnarray}
where $B_m, D_n$ are particles (different from the $A$ particles),
labeled by the indices $m = 1, \ldots, M$ and $n = 1, \ldots, N$. In
this description, we assume that the $A$ particles are produced in a
production region, they propagate and  they interact with (one or more of) the $B_m$ particles, which are located in a detector away from the source. The interaction of the $A$ particles with the $B_m$ particles produces the products particles $D_n$, which are the ones that are being detected. Relevant observables are the $D_n$-particles' time of detection, position, momentum and so on. These observables are determined through macroscopic pointer variables in the detector.

\paragraph{The Hilbert space.}

 The Hilbert space ${\cal H}_{tot}$, in which the process Eq. (\ref{proc}) are described,  is expressed as a tensor product
 ${\cal H}_{tot} = {\cal
H}_A \otimes {\cal H}_{r}$. ${\cal H}_A$ is the Fock space ${\cal
F}({\cal H}_{1A})$ corresponding to the $A$-particles. Explicitly, we write
\begin{eqnarray}
{\cal F}({\cal H}_{1A}) = {\bf C} \oplus H_{1A} \oplus
(H_{1A}\otimes H_{1A})_{S, A} \oplus (H_{1A}\otimes H_{1A} \otimes
H_{1A})_{S, A} \oplus \ldots,
\end{eqnarray}
where $S$ refers to symmetrization (bosons) and $A$ to
antisymmetrization (fermions) and $H_{1A}$ is the Hilbert space describing a
single $A$ particle. The single-particle Hilbert space is a direct
sum $\oplus_{i}{\cal H}_{i}$ of the mass-eigenspaces ${\cal H}_{i}$.
The degrees of freedom of the $B_m $ and $D_n$ particles are incorporated into the Hilbert space ${\cal H}_r$. In general, ${\cal H}_r$ is a tensor product of Fock spaces, one for each field other than $A$, participating in the process Eq.  (\ref{proc}).

\paragraph{Events.}
In order to apply the formalism of Sec. 5.1, it is
necessary to identify the subspaces ${\cal H}_{\pm}$ that define the transition under consideration. Since now the detection proceeds through the
process Eq. (\ref{proc}) and the $A$ particles are not directly
observable, the transition corresponds to subspaces of ${\cal H}_r$. The Hilbert space ${\cal H}_r$
is decomposed as ${\cal H}_0 \oplus {\cal H}_{prod}$, where  ${\cal H}_0$
is the subspace of states prior to the decay $A + B_m \rightarrow
D_n$, and $H_{prod}$ is the subspace corresponding to states of the
decay products. This means that we identify ${\cal H}_-$ with ${\cal H}_{0}$ and ${\cal H}_+$ with ${\cal H}_{prod}$.

The following are examples of such subspaces for specific detection schemes.

\begin{itemize}{}
\item  If the $K_L$ boson  is detected by means of the decay
$K_L \rightarrow 3 \pi^0$, ${\cal H}_r$ is identified with the Fock
space ${\cal F}_{\pi_0}$ for $\pi^0$ particles, ${\cal H}_0$ is the
vacuum subspace of ${\cal F}_{\pi_0}$ and  ${\cal H}_{prod}$ is the
subspace of ${\cal F}_{\pi_0}$ with non-zero particle number.

\item For the detection of neutrinos through the reaction $\bar{\nu}_e + p \rightarrow n + e^-$, we have only zero- and one-particle states for each of the $p, n, e^-$ particles. Therefore, the relevant Hilbert space for each particle $p$, $n$ and $e^-$ is the direct sum of the corresponding vacuum subspace ${\cal H}^{vac}_{n, p, e}$ and the single-particle Hilbert space ${\cal H}^1_{n, p, e}$. The Hilbert space ${\cal H}_r$ is then the tensor product $({\cal H}^{vac}_p \oplus {\cal H}^{1}_p)\otimes ({\cal H}^{vac}_n \oplus {\cal H}^{1}_n) \otimes ({\cal H}^{vac}_e \oplus {\cal H}^{1}_e)$. If the detector records product electrons, the subspace ${\cal H}_{prod}$ equals $({\cal H}^{vac}_p \oplus {\cal H}^{1}_p)\otimes ({\cal H}^{vac}_n \oplus {\cal H}^{1}_n) \otimes {\cal H}^{1}_e$ and ${\cal H}_0 = ({\cal H}^{vac}_p \oplus {\cal H}^{1}_p)\otimes ({\cal H}^{vac}_n \oplus {\cal H}^{1}_n) \otimes {\cal H}^{vac}_e$. If both product particles are detected then ${\cal H}_{prod} = ({\cal H}^{vac}_p \oplus {\cal H}^{1}_p)\otimes {\cal H}^{1}_n \otimes  {\cal H}^{1}_e$.

\item If the neutrino is detected through scattering, for example in a reaction $\nu + e^- \rightarrow \nu + e^-$, then the Hilbert space ${\cal H}_r$ is identified with the single-particle Hilbert space ${\cal H}^1_e$ of the electrons, which splits into two subspaces ${\cal H}_0 \oplus {\cal H}_{prod}$. ${\cal H}_0$ corresponds to the spectral projector of the Hamiltonian in the range $[0, E_0]$, and ${\cal H}_{prod}$ to the spectral   projector of the Hamiltonian in the range $[E_0, \infty)$. $E_0$ is the maximum energy in the initial state of the electrons.

\end{itemize}

We will denote as $\hat{P}$ the projector  $\hat{ P}: {\cal H}_r
\rightarrow {\cal H}_{prod}$ and $\hat{Q} = 1 - \hat{P}$ its
complement. The corresponding projectors on ${\cal H}_{tot}$ are $1
\otimes \hat{P}$ and $1 \otimes \hat{Q}$.

 Measurements carried out on the product particles  correspond to a family of positive operators
$1 \otimes \hat{\Pi}_{\lambda}$ onto ${\cal H}_{prod}$, where
$\hat{\Pi}_{\lambda}$ are positive operators on ${\cal H}_r$ corresponding to
different values $\lambda$ of the measured observables. They satisfy the completeness relation $\int d \lambda \hat{\Pi}_{\lambda} = \hat{P}$.

\paragraph{Dynamics.}

We assume a Hamiltonian of the form
\begin{eqnarray}
\hat{H} = \hat{H}_A \otimes \hat{1} + 1\otimes \hat{H}_{r} +
\hat{H}_I,
\end{eqnarray}
where $\hat{H}_A$
 is the Hamiltonian for free  $A$
particles, $\hat{H}_{r}$ is the Hamiltonian for the $B_m$ and $D_n$
particles, and $\hat{H}_I$ is the interaction term.

In the following, we will consider a single-particle initial state
for the $A$ particles, so that the Hamiltonian $\hat{H}_A$ is
restricted on the ${\cal H}_{1A}$ subspace, i.e.,
\begin{eqnarray}
\hat{H}_A = \sum_i \sqrt{m_i^2 + {\bf \hat{p}}^2} \hat{P}_i,
\end{eqnarray}
where ${\bf \hat{p}}$ is the $A$-particle momentum, and $\hat{P}_i$ the projector onto the eigenspace corresponding to particle mass $m_i$. For ease of notation,   we  ignore the A-particles' spin degrees of freedom since these do not  significantly contribute the results pertaining to oscillation probability.

We assume that any particles $B_m$ that
are present prior to detection are almost stationary. This condition
defines our reference frame, that in most cases coincides with the
laboratory frame. Assuming that the initial state for the $B_m$
particles has support to values of momentum much smaller than their
rest mass, the restriction of  $\hat{H}_r$ in the subspace ${\cal
H}_0$ is a constant. It is convenient to set this constant equal to
zero. Then,

\begin{eqnarray}
\hat{H}_r = \left[\epsilon_{th} + \sum_n (\sqrt{M_{D_n}^2 +
\hat{{\bf p}}_n^2} - M_{D_n})\right] \hat{P},
\end{eqnarray}
 where
 \begin{eqnarray}
 \epsilon_{th} = \sum_n M_{D_n} - \sum_m M_{B_m}
 \end{eqnarray}
  is
the threshold energy of the $A$ particles for the process $A + B_m
\rightarrow D_n$. In the above, $M_{B_m}$ and $M_{D_n}$ are the
masses of the particles $B_m$ and $D_n$ respectively and $\hat{{\bf
p}}_n$ are the momentum operators for the $D_n$ particles.

We specialize to the case that the $A$-particle is annihilated in the reaction through which it is detected. We will therefore consider an interaction Hamiltonian of the  form
\begin{eqnarray}
\hat{H}_I = \sum_i \int d^3x \left[\hat{b}_i({\bf x}) U_{ \alpha i}
\hat{J}_{\alpha}^{+}({\bf x}) + \hat{b}^{\dagger}_{i}({\bf x})
U^*_{ \alpha i}\hat{J}^-_{\alpha}({\bf x})\right], \label{hi}
\end{eqnarray}
 where $b_i, b^{\dagger}_i$ are annihilation and
creator operators on ${\cal H}_{A}$, $i$ labels mass eigenstates,
 $J^{\pm}_{\alpha}({\bf x})$ are current operators of flavor
$\alpha$ defined on ${\cal H}_{r}$, and $U_{\alpha i}$ is the mixing
matrix. A more general case (relevant to bosons) would involve a
kernel instead of a constant for the mixing matrix $U_{\alpha i }$,
reflecting the fact that the mixing coefficients may depend on
momentum. Here, however, we shall consider initial states sharply
concentrated in momentum, for which a constant value of $U_{
\alpha i}$ provides a good leading-order approximation.  The case of detection through scattering involves a different interaction Hamiltonian quadratic to the $A$-field variables  and it will be treated in a different publication.

The current operator $\hat{J}_{\alpha}^{\pm}$ involves products of
annihilation operators for the $B$ particles and creation operators
for the $D$ particles. Since no $A$-particles are created during the
detection process, the initial state $|\phi_0 \rangle$ in ${\cal
H}_r$ must satisfy
\begin{eqnarray}
\hat{J}_{\alpha}^-({\bf x}) |\phi_0 \rangle = 0.
\end{eqnarray}

\paragraph{Amplitudes.}

We assume an initial factorized state for the total system $|\psi_0
\rangle \otimes | \phi_0 \rangle$. The state $|\psi_0 \rangle$
 on ${\cal H}_{A}$ is  a single-particle
state
\begin{eqnarray}
| \psi_0 \rangle = \sum_i \int d^3 x \, \hat{b}_{i}^{\dagger}({\bf
x}) \psi_{i0}({\bf x})|0 \rangle_A,
\end{eqnarray}
 where $| 0 \rangle_A$ is the
vacuum of the Fock space ${\cal H}_A$. We only examine
single-particle states for the $A$ particle in this paper, so the choice of fermionic or bosonic statistics makes no difference. The state $|\phi_0 \rangle$ is, in general, a many-particle  state, since the detector contains a large number of $B$-particles and one cannot specify {\em a priori} which one participates in the reaction.

We now construct the amplitude $| \psi_0, \phi_0; \alpha, \lambda, [t, t +
\delta t] \rangle$, for the detection of a particle
produced from a reaction through flavor $\alpha$, at the time-interval
$[t, t+ \delta t]$ and with a measurement outcome $\lambda$.  From Eq. (\ref{ampl5a}), we obtain
\begin{eqnarray}
| \psi_0, \phi_0; \alpha, \lambda, [t, t + \delta t] \rangle =
- i \delta t e^{ - i \hat{H}T} \hat{C}_{\alpha, \lambda, t} |\psi_0 \rangle
\otimes  |\phi_0 \rangle,
\end{eqnarray}
in terms of a class operator $\hat{C}_{\alpha, \lambda, t}$.

 Since $
[\hat{H_A} \otimes 1 + 1 \otimes \hat{H}_r, 1 \otimes \hat{Q}] = 0$,
we can use  Eq. (\ref{perturbed}) in order to obtain the amplitude
 to leading order in perturbation theory. We find
\begin{eqnarray}
\hat{C}_{\alpha, \lambda, t} \left( |\psi_0 \rangle \otimes  |\phi_0
\rangle\right) \sum_i \int d^3 x \left(U_{\alpha i} \psi_i({\bf x},
t) | 0 \rangle_A\right) \otimes \left( e^{i \hat{H}_r t}
\sqrt{\hat{\Pi}_{\lambda}} \hat{J}^+_{\alpha}({\bf x})|\phi_0\rangle
\right), \label{ampl8}
\end{eqnarray}
where $\psi_i({\bf x}, t)$ is the solution of Schr\"odinger's
equation for the $A$-particle Hamiltonian, with initial condition
$\psi_i ({\bf x}, t) = \psi_{i0}({\bf x})$.

\paragraph{Probabilities.}
We next employ the amplitudes Eq. (\ref{ampl8}) for the construction of a probability density, using the methodology described in Sec. 5.1. To this purpose, we specify the temporal resolution $\tau_{dec}$ of the detection scheme and we employ Eq. (\ref{povm2}). Oscillations are observed in relation to the  distance of the detector from the source and not directly with respect to the values of time of arrival---whether the time of arrival can be distinguished or not.  Hence, we  consider the partial POVM Eq. (\ref{partialpovm}), obtained  by integrating over the time of arrival in  Eq. (\ref{povm2}).



The probability that a decay through flavor $\alpha$ has
happened at some time in $[0, T]$ and that the value $\lambda$ for
an observable of the product particles has been found equals
\begin{eqnarray}
p_{\alpha}(\lambda) = \sum_{ij}\int_0^T dt \int_0^T dt' \int d^3x
d^3x' \psi^*_{i}({\bf x'},t') \psi_{j}({\bf x},t)
 U^*_{\alpha i} U_{ \alpha j} G_1(t'-t) R^{\alpha}_{\lambda}({\bf x},
{\bf x'}, t-t') , \label{prob3}
\end{eqnarray}
where
\begin{eqnarray}
R_{\lambda}^{\alpha}({\bf x}, {\bf x'}, t\!-\!t') =
 \langle \phi_0|\hat{J}^-_{\alpha}({\bf x'}) \sqrt{\hat{\Pi}_{\lambda}} e\!^{-i\hat{H}_r(t'\!-t)}
  \sqrt{\hat{\Pi}_{\lambda}} \hat{J}^+_{\alpha}({\bf x})|\phi_0\rangle.
  \label{R1}
\end{eqnarray}
and $G_1(s'-s)$ is the kernel Eq. (\ref{kernel1b}), for $\tau = \tau_{dec}$, that carries the dependence on the resolution scale $\tau_{dec}$ of the detector.

\paragraph{Observables.} In order to complete the description of the
measurement scheme, we must specify the observables that are being
recorded at the detection of the decay particles. This is encoded
into the specification of the positive operators
$\hat{\Pi}_{\lambda}$ in Eq. (\ref{R1}). In the most general case,
the positive operators $\hat{\Pi}_{\lambda}$ correspond
to joint measurements of phase-space variables, i.e., the apparatus records the position and momentum with an accuracy corresponding to a phase space volume of size much larger than $\hbar^{3/2}$ \cite{Omn1, Omn4}. However, the degree of coarse-graining associated to the sampling of momentum is much larger than the coherences of momentum in the quantum states, relevant to particle oscillations, so for simplicity we specialize here to the case of sampling only with respect to position. In Sec. 5.4.2, we show that this choice does not affect the main conclusions.

We  assume that the positive operators
$\hat{\Pi}_{\lambda}$ correspond to the sampling of the position
${\bf X}$ of one of the product particles, with an accuracy of order
$\delta$. For example, we can use  Gaussian  operators of the form
\begin{eqnarray}
\hat{\Pi}_{\bf X} = \int d^3 {\bf X'} e^{ - \frac{|{\bf X} - {\bf
X'}|^2}{ 2 \delta ^2}} |{\bf X'} \rangle \langle {\bf X'}|\otimes 1,
\label{position}
\end{eqnarray}
where the tensor product with unity refers to the remaining degrees
of freedom in ${\cal H}_{prod}$.

The vector  $\hat{J}^+_{\alpha}({\bf x})|\phi_0 \rangle$ in Eq.
(\ref{R1}) corresponds to the state of the product particles for
decays that have taken place in a neighborhood of the point ${\bf
x}$. In order for the operators $\hat{\Pi}_{\bf X}$ to provide a
definite determination of the locus of the decay event with an
accuracy $\delta$, it is necessary that they have
$\hat{J}^+_{\alpha}({\bf x})|\phi_0 \rangle$ as approximate
eigenstates, i.e.,
\begin{eqnarray}
\hat{\Pi}({\bf X}) \hat{J}^+_{\alpha}({\bf x})|\phi_0 \rangle \simeq
0 , \mbox{if} \;|{\bf x} - {\bf X}| >> \delta \nonumber \\
\hat{\Pi}({\bf X}) \hat{J}^+_{\alpha}({\bf x})|\phi_0 \rangle \simeq
\hat{J}^+_{\alpha}({\bf x})|\phi_0 \rangle , \mbox{if} \; |{\bf x} -
{\bf X}| << \delta \label{appr}
\end{eqnarray}

 Heuristically, this means that the width $\delta$ of the sampling
 is much larger than any length scale characterizing the
 interaction. As we discussed in Sec. 3.1,
the accuracy $\delta$ of a position measurement is macroscopic, because a
definite recording of position is only possible at a scale coarser
than the  fluctuations of the corresponding pointer variable. On the other hand, the currents $\hat{J}^+_{\alpha}({\bf x})$ reflect
 microscopic (in fact sub-atomic) length-scales. Hence, there is a sharp
 separation of scales and the conditions Eq. (\ref{appr}) are expected
 to hold. Moreover, at this level of coarse-graining, the
 localization of a single product particle determines the location of
 all other product particles at the moment of detection;
 their wave-functions after the decay have the same support in position
 as $\hat{J}^+_{\alpha}({\bf x})|\phi_0 \rangle$.

While $\delta$ is much larger than any microscopic scale
characterizing the decay event, a particle oscillation experiment is
only meaningful if $\delta$ is much smaller than the  length scales of particle oscillations. Therefore, to a first approximation, we can substitute
$\psi_{\alpha}({\bf x}, t) $ by $\psi_{\alpha}({\bf X}, t) $ in Eq.
(\ref{prob3}).

There are three length scales entering the expression for the particle oscillation probability:  a microscopic scale
$l_{mic}$ associated to the decay process,  a macroscopic scale $\delta$
associated to the detection,  and  a macroscopic scale $L$ characterizing the
distance of the detector from the source.  These scales satisfy the condition  $l_{mic} <<
\delta << L$, that results into   the kernel $R$  being approximately proportional to $\delta ({\bf x}, {\bf X}) \delta({\bf x'}, {\bf X})$.
In order to estimate the factor of proportionality and its time-dependence we note that, by virtue of Eq. (\ref{appr}), the kernel $R$ becomes
\begin{eqnarray}
R_{{\bf X}}^{\alpha}({\bf x}, {\bf x'}, t\!-\!t')\! \simeq\!
 \langle \phi_0|\hat{J}^-_{\alpha}({\bf x'})  e\!^{-i\hat{H}_r(t'\!-t)} \hat{J}^+_{\alpha}({\bf x})|\phi_0\rangle,
\end{eqnarray}
where ${\bf x}$ and ${\bf x'}$ are assumed to lie within a distance of order $\delta$ from ${\bf X}$. The vectors $\hat{J}^+_{\alpha}({\bf x})|\phi_0$  are then proportional to $|{\bf x} \rangle_1  \otimes |{\bf x} \rangle_2 \otimes \ldots \otimes |{\bf x} \rangle_n$, where by $|{\bf x} \rangle_i$ we mean a state for the product particle labeled by $n$, localized around the point $x$. At a scale that  we can approximate $R_{\bf X}$ by a local kernel, the matrix elements are given by
\begin{eqnarray}
 \langle \phi_0|\hat{J}^-_{\alpha}({\bf x'})  e\!^{-i\hat{H}_r(t'\!-t)} \hat{J}^+_{\alpha}({\bf x})|\phi_0\rangle \simeq K_1 e^{i \epsilon_{th}(t\!-\!t')}
 \times \prod_n \left(\int_{C_{X, \delta}} dx \int_{C_{X, \delta}} dx' \; \langle {\bf x} | e\!^{-i\sqrt{{\bf \hat{p}}_n^2 + M_{D_n}^2} (t'\!-t)}| {\bf x'} \rangle \right),
\end{eqnarray}
where $K_1$ is a constant, $C_{X, \delta}$ is a region around ${\bf X}$ of width $\delta$, and $\langle {\bf x} | e\!^{-i\sqrt{{\bf \hat{p}}_n^2 + M_{D_n}^2} (t'\!-t)}| {\bf x'} \rangle$ is the propagator for a relativistic particle of mass $M_{D_n}$. Substituting the integration over $C_{{\bf X}, \delta}$ with an integral with a Gaussian smearing function, we obtain the local approximation to the kernel

\begin{eqnarray}
 R_{{\bf X}}^{\alpha}({\bf x}, {\bf x'},\! t\!-\!t')
\! \simeq K_1 \delta^3({\bf X}, \!{\bf x}) \delta^3({\bf X}, \!{\bf
x'}) e^{i \epsilon_{th}(t\!-\!t')} \! G_2(t'\!-\!t). \hspace{-0.2cm}
\label{pi2}
\end{eqnarray}

 The function $G_2(s)$ arises   from the smearing of the free particle propagators, in a region of width $\delta$
\begin{eqnarray}
 G_2(s) = \prod_n \int d^3 p \; e^{ - i
(\sqrt{M_{D_n}^2 + {\bf p}^2} - M_{D_n})s - \delta ^2 {\bf p}^2}.
\end{eqnarray}

The index $n$ runs over all product particles $D_n$. If $M_{D_n}
\delta >>1$, the saddle-point approximation applies and
\begin{eqnarray}
 G_2(s) =\prod_n \left(\frac{ M_{D_n}}{2 \pi i (s - i M_{D_n}\delta^2/2) }\right)^{3/2}. \label{ff}
\end{eqnarray}

The function $G_2(s' - s)$ suppresses contributions from the amplitudes $\psi_{\alpha}({\bf x}, s)$ at different times $s$ and $s'$ only if
$|s - s'| > \tau_{sup}$, where the suppression time scale $\tau_{sup} =   M_{min} \delta^2$ is determined by the mass $M_{min}$ of the lightest product particle. The lowest conceivable value of $\tau_{sup}$ is obtained by setting
 $M_{min} = m_{e}$ and taking for $\delta$ the atomic time-scale $0.1 nm$. We then obtain $\tau_{sup} > 10^{-16} s$. A more realistic value for $\tau_{sup}$ is obtained by placing $\delta$ in the macroscopic regime, since it corresponds to the macroscopic accuracy in the determination of a particle position. For $\delta \sim 100 nm$, we find $\tau_{sup} \sim 10 ns$ \cite{comment3}.

In the regime relevant for particle oscillations, the precise form of the kernel $G_2(s)$ does not affect significantly the particle detection probability. The most important information contained in $G_2(s)$  is the suppression time-scale $\tau_{sup}$ than is uniquely determined on dimensional grounds. Hence, none of the approximations used in the derivation of Eq. (\ref{ff}) affects the main results.

Eqs. (\ref{prob3}) and (\ref{pi2}) define a probability density for
the particle's position  ${\bf x}$, at scales much larger than
$\delta$,
\begin{eqnarray}
p_{\alpha}({\bf x})\! =\! K  \int_0^T dt
\int_0^T dt' {\cal A}_\alpha(t, {\bf x}) {\cal
A}^*_{\alpha}(t', {\bf x}) e^{- i \epsilon\!_{t\!h}\!(t'\!-\!t)}
F(t', t)\hspace{0.1cm} \label{main}
\end{eqnarray}
where ${\cal A}_{\alpha} (t, x) = \sum_i U_{\alpha i } \psi_{i}(t,
x)$ is the probability amplitude corresponding to the flavor
$\alpha$, and
\begin{eqnarray}
F(t'- t) = G_1(t'-t) G_2(t'-t).
\end{eqnarray}
The probability in Eq. (\ref{main}) can be explicitly
computed for any type of oscillating particle provided we specify
an initial state.

\subsubsection{Wavelength of particle oscillations}

Next we specify a class of initial states for the $A$ particles relevant to the description of oscillation experiments.
It is convenient to assume that the solid angle connecting the
 production region to the detector is very small  so that
only particles with momentum along the axis that connects the two
regions are detected, and the problem becomes  one-dimensional.
 We employ a $\beta$-flavor superposition of Gaussian mass-eigenstates, centered around ${\bf x} = 0$,

\begin{eqnarray}
\psi_0( x) = \sum_{i} U^*_{\beta i} \frac{1}{(\pi \sigma^2)^{1/4}}
e^{-\frac{x^2}{2\sigma^2} + i p_{i} x}, \label{psi0}
\end{eqnarray}
where $\sigma$ is the spread of the Gaussian that we take to be a free parameter. The dependence on the mixing matrix $U^*_{\beta i}$ is a consequence of the  creation of the $B$-particles through an $\alpha$-flavor current. The  Gaussian form of the initial state is chosen for convenience. In Sec. 5.4.2, we will show that the results obtained from a state of the form Eq. (\ref{psi0}) hold for a much larger class of initial states.

For the initial state Eq. (\ref{psi0}) the amplitudes ${\cal A}_{\alpha}(t, x)$ in Eq. (\ref{main}) are
\begin{eqnarray}
{\cal A}_{\alpha}(t, x) = \sum_{i} U^*_{\beta i}U_{ \alpha i} (4 \pi
\sigma^2)^{1/4} \; \int \frac{dp}{2\pi} e^{- \frac{\sigma^2}{2} (p - p_{i})^2 +i
p x - i E_{i}(p) t - \Gamma_{i}(p)t}, \label{AA}
\end{eqnarray}
where $E_{i}(p) = \sqrt{m_{i}^2 + p^2}$ and $\Gamma_{i}(p)$ are the
decay rates in  the different mass eigenspaces \cite{decayr}.

We calculate the amplitude in the regime where wave-packet dispersion is negligible. To this end, we employ the
approximation of expanding $E_{i}(p)$ to first order in
$p - p_{i}$ and $\Gamma_{i}(p)$ to zero-th order in $p- p_{i}$,
i.e.,
\begin{eqnarray}
E_{i}(p) = E_{i} + v_{i} (p - p_i); \hspace{1cm} \Gamma_{i} (p) =
\Gamma_{i}, \label{approx}
\end{eqnarray}
where $E_{i} = E_{i}(p_{i})$, $v_{i} = p_{i}/E_{i}$, and $\Gamma_{i}
= \Gamma_{i}(p_{i})$.

We then obtain
\begin{eqnarray}
{\cal A}_\alpha(t, x) = \frac{1}{(\pi \sigma^2)^{1/4}}\sum_{i}U^*_{\beta i}U_{ \alpha i}\;
e^{-\frac{(x - v_{i}t)^2}{2 \sigma^2} + i p_{i}x - iE_i t -
\Gamma_{i}t}. \label{ampl}
\end{eqnarray}
Substituting Eq. (\ref{ampl}) in Eq. (\ref{main})  and letting $T \rightarrow
\infty$, we obtain the probability density for the detection of
flavor $\alpha$ at $x = L$
\begin{eqnarray}
p_{\alpha}(L) = K \left( \sum_{i} S_{i}  e^{-2
\frac{\Gamma_{i}}{v_i}L} + \sum_{i < j} 2 e^{-
(\frac{\Gamma_{i}}{v_i} +
\frac{\Gamma_j}{v_j})L}  Re (T_{ij} \;  e^{
 i k_{ij}
L}) \right), \label{fin}
\end{eqnarray}
where
\begin{eqnarray}
S_i = |U^*_{\beta i}U_{ \alpha i}|^2 \frac{1}{\sqrt{2 v_i^2}} e^{\frac{2 \sigma^2 \Gamma_i^2}{v_i^2}}
\int_{-\infty}^{\infty} d \xi F(\xi) e^{- \frac{v_i^2}{2 \sigma^2} \xi^2 - i (E_i - \epsilon_{th}) \xi}  \label{si}\\
T_{ij} = U^*_{\beta i}U_{\alpha i } U_{\beta j } U^*_{\alpha j} \frac{1}{\sqrt{v_i^2 + v_j^2}} \int_{-\infty}^{\infty} d \xi F(\xi)
e^{\frac{\sigma^2[(v_i^2c - v_j^2) \xi - i (E_i - E_j) - (\Gamma_i + \Gamma_j)]^2}{2 (v_i^2 + v_j^2)}} \nonumber \\
\times e^{- \frac{v_i^2 +v_j^2}{8 \sigma^2} \xi^2 - \frac{1}{2} (\Gamma_i - \Gamma_j) \xi - i (\frac{E_i + E_j}{2} - \epsilon_{th})\xi}, \label{tij}
\end{eqnarray}
and
\begin{eqnarray}
 k_{ij} = \frac{ E_i
 - \epsilon_{th}}{v_i} - \frac{ E_j
 - \epsilon_{th}}{v_j} - (p_i - p_j). \label{k1}
\end{eqnarray}

In the derivation of Eqs. (\ref{fin}--\ref{k1}) we have used the fact that $ L >> \sigma$, that allows us to take the range of time integration  into
the whole real axis.

The expressions Eq. (\ref{si}) and Eq. (\ref{tij}) simplify significantly in the physically relevant regime where the time-scale
 of the oscillations is much smaller than the characteristic timescales $\tau_{sup}$ and $\tau_{dec}$. We saw that the physically relevant values of $\tau_{sup}$ is of the order of nanoseconds, while the resolution time of a solid state device is also of the order of nanoseconds. Hence,
 the timescales corresponding to Eq. (\ref{k1}) are many orders of magnitude smaller, either for neutrinos or for neutral bosons. This condition also implies that the energy and decay constant differences $|E_i - E_j|$, $|v_i - v_j|$ and $|\Gamma_i - \Gamma_j|$ are very small in comparison to $E_i$, $v_i$ and $\Gamma_i$, respectively . We can therefore take their values, in all non-oscillating terms of Eq. (\ref{si}) and Eq. (\ref{tij}), as approximately equal: $E_i \simeq E$, $v_i = v$ and $\Gamma_i \simeq \Gamma$. We then obtain
\begin{eqnarray}
S_i &=& |U^*_{\beta i}U_{ \alpha i }|^2  \frac{(4 \pi \sigma^2)^{\frac{1}{4}}}{v} F(0)
e^{\frac{-2 \sigma^2 [  (E - \epsilon_{th})^2 - \Gamma^2]}{v^2}} \label{si2}\\
T_{ij} &=& U^*_{\beta i}U_{\alpha i } U_{\beta j } U^*_{\alpha j}  \frac{(4 \pi \sigma^2)^{\frac{1}{4}}}{v} F(0) e^{\frac{-2 \sigma^2 [  (E - \epsilon_{th})^2 - \Gamma^2]}{v^2}}
 e^{i \frac{\Gamma_i (E_i - \epsilon_{th}) }{v_i^2} - i \frac{\Gamma_j (E_j - \epsilon_{th}) }{v_j^2}} \label{tij2}
\end{eqnarray}

The  oscillating term at the end of Eq. (\ref{tij}) does not depend on $L$ and the corresponding phase is much smaller than $k_{ij}L$ (it vanishes for neutrinos) so we can ignore it. Then, Eq. (\ref{fin})
becomes

\begin{eqnarray}
p_{\alpha}(L) = K' \left( \sum_{i}  |U^*_{\beta i}U_{\alpha i }|^2  e^{-2 \frac{\Gamma_{i}}{v_i}L} + \sum_{i < j} 2 Re \left( U^*_{\beta i}U_{\alpha i } U_{\beta j } U^*_{\alpha j} e^{-(\frac{\Gamma_{i}}{v_i} + \frac{\Gamma_j}{v_j})L}    e^{  i k_{ij} L}) \right) \right), \label{fin2}
\end{eqnarray}
where terms in Eqs. (\ref{si2}, \ref{tij2}) have been absorbed into a redefined  constant $K'$. Eq. (\ref{fin2}) is similar in form to the common expression for particle oscillations Eq. (\ref{sqamp}) and it allows a direct comparison: it predicts a different value for the oscillation  wavelength. For example, if $\epsilon_{th}/E_i << 1$ then  $k_{ij} = m_i^2/p_i - m_j^2/p_j$, i.e., the value of $k_{ij}$ is twice that of the standard expression.

In general, assuming that $|p_i - p_j| << p$ and $|m_i - m_j| <<p$, where $p$ is the mean momentum,  Eq. (\ref{k1}) yields
\begin{eqnarray}
k_{ij} = (1  - \frac{\epsilon_{th}}{2p})(m_i^2 - m_j^2)/E \label{keth0}.
\end{eqnarray}
In the ultra-relativistic limit Eq. (\ref{keth0}) yields
\begin{eqnarray}
k_{ij} = (1  - \frac{\epsilon_{th}}{2E})(m_i^2 - m_j^2)/E. \label{keth}
\end{eqnarray}
  The oscillation wavelength carries
a strong dependence on the threshold energy; this dependence does
not arise {\em at all} in the standard treatment.

It is important to emphasize the robustness of this result. While the first-principles treatment leading to Eq. (\ref{fin2}) is  involved Eq. (\ref{fin2}) itself can be recovered  by simple heuristic arguments. The key points are (i) that the summation over different times takes place at the level of amplitudes, (ii) that the time-scales characterizing the oscillation are much smaller than the characteristic time-scales of the measurement scheme, and (iii) that the presence of an energy threshold involved in the detection is  equivalent to the presence of a constant potential
$V = - \epsilon_{th}$, in which the $A$-particle propagates prior to
detection.  With these three conditions, Eq. (\ref{k1}) can be derived even with the plane wave method (Sec. 4.1.1).

\paragraph{Separation of timescales.}
The key observation leading to this result is the large separation of the time-scales relevant to the problem. The terms responsible for the difference between the present result and the usual derivations arise from the contribution of the interference between different times of arrival in Eq. (\ref{main}). These take place at a time-scale of order $\delta t_{ij} \sim |\frac{L}{v_i} - \frac{L}{v_j}|$. In the ultra-relativistic regime relevant to neutrinos, $\delta t_{ij} \sim L \frac{\Delta m_{ij}^2}{E^2} $, i.e., $\delta t_{ij}$ is a small fraction of the classical arrival time $t_{cl} \simeq L$. For $L \sim 1 km$,  $E \sim 100 MeV$, and $\Delta m_{ij}^2 \sim 1 eV$, we find $\delta t_{ij} \sim 10^{-21}s$. On the other hand, the characteristic suppression timescale  is $\tau_{sup} \sim m_e \delta^2$, that is of the order of $10^{-8}s$, and the temporal resolution  of the detector $\tau_{dec}$ is  of the order of $10^{-9}s$. There is a sharp separation in the orders of magnitude between the decoherence time-scale and the interference time-scale,  which means that time-of-arrival interferences are not suppressed and their contribution to the oscillation wavelength persists.


\subsection{Critical analysis of the quantum measurement approach to particle oscillations}

We proceed to an analysis of the method presented in Secs. 5.1 and 5.2 in order to show that it  resolves the problems  in the theoretical treatments of particle oscillations as presented in Sec. 4 and that it  does not suffer from  the problems associated to derivations of non-standard formulas for particle oscillations. We will also discuss the points at which the  method presented here needs further improvement.

The quantum measurement approach to particle oscillations was constructed with the avowed aims of providing a derivation of probabilities relevant to particle oscillation experiments that is based on the first principles of quantum theory. It also incorporates a fully quantum treatment of the time of arrival as a physical observable. To this end, we go beyond the modeling of experiments by von Neumann measurements that are inappropriate for the treatment of particle oscillations.

The above aims are fulfilled by the general framework for quantum measurements developed in Sec. 5.1. In this framework, detection time   is treated as a physical observable thus incorporating the crucial distinction between time as a parameter of Schr\"odinger's equation and time of arrival.   The end result is a class of POVMs that are applicable to a large class of measurement situations of which von Neumann measurements are a special case.

 In Sec. 5.1, we present our method in its most   general and abstract context. The method  applies to any physical system and the only inputs are the Hamiltonian operator and the specification of observables that are being measured. In Sec. 5.2, the method is applied to the special case of particle oscillations. The result is a probability distribution $p_{\alpha}(t, L)$ for detection events depending on  time of arrival $t$ and distance $L$. Since the time of arrival $t$ is treated as an observable, integration of the probabilities over $t$ is well-defined and unambiguous. It leads to a probability density Eq. (\ref{fin2}) for detection, solely as a function of the distance $L$.

The probability density Eq. (\ref{fin2}) is constructed through a POVM, hence, it is a linear functional of the initial state $\hat{\rho}_0$. The construction thus respects the convexity properties of quantum  states and the resulting probabilities have  no ambiguity in their statistical interpretation. In particular, the  constant $K'$ that appears in Eq.  (\ref{fin2}) is not inserted by hand for normalization purposes, but it appears as a result of an evaluation of the kernel Eq.  (\ref{R1}). Its precise value  can be fully computed once explicit expressions for  the positive operators $\hat{\Pi}_{\lambda}$ and the initial state have been provided.


In the physically relevant regime of parameters, the resulting probabilities take a particularly simple form that could also be obtained by other means. For example, an equation analogous to Eq. (\ref{fin2}) can be obtained by the plane wave method using an unequal-times prescription. The key difference is that our result is a last step of a precise procedure, starting from first principles and making no assumptions unwarranted by quantum theory. Hence, even if our expression for oscillation wavelength (partially) coincides with other non-standard oscillation formulas, the methodology and the underlying assumptions are completely different. In particular, our derivation requires no  assumption about the relation of energies $E_i$ and momenta $p_i$, in the different mass eigenspaces of the initial state. In our opinion, such assumptions about the initial state require an artificial fine-tuning of parameters that cannot be justified by the physics of the production process. A reliable physical prediction should be insensitive to tiny details in modeling of the initial state. Our results satisfy this property: we show in Sec. 5.4.2
 that Eq. (\ref{fin2}) is valid not only for initial states of the form Eq. (\ref{psi0}) but for a large class of initial states sharing  some broad qualitative characteristics.

 The use of the unequal-times prescription led to a prediction of oscillations for recoil particles, for instance the muon in $\pi \rightarrow \mu \nu$ or the $\Lambda$ in $\pi^-p \rightarrow \Lambda K^0$. Here, in contrast, we defined the POVM for particle oscillations in terms of the detection of {\em product} particles, since the neutrino is not directly observed. Hence, we can apply Eq. (\ref{fin2}) also for the detection of the $\Lambda$ particle. This has a single mass eigenspace, therefore no off-diagonal terms and, trivially, no oscillations appear.

 Eq. (\ref{k1}) for the oscillation wavenumber followed from a lengthy calculation from first principles of the quantum formalism. Nonetheless, its difference from the  standard oscillation formula can be simply explained: the standard result involves an assumption  that the contribution of time-of-arrival interferences to the total probability is suppressed. This assumption is implemented in different ways in the literature, but it underlies all existing derivations. We do not make such an assumption here. A regime corresponding to suppressed time-of-arrival interferences also exists in our formalism but it does not correspond to the physical range of parameters. To see this
  consider Eq. (\ref{main}). If the kernel $F(s-s')$ can be substituted by a delta function the resulting probabilities are proportional to $\int dt |{\cal A}_{\alpha}(t, L)|^2$,  and this expression leads to the standard result for the oscillation wavelength. However, this condition requires that either the decoherence time-scale $\tau_{dec}$ or the suppression time-scale $\tau_{sup}$ in $F(s-s')$  is much smaller than $\delta t_{ij}$, which is not the case for particle oscillation experiments.


\paragraph{Different experimental prediction.} The quantum measurement approach
to particle oscillations makes a definite physical prediction, according to
Eq. (\ref{keth}), the product $k_{ij}E$  increases with energy. In particular, it varies by as much as 50\% from its smallest value, for energies near the threshold $\epsilon_{th}$, to its largest value, at the limit  $E >> \epsilon_{th}$. In contrast, the standard expression predicts a constant value of $k_{ij}E$. Previous non-standard oscillation formulas also predict a constant value of $k_{ij}E$ and they cannot be distinguished from the standard one, in absence of an independent determination of neutrino masses.

\begin{figure}[tbp]
\includegraphics[height=4cm]{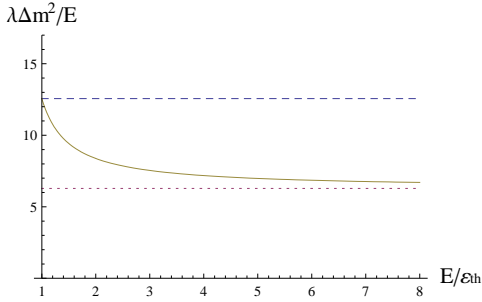} \caption{ \small The dependence of the oscillation wavelength $\lambda$ on energy $E$ is shown in this plot, where the ratio $\lambda \Delta m^2/E$ is plotted as a function of $E/\epsilon_{th}$. $\Delta m^2$ is the neutrino squared masses difference and $\epsilon_{th}$ the energy threshold for the detection process. The dashed line represents the standard expression, the dotted line the non-standard expression Eq. (\ref{wlength2}) and the solid line our result, Eq. (\ref{keth}). }
\end{figure}

The dependence of the oscillation wave-number $k_{ij}$ on energy, in Eq. (\ref{keth}),  involves a single unknown parameter, because the threshold energy for each detection process is a known quantity. Hence, Eq. (\ref{keth}) can be directly compared with the standard expression in terms of best-fit to existing data. Moreover, if the accuracy of energy sampling $\sigma_E$ is  not larger than $\epsilon_{th}$, then the corresponding distributions of number of neutrino detections as a function of energy, have a significant statistical distance. As such, the two distributions can be sharply distinguished given a sufficient number of events.

 One way to see this is by considering the two-neutrino case where the only unknown   parameter is the difference $\Delta m^2$ of squared neutrino masses. We further simplify the description by ignoring the dependence on energy of the pre-factor $K'$  on energy, so that the energy dependence of the number of events is dominated by the behavior of the oscillating term.

Then, according to the standard oscillation formula, the number of events as a function of energy is given by a distribution
\begin{eqnarray}
n_1(E) \sim \sin^2\frac{\Delta m^2 L}{4E}, \label{osc1}
\end{eqnarray}
while we predict a different distribution
\begin{eqnarray}
n_2(E) \sim \sin^2\left[\frac{\Delta m^2L}{2E}(1 - \frac{\epsilon_{th}}{2E})\right]. \label{osc2}
\end{eqnarray}

Since $\Delta m^2$ is unknown, we define a parameter $\mu^2$, as the value of the product $k E$ at the limit $E \rightarrow \infty$.
Hence, $\mu^2 = \frac{1}{2} \Delta m^2$ in Eq. (\ref{osc1}) and $\mu^2 = \Delta m^2$ in Eq. (\ref{osc2}). The event count at high energies  determines a value of $\mu$, and  the predictions of
 Eqs. (\ref{osc1}) and (\ref{osc2}) can be compared at low energies (as $E$ approaches $\epsilon_{th}$) where they diverge.

 Defining $\alpha = \frac{\mu^2 L}{2\epsilon_{th}}$ and introducing the scaled energy $x = E/\epsilon_{th}$, we
compare the probability distributions
\begin{eqnarray}
p_1(x) &=& C_1 \sin^2\frac{\alpha}{x},\label{p1} \\
p_2(x) &=& C_2 \sin^2 \frac{\alpha(1 - \frac{1}{2x})}{x}. \label{p2}
\end{eqnarray}
The constants $C_1$ and $C_2$ above have been introduced for the purpose of normalization over recorded events in any specified energy range.

 The difference in the two distributions at low energies is highlighted in Figs. 2 and 3.  Fig. 2 is a plot of the two distributions as a function of $x$ for different values of $\alpha$. Fig.3 plots the statistical distance of the two distributions as a function of $\alpha$.

\begin{figure}[tbp]
\includegraphics[height=3.4cm]{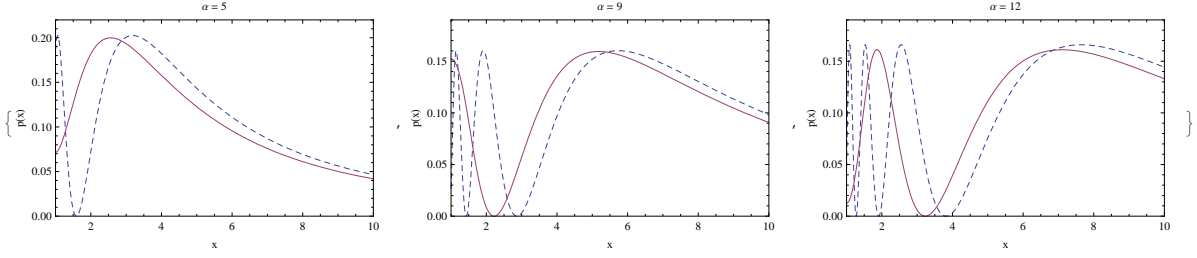} \caption{ \small Heuristic plots of the distributions $p_1(x)$ Eq. (\ref{p1}) (dashed line) and $p_2(x)$ Eq. (\ref{p2}) (solid line) as a function of $x = E/\epsilon_{th}$, for  different values of the parameter $\alpha =\frac{\mu^2 L}{2\epsilon_{th}}$. The distributions are normalized to 1 for all events with $x \in [1, 10]$. }
\end{figure}

\begin{figure}[tbp]
\includegraphics[height=3.4cm]{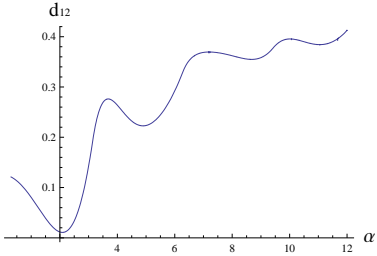} \caption{ \small The statistical distance $d_{12} = 8(1 - \int dx \sqrt{p_1(x) p_2(x)}) $ of the two distributions $p_1(x)$ and $p_2(x)$ provides a measure of their distinguishability. Here it is plotted as a function of $\alpha =  \frac{\mu^2 L}{2\epsilon_{th}}$. There is an overall increase  of the statistical distance with $\alpha$. }
\end{figure}

At low energies the divergence of the present oscillation formula from the standard one is stronger. This fact is rather significant, because it is at low energies that the discrepancy of the number of observed neutrino events has been observed, in the MiniBooNE experiment. The result obtained in this paper is a possible explanation of this discrepancy that does not require the introduction of physics beyond the Standard Model.

\paragraph{Open issues in the formalism.} The main advantages of the quantum measurement approach to particle oscillations are that it  makes no assumptions extrinsic to quantum theory, and, by employing sound probabilistic reasoning, and leads to well-defined probabilities.

There are, however, two points where further improvement is needed.
First, we  need to guarantee the uniqueness of the results, namely, we must show that alternative methods for constructing joint probabilities for time of arrival and position would lead to the same conclusions when applied to  particle oscillations. We believe that the simplicity and robustness of our final result is a strong indication that it would persist in alternative treatments. Nonetheless, several different approaches for the   quantum treatment of the time of arrival exist, and their generalization to the particle-oscillation context could provide an independent theoretical confirmation. It is to be noted, however,  that many of these approaches to  the time of arrival---especially ones relying on probability currents---cannot guarantee positive probabilities when the time-of-arrival interferences are significant (see, Sec. 4.2.6). Hence, they may not be applicable to particle oscillations. In our opinion, the most promising candidates for an alternative treatment of the time of arrival are the axiomatic method of Kijowski \cite{Kij74} and the decoherent histories approach of Halliwell and Yearsley \cite{HalYea}.

An unsatisfactory point of our approach involves the probability  of no-detection, Eq. (\ref{nodet}). While it is mathematically well-defined, and its meaning is physically clear, i.e., not all neutrinos generated at the source will be detected, the definition Eq. (\ref{nodet}) follows from the normalization of the POVM and it is not obtained by independent arguments. The problem with an independent definition of this probability is that the natural candidate, namely, $\lim_{t \rightarrow \infty} Tr(\hat{S}_t \hat{\rho}_0 \hat{S}_t^{\dagger})$, is identically equal to 1 \cite{AnSav}. This fact is a special case of the quantum Zeno effect \cite{MiSu} that is generic in quantum theory. The Zeno effect creates problems for any {\em direct} definition of the probability of no-detection. The usual construction of the probability of no-detection proceeds through the computation of a state's
persistence amplitude. This method is reliable for the study of exponential decays but it does not constitute a fundamental definition, because  it cannot guarantee positive probabilities \cite{expon}.
The definition Eq. (\ref{nodet}) for the probability of no detection has been shown to  be physically sound in other applications of the formalism \cite{AnSav, AnSav2} and it reproduces the standard result in cases where independent methods for its determination exist (exponential decays of unstable states \cite{An08}). A better definition is desirable for the conceptual completeness of the method but the present one, Eq. (\ref{nodet}), is mathematically and physically sound and it does not curtail the applicability of our formalism.

\subsection{Application of the formalism to more general settings}

Starting from Eq. (\ref{povm2}), Eq.  (\ref{k1}) in Sec. 5.2  was derived using three approximations.
\begin{itemize}

\item The local approximation Eq. (\ref{pi2}) for the kernel $R$ in Eq. (\ref{prob3}).

\item The choice Eq. (\ref{psi0}) for the initial state of the $A$-particles.

\item The evaluation of the amplitudes ${\cal A}_{\alpha}(t, {\bf x})$, in the limit of negligible dispersion.
\end{itemize}

We  now proceed to show that our result, in particular, the expression for the oscillation wavelength is insensitive to the above approximations.

\subsubsection{More general samplings }

We first examine the case of  samplings $\hat{\Pi}_{\lambda}$ on the detected particles, that are more general than the ones considered in Sec. 5.2.1. In the process, we will also elaborate on the validity of the   local approximation Eq. (\ref{pi2})  that was employed in the derivation of Eq. (\ref{k1}).

To this end, we return to Eq.
 (\ref{prob3}) and, instead of position samplings on one product particle, we assume that the particle's momentum is also measured. The positive operators $\hat{P}_{\lambda}$ will therefore correspond to (coarse) joint measurements of phase space observables, i.e., the position ${\bf X}$ and momentum ${\bf P}$ of the recorded particle.

 An important property of phase space POVMs is that they are practically equivalent at the limit of large coarse-graining  \cite{Omn1, Omn4}. This limit refers to properties of macroscopic observables, such as the pointer variables in a measurement scheme. In particular, if $\hat{\Pi}_U$ and $\hat{\Pi}_U'$ are two different positive operators, corresponding to the same phase space region of volume $[U] >> \hbar^{n/2}$, then
 \begin{eqnarray}
\frac{ Tr|\hat{\Pi}_U - \hat{\Pi}_U'|}{Tr |\hat{\Pi}_U|} = O \left(\frac{\hbar^{n/2}}{[U]} \right) \label{microl}
 \end{eqnarray}
  where $n$ is the dimensionality of the phase space \cite{microloc}.

We  consider Eqs. (\ref{prob3}) and (\ref{R1}) with $\lambda$ corresponding to the pair $({\bf X}, {\bf P})$  and $\bar{\Pi}_{\lambda}:= \hat{\Pi}_{{\bf X},{\bf P}}$, a POVM that corresponds to joint position and momentum measurements with respective accuracies $\delta$ and $s_p$, such that $s_p \delta  >> \hbar$.
We evaluate Eq. (\ref{prob3}) for the class of initial states Eq. (\ref{psi0}),
\begin{eqnarray}
p_{\alpha}({\bf X}, {\bf P}) = \! \int_0^{\infty} \!\!\!\! dt \int_0^{\infty} \! \!\! \!dt' \! \! \int \!\! d^3x
d^3x' {\cal A}^*_{\alpha}({\bf x'},t', {\bf p}) {\cal A}_{\alpha}({\bf x},t, \bf{p})  G_1(t'-t) e^{i\epsilon_{th} (t'-t)} Q^{\alpha}_{{\bf X}, {\bf P}}({\bf x},
{\bf x'}, t-t'),
\end{eqnarray}
where
\begin{eqnarray}
Q^{\alpha}_{{\bf X}, {\bf P}}({\bf x},
{\bf x'}, s) =  \langle \phi_0|\hat{J}^-_{\alpha}({\bf x'}) \sqrt{\hat{\Pi}_{{\bf X}, {\bf P}}} e^{-i\sqrt{{\bf \hat{p}}^2 + M_D^2}(s)}
  \sqrt{\hat{\Pi}_{\lambda}} \hat{J}^+_{\alpha}({\bf x})|\phi_0 \rangle. \label{Q}
\end{eqnarray}

The amplitudes ${\cal A}_{\alpha}({\bf x},t, \bf{p})$ in Eq. (\ref{Q}) are identical to the ones of Eq. (\ref{AA}), however we have inserted explicitly their dependence on the mean momentum ${\bf p}$ of the initial state Eq. (\ref{psi0}). We have assumed that the differences in initial momentum, between different mass eigenstates, are many orders of magnitude smaller than the sampling accuracy $s_p$, so that without loss of generality they may be considered equal.

In Sec. 5.2.1 we showed that  for position samplings the kernel   $Q^{\alpha}_{{\bf X}, {\bf P}}({\bf x},
{\bf x'}, s)$ decays to zero for values of $s$ much larger than a suppression timescale $\tau_{sup}$. This time-scale arose from the diffusion of a state localized in an area of size $\delta$, under the action of the free particle Hamiltonian. This time-scale is not affected by the momentum samplings because momentum commutes with the free particle Hamiltonian.

We now consider that the detector lies at a distance $L$ from the source, such that the scale $\delta t_{ij}$ for the time-of-arrival interferences is many order of magnitudes smaller than $\tau_{sup}$. Hence, as in the derivation of Eq. (\ref{fin2}), we  evaluate   $Q_{\bf XP}({\bf x}, {\bf x'},s )$ at the limit $s  \rightarrow 0$. We  obtain

\begin{eqnarray}
p_{\alpha}({\bf X}, {\bf P}) =  \int_0^{\infty} dt \int_0^{\infty} dt' \int d^3x
d^3x' {\cal A}^*_{\alpha}({\bf x'},t', {\bf p}) {\cal A}_{\alpha}({\bf x},t, \bf{p}) e^{i\epsilon_{th}(t'-t)}  Q^{\alpha}_{{\bf X}, {\bf P}}({\bf x},
{\bf x'}), \label{prob9}
\end{eqnarray}
where we wrote $ Q^{\alpha}_{\bf XP}({\bf x}, {\bf x'}) =Q^{\alpha}_{{\bf X}, {\bf P}}({\bf x},
{\bf x'}, 0) $. The kernel $Q^{\alpha}_{{\bf X}, {\bf P}}$ defines a positive operator on the Hilbert space ${\cal H}_A$ of the oscillating particle. This operator does not correspond to a POVM because it is not normalized to unity. However, it does inherit the localization properties of the POVM $\hat{\Pi}_{\bf XP}$. In particular, if $|{\bf x} - {\bf X}| >> \delta $ and  $|{\bf x} - {\bf X}| >> \delta $, then $Q^{\alpha}_{\bf XP}({\bf x}, {\bf x'}) \simeq 0$ since the scale $\delta$ is macroscopic and hence much larger than the time scale associated to the microscopic currents. Moreover, from Eq. (\ref{Q}), we see that if the detected particle's wave-function (as described by $\hat{J}^+({\bf x}|\phi_0\rangle)$) is not localized on values of momentum within an accuracy $s_p$ from ${\bf P}$, the kernel $Q^{\alpha}$ again vanishes. If the momenta of all product particles are measured, then the momentum ${\bf p}$ of the incoming particle can be inferred, by conservation. Hence,  the
 action of $Q^{\alpha}$ on the vectors of ${\cal H}_A$  leads to momentum localization  with spread $s_P$. If some of the product particles are undetected then the measurement does not allow the inference of the A-particle's momentum with the  accuracy $s_p$, but  with a larger spread $s_p'$ depending on the details of the interaction.

 Consequently,  $Q^{\alpha}$ suppresses the values of the integral Eq. (\ref{prob9}), unless ${\bf x}$ and ${\bf x'}$ lie within distance of order $\delta$ from ${\bf X}$, and ${\bf p}$ lies within distance of order $s_p$ from ${\bf P}$. For the amplitudes in Eq. (\ref{prob9})  there is no coherence for the position variable at the scale $\delta$ (i.e.,  $\sigma << \delta$) and no coherence of the momentum variable at the scale of $s_p$ (i.e., $\sigma^{-1} << s_p$). Since $  s_p \delta>> \hbar$,
  this implies that   Eq. (\ref{prob9}) is well approximated by an expression of the form
 \begin{eqnarray}
 p_{\alpha}({\bf X}, {\bf P}) = K({\bf X}, {\bf P})  \int d^3 x \int d^3 p \chi_{\delta} (|{\bf x} - {\bf X}|) \chi_{s_p}' (|{\bf p} - {\bf P}|) \\ \nonumber
 \times \int_0^{\infty} dt \int_0^{\infty} dt' \int d^3x
d^3x' {\cal A}^*_{\alpha}({\bf x},t', {\bf p}) {\cal A}_{\alpha}({\bf x},t, \bf{p}) e^{i\epsilon_{th}(t'-t)}, \label{smearprob}
 \end{eqnarray}
where $\chi_{\delta}(x), \chi'_{s_p}(p)$ are  positive functions localized around $x = 0$ with width $\delta$, and $p = 0$ with width $s_p$, respectively. The multiplicative constant $K({\bf X}, {\bf P})$ has been inserted so that we can choose $\chi_{\delta}(x)$, $\chi'_{s_p}(p)$ to be normalized as
\begin{eqnarray}
\int d^3x \chi_{\delta}(x) = 1 \;\; \; \int d^3p \chi_{\delta}(p) = 1.
\end{eqnarray}

Eq. (\ref{smearprob}) shows that the effect of the uncertainties in the position and momentum samplings essentially correspond to  a smearing of the probability distribution Eq. (\ref{fin2}), at a scale of $\delta$ for position and $s_p$ for momentum.
If the relative sampling errors $\delta/L$ and $\sigma_p/P$ are much smaller than unity, then the oscillatory behavior of the probability density Eq. (\ref{fin2}) remains unchanged.

\subsubsection{General initial state}

In our previous derivation of the probabilities for particle oscillations we assumed a superposition of Gaussian  initial states Eq. (\ref{psi0}). However, in the Appendix B, we show that the initial state for particle oscillation experiments  is a mixed state of the form Eq. (\ref{init}), that can be expressed as a convex combination of states $\hat{\rho}_{\xi 0}$;  $\xi$ corresponds to quantities that are distinguished macroscopically in the production process.
The macroscopic variables $\xi$  allow for a coarse determination of the particles' position and momentum. We assume that the initial states of the target particles in the production region (see Appendix B) are quasi-classical so that no spatial quantum coherence is preserved in $\hat{\rho}_{\xi0}$. This means that we can express $\hat{\rho}_{\xi0}$ as a  combination of localized pure states such as Eq. (\ref{psi0}), i.e.,

\begin{eqnarray}
\hat{\rho}_{\xi 0} = \sum_{ij} U_{ \beta i } U^*_{\beta j} \int dq dp f_{\xi}(q, p) |q p\rangle_i {}_j\langle q p |, \label{roxi}
\end{eqnarray}
where
\begin{eqnarray}
\langle x|q p \rangle_i = \frac{1}{(2\pi \sigma^2)^{1/4}} e^{ - \frac{(x - q)^2}{4 \sigma^2} + i p x},
\end{eqnarray}
 the index $i$ labels the mass eigenspaces. The  function $f_{\xi}(q, p)$ is non-negative and it corresponds to a distribution of inferred values of position $q$ and momentum $p$ for the oscillating particle \cite{psymbol}.
In Eq. (\ref{roxi}) we have implicitly assumed that the uncertainties in the initial momentum $p$ due to incoherent mixing are much larger than the differences $|p_i - p_j|$ of momenta in Eq. (\ref{psi0}).

Since the probabilities Eq. (\ref{main}) are convex functions of the initial states, we can write
\begin{eqnarray}
p_{\alpha}(L) = \int dq dp f_{\xi}(q, p) p_{\alpha}^{(q,p)}(L),
\end{eqnarray}
where $p_{\alpha}^{(q,p)}(L)$ are the probabilities Eq. (\ref{fin2}) computed for an initial state characterized by mean position and momentum $q$ and $p$ respectively.  The Hamiltonian for free particles is translationally invariant, hence, the variation of $q$ can be absorbed in a redefinition of $L \rightarrow L + q$.  For the regime where Eq. (\ref{fin2}) holds,

\begin{eqnarray}
p_{\alpha}(L) = \int dq dp f_{\xi} (q, p) K'(q, p) \left[ \sum_i  |U^*_{\beta i}U_{\alpha i }|^2  e^{-2 \frac{\Gamma_i}{v_i}(L+q)} \right. \nonumber \\
\left.+ \sum_{i < j} 2 Re \left( U^*_{\beta i}U_{\alpha i } U_{ \beta j} U^*_{\alpha j} e^{-(\frac{\Gamma_{i}}{v_i} + \frac{\Gamma_j}{v_j})(L+q)}    e^{  i k_{ij} (L+q)}
 \right) \right], \label{meanosc}
\end{eqnarray}
where $k_{ij}$ is computed from Eq. (\ref{k1}) for $p_i = p_j = p$.

We note that the maximum scale of variation for $q$ is set by the dimension $d$ of the production region that satisfies $ d << L$. We also assume that the uncertainty in the specification of the initial momentum $\Delta p << \bar{p}$, where $\bar{p}$ is the mean momentum corresponding to the probability distribution $f_{\xi}(q, p)$. The oscillating phases $\phi_{ij} = k_{ij}(L+q)$ is  a variable of $q$ and $p$. The mean value $\bar{\phi} = \langle k_{ij} (L+q)\rangle$ with respect to $f_{\xi}$ approximately equals $\bar{k}_{ij} L$, where $\bar{k}_{ij}$ is the oscillation wave-number evaluated for the mean momentum $ \bar{p}$. The corresponding spread in the phase is
\begin{eqnarray}
\Delta \phi_{ij} < L \Delta k_{ij} + \Delta q \bar{k}_{ij} < \bar{\phi}_{ij} (\frac{d}{L} + \frac{\Delta p}{p}) << \bar{\phi}_{ij}. \label{spread}
\end{eqnarray}
 This implies that the substitution of the mean value of $e^{i \phi_{ij}}$ with
  $e^{i \bar{\phi}_{ij}}$ is a good approximation\footnote{ The `mean value' and `deviation' of quantities such as $\phi_{ij}$ refers to the distribution
  of the parameters of Eq. (\ref{psi0}) in the weighting function. These objects describe features of the initial state
   and {\em not}  statistics of measurement outcomes.}. Hence there is no significant modification
    in the oscillation phase when the effects of incoherent mixing is taken into account.

To summarize, we have shown that    Eq. (\ref{k1})  for the oscillation wave-number remains valid for a large class of initial states of the form Eq. (\ref{roxi}).
The latter exhibit no spatial quantum coherence at  scales larger than   the width $\sigma$ of the Gaussian states Eq. (\ref{psi0}).  Hence, the effect of an initial state Eq. (\ref{roxi}) to the detection probability is formally similar to that of sampling uncertainties  sampling, i.e., they both induce a smearing of the distribution probability by {\em classical} uncertainties. Thus, both sampling uncertainties and  uncertainties from the initial condition contribute to the random errors of the measurement:  they can only degrade the oscillation pattern of Eq. (\ref{fin2}), and not modify its main features.


\subsubsection{Dispersion effects}

In order to examine the effect of dispersion on the probabilities, we calculate the amplitude  Eq. (\ref{psi0}), by expanding $E_i(p)$ to second order and $\Gamma_i(p)$ to first order around $p_i$, i.e.,
\begin{eqnarray}
E_i(p) = E_i + v_i(p - p_i) + \frac{1}{2} \mu_i (p - p_i)^2
\Gamma_i(p) = \Gamma_i + \Delta_i (p - p_i)
\end{eqnarray}
where
\begin{eqnarray}
E_i = \sqrt{m_i^2 + p_i^2}, \hspace{0.5cm} \Gamma_i = \Gamma_i(p_i),  \hspace{0.5cm} v_i = \frac{p_i}{E_i},  \hspace{0.5cm}
\mu_i = m_i^2/E_i^3 , \hspace{0.5cm} \Delta_i = (d \Gamma_i/ d p)(p_i).
\end{eqnarray}

We then obtain
\begin{eqnarray}
{\cal A}_{\alpha}(t, x) = \sum_{i} U^*_{\beta i}U_{\alpha i } \frac{
(\sigma^2/\pi)^{1/4}}{\sqrt{\sigma^2 + i \mu_i t}} e^{i (p_ix - E_i t) - \Gamma_it} \nonumber \\
 \times \exp \left[ - \frac{\sigma^2}{2(\sigma^4 + \mu_i^2 t^2)} \left( (x - v_i t - \frac{\Delta_i \mu_i t^2}{\sigma^2})^2 + \frac{\Delta^2_i \mu_i^2}{\sigma^4} t^4 + \sigma^2 \Delta_i^2 t^2 \right) \right. \nonumber \\
  \left.
  + \frac{i \mu_i t}{2(\sigma^4 + \mu_i^2 t^2)} \left( (x - v_i t - \frac{\sigma^2 \Delta_i}{\mu_i})^2 - \frac{\sigma^4 \Delta_i^2}{\mu_i^2} + \Delta_i t^2 \right) \right].
 \label{AA2}
\end{eqnarray}

The effects of the dispersion become significant for time-scales such that $ \sigma^2 \sim \mu_i t$. Since the amplitudes Eq. (\ref{AA2}) are peaked for times $t \simeq L/v_i$ the no-dispersion regime corresponds to the condition $\frac{m_i^2 }{E_i^3 \sigma^2} \frac{L}{v_i} << 1$. The spread of the Gaussian $\sigma$ is a free parameter that may take any value between the atomic and the weak timescale. For $\sigma \sim 10^{-14} m $, $E_i \sim 100 MeV$, the non-dispersive regime provides an adequate description if $L << 10^{10} km$. A value of $\sigma \sim 10^{-19}m $ is required, in order to see a failure of the dispersive regime at a distance $L$  relevant to
neutrino oscillation experiments.

\subsubsection{Particle detection through several processes}

We finally consider the case of a set-up where oscillating particles are detected by different processes. We show that the probability distributions for each process are added consistently into the expression for the total oscillation probability.

Let us label by the index $n = 1, \ldots , N$ the different channels through which oscillating particles can be detected. Each channel  corresponds to a different Hilbert space ${\cal H}_n$, that describes the non-oscillating particles  participating in the $n$-th process. The construction in Sec. 5.2.1. then can be used with the following changes.
\begin{enumerate}{}
\item The Hilbert space ${\cal H}_r$ is the tensor product $\otimes_n {\cal H}_n$.
\item The initial state $|\phi_0\rangle$ of particles on the detector is of the form $|\phi_0^{(1)} \rangle \otimes \ldots \otimes |\phi_0^{(N)} \rangle$, in terms of states $|\phi_0^{(n)} \rangle \in {\cal H}_n$.

\item The Hamiltonian $\hat{H}_r$ on the Hilbert space ${\cal H}_r$ splits as
\begin{eqnarray}
\hat{H}_r = \hat{H}_{(1)} \otimes 1 \otimes \ldots \otimes 1 + \ldots + 1 \otimes 1 \otimes \ldots \otimes \hat{H}_{(n)},
\end{eqnarray}
in terms of Hamiltonian operators $\hat{H}_{(n)}$ on ${\cal H}_n$.

    \item The  currents $\hat{J}^{\pm}_{\alpha}({\bf x})$ in Eq. (\ref{hi}) for the interaction Hamiltonian are of the form
    \begin{eqnarray}
    \hat{J}^{\pm}_{\alpha}({\bf x}) = \hat{J}^{(1)\pm}_{\alpha}({\bf x}) \otimes 1 \otimes \ldots \otimes 1 + 1 \otimes\hat{J}^{(2)\pm}_{\alpha}({\bf x}) \otimes \ldots \otimes 1 + \ldots + 1 \otimes 1 \otimes \ldots \otimes \hat{J}^{(N)\pm}_{\alpha}({\bf x}),
    \end{eqnarray}
in terms of current operators $\hat{J}^{(n)\pm}_{\alpha}({\bf x})$ on ${\cal H}_n$.

\item The positive operators $\hat{\Pi_{\lambda}}$, corresponding to  measurements of position of product particles, are of the form
\begin{eqnarray}
\hat{\Pi}_{\lambda} = \hat{\Pi}_{\bf X}^{(1)} \otimes 1 \otimes \ldots \otimes 1 + \ldots + 1 \otimes 1 \otimes \ldots \otimes \hat{\Pi}^{(N)}_{\bf X},
\end{eqnarray}
in terms of positive operators $\hat{\Pi}^{(n)}_{\bf X}$ on ${\cal H}_n$ such that $\hat{\Pi}^{(n)}_{\bf X} |\phi^(n)_0\rangle = 0$.
\end{enumerate}

Then Eq. (\ref{prob3}) follows,  with the kernel $R_{\lambda}^{\alpha}({\bf x}, {\bf x'}, t\!-\!t')$ expressed as
\begin{eqnarray}
R_{\lambda}^{\alpha}({\bf x}, {\bf x'}, t\!-\!t') = \sum_n
 \langle \phi_0^{(n)}|\hat{J}^{(n)-}_{\alpha}({\bf x'}) \sqrt{\hat{\Pi}^{(n)}_{\bf X}} e^{-i\hat{H}_{(n)}(t'\!-t)}
  \sqrt{\hat{\Pi}^{(n)}_{\bf X}} \hat{J}^{(n)+}_{\alpha}({\bf x})|\phi_0^{(n)}\rangle.
  \label{Rn}
\end{eqnarray}

The remaining analysis of Sec. 5.2 passes unchanged, leading to
\begin{eqnarray}
p_{\alpha}(L) = \sum_n K'_n \left( \sum_{i}  |U^*_{\beta i}U_{\alpha i }|^2  e^{-2 \frac{\Gamma_{i}}{v_i}L} + \sum_{i < j} 2 Re \left( U^*_{\beta i}U_{ \alpha i} U_{\beta j } U^*_{\alpha j} e^{-(\frac{\Gamma_{i}}{v_i} + \frac{\Gamma_j}{v_j})L}    e^{  i k^{(n)}_{ij} L}) \right) \right). \label{finn}
\end{eqnarray}
 This is a convex combination of the probability densities Eq. (\ref{fin2}), where  each process is weighted by different multiplicative coefficients $K'_n$. The  oscillation pattern of each term is modulated by  wave-numbers $k_{ij}^{(n)}$ as in Eq. (\ref{k1}), differing only by the values $\epsilon^{(n)}_{th}$ of the process' threshold. In other words, the detection channels are statistically independent and the resulting probabilities can be added.

\begin{appendix}
\section{Coarse-graining in time-of-arrival measurements }

In order to demonstrate how the decoherence time scale used in the definition of the POVM Eq. (\ref{povm2}) arises, we employ the formalism of Sec. 5.1 for a particle detection by a macroscopic apparatus.
 We also examine special cases for the POVM Eq. (\ref{povm2}), showing that it recovers von Neumann measurements in a specific regime.

  We assume that  the particle is annihilated or is absorbed by the apparatus, as part of the detection process. It is therefore convenient to employ the Fock space
\begin{eqnarray}
{\cal F}_s = {\bf C} \oplus {\cal H}_s \oplus ({\cal H}_s  \otimes {\cal H}_s)_S \oplus \ldots
\end{eqnarray}
associated to the Hilbert space ${\cal H}_s$ of a single particle; $S$ denotes symmetrization. We denote the Hilbert space describing the apparatus's degrees of freedom as ${\cal H}_{app}$.
The Hilbert space of the combined system is then ${\cal F}_s \otimes {\cal H}_{app}$.

The Hamiltonian of the combined system is $\hat{H}_s \otimes 1 + 1 \otimes \hat{H}_{app} + \hat{H}_{int}$,
where $\hat{H}_s$ describes the dynamics of the
microscopic system and $\hat{H}_{app}$ describes the dynamics of the apparatus. The general form of an interaction Hamiltonian describing the absorption of the microscopic particle in the process of detection is
\begin{eqnarray}
\hat{H}_{int} = \sum_i\int d^3 x (\hat{a}_i({\bf x}) \hat{J}_i^+({\bf x}) + \hat{a}_i^{\dagger} ({\bf x}) \hat{J}_i^-({\bf x})),
\end{eqnarray}
where $\hat{a}_i({\bf x}), \hat{a}^{\dagger}_i({\bf x})$ are the annihilation and creation operators on ${\cal F}_s$,
$i$ labels non-translational degrees of freedom (e.g., spin), and  $\hat{J}^{\pm}({\bf x})$ are current operators
on ${\cal H}_{app}$ with support in the region $D$ where the detector is
located. We assume that  the microscopic system is prepared in the  single-particle state $|\psi_0\rangle$ and
that the initial state of the apparatus $| \Psi_0\rangle$ is stationary. We set the scale of energy so that $\hat{H}_{app}|\Psi_0 \rangle = 0$.
 Since the particle is absorbed by the detector we require that $\hat{J}^-({\bf x}) |\Psi_0 \rangle = 0$.

The pointer variables correspond to a POVM $1 \otimes \hat{\Pi}_{\lambda}$, where $\hat{\Pi}_{\lambda}$ is a POVM
on ${\cal H}_{app}$ defined so that $\int d \lambda \hat{\Pi}_{\lambda} = \hat{P}$ is the projector on the
excited states of the detector. In order to ensure that no detection events occur unless a particle enters the detector, we must assume that $[\hat{H}_{app}, \hat{P} ]   = 0$. It follows that $[\hat{P}, \hat{H}_s \otimes 1 + 1 \otimes \hat{H}_{app}] = 0$. Hence,  Eq. (\ref{perturbed}) applies,

\begin{eqnarray}
\hat{C}_{\lambda, t} \left(|\psi_0 \otimes \rangle |\Psi_0\rangle \right)= \sum_i \int d^3 x \left(\psi_i({\bf x}, t) |0 \rangle\right) \otimes \left( e^{i \hat{H}_{app}t}\sqrt{\hat{\Pi}_{\lambda}} \hat{J}_i({\bf x}) |\Psi_0\rangle \right) ,
\end{eqnarray}
where $\psi({\bf x}, t)$ is the evolution of $|\psi_0\rangle$ under $e^{- i \hat{H}_st}$ in the position representation.

Then,
\begin{eqnarray}
Tr \left(\hat{C}_{\lambda, s} \hat{\rho}_0 \hat{C}_{\lambda, s'}^{\dagger} \right) = \sum_{ij}\!\!\int \! \! d^3 x d^3 x' \psi_i({\bf x}, s) \psi^*_j({\bf x'}, s')
\langle \Psi_0| \hat{J}_j^-({\bf x'}) \sqrt{\hat{\Pi}_{\lambda}} e^{- i \hat{H}_{app}(s' - s)} \sqrt{\hat{\Pi}_{\lambda}} \hat{J}_i^+({\bf x})|\Psi_0\rangle. \label{crc}
\end{eqnarray}

In order to obtain a probability density $p(\lambda, t)$ we
employ a smearing function $\sqrt{f_{\tau}(s-t)}$ on both arguments $s, s'$ of Eq. (\ref{crc}); $\tau$ corresponds to the temporal resolution of the apparatus. We note that the
$s$ and $s'$ dependence in Eq. (\ref{crc}) arises from two independent terms: $\psi_i({\bf x}, s)$ and the currents $\hat{J}^+{\pm}({\bf x}, s)$. In the former, the time dependence
refers to the evolution of the microscopic system prior to detection, while in the latter it refers to properties of the macroscopic apparatus. The temporal resolution of the apparatus is determined by the latter terms.

In order to identify the temporal resolution of the apparatus, we examine the expectation value
\begin{eqnarray}
P(s) = \langle \Psi_0| \hat{J}_j^-({\bf x'}) \sqrt{\hat{\Pi}_{\lambda}} e^{- i \hat{H}_{app}s} \sqrt{\hat{\Pi}_{\lambda}} \hat{J}_i^+({\bf x})|\Psi_0\rangle
\end{eqnarray}
from Eq. (\ref{crc}). This satisfies
\begin{eqnarray}
|P(s)| \leq |\langle \Psi_0| \hat{J}_j^-({\bf x'}) \sqrt{\hat{\Pi}_{\lambda}} e^{- i \hat{H}_{app}s} \sqrt{\hat{\Pi}_{\lambda}} \hat{J}_i^+({\bf x})|\Psi_0\rangle| \cdot ||\hat{\Pi}_{\lambda}^{1/4} e^{-i \hat{H}_{app}s} \hat{\Pi}_{\lambda}^{1/4}||_2, \label{bbound}
\end{eqnarray}
where $||\cdot||_2$ stands for the Hilbert-Schmidt norm. The $s$-dependent terms in the right-hand side of Eq. (\ref{bbound}) equal
$\sqrt{Tr \left(\sqrt{\hat{\Pi}_{\lambda}} e^{- i \hat{H}_{app}s} \sqrt{\hat{\Pi}_{\lambda}} e^{ i \hat{H}_{app}s}\right)}$ and they can be written as
\begin{eqnarray}
\sqrt{\frac{Tr[e^{ i \hat{H}_{app}s} \hat{\rho}e^{- i \hat{H}_{app}s} \hat{\rho}]}{Tr\hat{\Pi}_{\lambda}}},
\end{eqnarray}
in terms of a density matrix $\hat{\rho} = \sqrt{\hat{\Pi}_{\lambda}}/\sqrt{Tr \hat{\Pi}_{\lambda}}$.

The object $Tr[e^{ i \hat{H}_{app}s} \hat{\rho}e^{- i \hat{H}_{app}s} \hat{\rho}]$ is known as the persistence probability of the state $\hat{\rho}$; its properties have been studied in relation to exponential decay \cite{expon}. Typically, a persistence probability is characterized by a decay time-scale $\tau_{dec}$, for $s >> \tau_{dec}$ the persistence probability becomes negligible. In the context of Eq. (\ref{crc}), this implies that
  interference between terms with different values of $s$ and $s'$ in Eq. (\ref{crc}) are suppressed if $|s - s'| >> \tau_{dec}$.
In particular, this implies that the interference terms in Eq. (\ref{add}) become negligible if the width of the intervals is much larger than $\tau_{dec}$. Restricted to such intervals,  Eq. (\ref{prob1}) defines a probability measure. Equivalently, one may employ Eq. (\ref{ampl6}) with  functions $f_{\tau}$ characterized by values of the smearing parameter $\tau >> \tau_{dec}$.

In order to estimate the decay time-scale $\tau_{dec}$, one should  consider specific models for the self-dynamics of the pointer variable. In quantum Brownian motion models, for example, one  describes the pointer variable
as the position of an harmonic oscillator, of effective mass $\mu_*$, in contact with a heat bath
at temperature $T$.  Taking the positive operators $\hat{\Pi}_{\lambda}$ to
 correspond to a sampling of the pointer variable with an accuracy of order $\delta$, we find that
$\tau_{dec} \sim \tau_{rel} (\mu_* T \delta^2)^{-1}$, where $\tau_{rel}$ is the relaxation time for the pointer variable \cite{PHZ93}.

In conclusion, a choice of $\tau$ in the smearing functions of the order of $\tau_{dec}$, as determined by the dynamics of the apparatus  guarantees that the interference terms in Eq. (\ref{add}) are negligible. Hence, $\tau_{dec}$ sets a scale of minimal temporal coarse-graining necessary for the quasi-classical attribution of definite macroscopic properties to the pointer variables.

\paragraph{Special cases.}

 If the states $\psi_i({\bf x}, s)$ depend on $s$ at a scale much smaller than the characteristic time-scales  in $\sqrt{\hat{\Pi}_{\lambda}} e^{- i \hat{H}_{app}t}
\sqrt{\hat{\Pi}_{\lambda}}$,
then we can approximate $\sqrt{\hat{\Pi}_{\lambda}}
e^{- i \hat{H} (s' -s)}  \sqrt{\hat{\Pi}_{\lambda}} \simeq \hat{\Pi}_{\lambda}$ in Eq. (\ref{crc}).  In this regime, the time of arrival cannot be distinguished. Therefore, we consider the partial probabilities  $p(\lambda)$ obtained by integrating $p(t, \lambda)$ in Eq. (\ref{ampl6}), over the time of arrival.  Then, Eq. (\ref{probdd}) applies and we obtain
\begin{eqnarray}
p(\lambda) = \frac{1}{2 \pi \tau^2} \sum_{ij} \int d^3x d^3 x'  \hat{Q}_{ij}(\lambda)({\bf x}, {\bf x'}) \left(\int_0^T ds \psi_i({\bf x}, s)\right)
\left(\int_0^T ds' \psi_j({\bf x'}, s')\right), \label{case2}
\end{eqnarray}
 where $\hat{Q}(\lambda)$ is a family of positive operators on the Hilbert space ${\cal H}_s$   with matrix elements
\begin{eqnarray}
\langle {\bf x'} |\hat{Q}_{ij}(\lambda) |{\bf x}\rangle = 2 \pi \tau^2 \langle \Psi_0| \hat{J}_j^-({\bf x'}) \hat{\Pi}_{\lambda} \hat{J}_i^+({\bf x}) |\psi_0\rangle, \label{QQ}
\end{eqnarray}
where the factor $2 \pi \tau^2$ was inserted into the definition of $\hat{Q}$ so that $\langle \psi|\hat{Q}(\lambda)|\psi\rangle$ has the correct dimensions for a probability density over $\lambda$.

If, on the other hand,  time-scales in $\psi_i({\bf x}, s)$ are much larger than $\tau_{dec}$, we can substitute the smearing functions $\sqrt{f(\tau)} $ in Eq. (\ref{smearing}) with  delta-functions, i.e., we set $\sqrt{f_{\tau}(t-s)} = (2 \pi \tau^2)^{1/4} \delta(t-s)$. We then obtain

\begin{eqnarray}
p(t, \lambda) = \frac{1}{\sqrt{2 \pi \tau^2}} \sum_{ij} \int d^3x d^3 x' \psi_i({\bf x}, t) \psi_j({\bf x'}, t) \hat{Q}_{ij}(\lambda) ({\bf x}, {\bf x'}). \label{case1}
\end{eqnarray}

The function  $p(t, \lambda)$ in Eq. (\ref{case1}) is a proper density with respect to the time of arrival $t$, so it can be integrated over $t$ yielding a probability density   $p(\lambda)$
\begin{eqnarray}
p(\lambda) = \frac{1}{\sqrt{2 \pi \tau^2}} \int_0^{\infty}  dt \langle \psi_t| \hat{Q}(\lambda) |\psi_t \rangle, \label{intprob4}
\end{eqnarray}
where $|\psi_t \rangle = e^{-i \hat{H}_st} |\psi_0 \rangle$.

In Eq. (\ref{intprob4}) the probability $p(\lambda)$ is defined as an integral over the single-time probabilities  $\langle \psi_t| \hat{Q}(\lambda) |\psi_t \rangle$. As we also argued in Sec. 4.2.3, this result applies only to the specific regime we considered here.

Von Neumann measurements can be recovered from Eq. (\ref{case1}), if
the initial states are sufficiently localized. This means that the distribution of the time of arrival  should be sharply peaked around  classical arrival time $t_{cl}$, with a spread indistinguishable by the temporal resolution of the detector. Indeed, setting $t = t_{cl}$ in Eq. (\ref{case1}),
    we obtain $p(\lambda) \sim \langle \psi_{t_{cl}}|\hat{Q}(\lambda)|\psi_{t_{cl}}\rangle$.

When we integrate $p(t, \lambda)$  over $\lambda$ in Eq. (\ref{case1}), we construct a probability density for the time of arrival $t$,
\begin{eqnarray}
p(t) \sim   \langle \psi_t| \hat{B} | \psi_t \rangle,
\end{eqnarray}
where $\hat{B} = \int d \lambda \hat{Q}(\lambda)$. In this regime,  $\hat{B}$ acts as a probability current operator. Its explicit form depends on the interaction between the microscopic system and the detector, i.e., on the currents $\hat{J}_{\pm}$. We also note that in the regime under consideration, decoherence has rendered  the time of arrival  a  quasi-classical variable, so that the expectation value of a current operator is positive-valued---see Refs. \cite{HalYea}.

We can approximate $\hat{B}$ with an operator whose Wigner function is of the form
\begin{eqnarray}
 W(x, p) = f(|p_x|) \delta(x - L), \label{wig}
 \end{eqnarray}
by assuming that the size $d$ of the detector is much smaller than the source-detector distance $L$ and that the detector is homogeneous, so that for ${\bf x}$ and ${\bf x'}$ lying in the detection region, $\langle \Psi_0| \hat{J}_j^-({\bf x'})  \hat{J}_i^+({\bf x}) |\psi_0\rangle$ is a function of $|{\bf x} - {\bf x'}|$.

The positive function $f$ in Eq. (\ref{wig}) depends on the details of the interaction. The standard probability current (modulo terms of higher order in $\hbar$) corresponds to $f(|p_x|) \sim |p_x|$. For $f$ a constant, we obtain a particularly simple form,
$p(t) \sim |\psi(L, t)|^2$.

\section{The role of the initial state in neutrino production}

In this section, we re-derive a fundamental property of quantum measurement theory, in a context relevant for neutrino oscillations. We consider the most general set-up for neutrino (or any particle) production and we show that all physical information that can be extracted by measurements is encoded in the reduced density matrix of the neutrinos. One  reason for including this derivation is that a large number of articles associate the standard expression of oscillation wavelength with special initial states of the neutrinos. The results in this section  allow us to argue that such special conditions are highly implausible in a treatment of the production process. Another reason is that this derivation demonstrates conclusively why a probability assignment, linear with respect to the initial density matrix, is necessary in any precise quantum mechanical treatment of particle oscillations.

 We consider a production reaction of the form
\begin{eqnarray}
a + b \rightarrow \nu + d_1 + d_2 +\ldots, \label{event1}
\end{eqnarray}
 where $a$ are the particles of an incoming beam, $b$ are the target particles, and $d_i$  product particles of the reaction other than the neutrinos. There are many $b$-type particles in the target, but each neutrino is produced in a reaction that only one of them participates. Hence, to each production event we can associate a unitary operator
 \begin{eqnarray}
 {\cal V}: {\cal H}_a \otimes {\cal H}_{b} \rightarrow {\cal H}_{\nu} \otimes {\cal H}_{d_1} \otimes {\cal H}_{d_2} \otimes \ldots, \label{un0}
  \end{eqnarray}
where ${\cal H}_a, {\cal H}_b, \ldots$  are single-particle Hilbert spaces for the $a, b, \ldots$ particles.

We define the corresponding Fock spaces ${\cal F}_a = {\bf C} \oplus {\cal H}_a \oplus ({\cal H}_a \otimes {\cal H}_a)_{S,A} \oplus \ldots...$, and similarly for other particles;  $S$ or $A$ stands for symmetrization or anti-symmetrization respectively. The total Hilbert space of the system is ${\cal H}_{tot} = {\cal F}_a \otimes {\cal F}_b \otimes {\cal F}_{\nu} \otimes {\cal F}_{d_1} \otimes {\cal F}_{d_2} \otimes \ldots$.

We typically assume that the reaction of a single particle in the beam corresponds to a single element of the statistical ensemble described by the quantum state. Hence, the $a$-particles are initially found in a single-particle state $|\psi_a \rangle_a$. The $b$-particles are assumed initially in an $N$-particle state $|\phi_{1}, \ldots \phi_{N} \rangle_b$. The initial state for the neutrinos  and the $d_i$ particles is the vacuum. Hence, the initial state for the system is
\begin{eqnarray}
|\psi_a \rangle \otimes |\phi_1, \ldots, \phi_N \rangle \otimes |0 \rangle_{\nu} \otimes |0 \rangle_{d_1} \otimes |0 \rangle_{d_2} \otimes \ldots
\end{eqnarray}

An $a$-particle interacts through the channel Eq. (\ref{un0}) with one of the $b$ particles, thus the quantum state after the interaction is
\begin{eqnarray}
| \Psi \rangle = |0 \rangle_a \otimes \sum_i \left(|\phi_1, \ldots, \displaystyle{\not} \phi_i, \ldots, \phi_N \rangle_b \otimes \left[{\cal V} (|\psi\rangle_a \otimes |\phi_i\rangle_b)\right]_{\nu, d_1, d_2, \ldots} \right),
\end{eqnarray}
where $|\phi_1, \ldots, \displaystyle{\not} \phi_i, \ldots \phi_N \rangle_b$ is the $(N-1)$-particle state obtained by removing the $i$-th entry from the $N$-particle state $|\phi_1, \ldots, \phi_N \rangle$.

 Reactions of the form Eq. (\ref{event1}) can often be identified macroscopically by the detection of some of  the produced particles $d_i$, and by the ensuing measurement of some of their properties (position, momentum, and so on) collectively denoted by $\xi$. The variables $\xi$ may be defined at different moments of time and they may include the times of detection for the $d_i$ particles. In Ref. \cite{Glash}, it is argued that the neutrino and the other produced particles are initially entangled  and that a detection of the latter, causing disentanglement, is a necessary condition for neutrino oscillations.

Let  $\Pi_{\xi}$ be a POVM on ${\cal H}_{tot}$ that corresponds to measured properties of the $d_i$ particles, as above. Then the ensemble of all neutrino production events splits into sub-ensembles characterized by different values of $\xi$. Each sub-ensemble is described by the density matrix
\begin{eqnarray}
\hat{\rho}_\xi = \frac{\sqrt{\hat{\Pi_{\xi}}} |\Psi\rangle \langle \Psi|  \sqrt{\hat{\Pi_{\xi}}}}{\langle \Psi| \hat{\Pi}_{\xi}|\Psi\rangle}.
\end{eqnarray}

Subsequent evolution of $\hat{\rho}_{\xi}$ proceeds under a unitary operator $\hat{U}_{tot}$ on ${\cal H}_{tot}$ for the dynamics of all degrees of freedom. To leading approximation, there is no interaction between neutrinos so $\hat{U}_{tot}$ factorizes as $\hat{U}_{\nu}\otimes\hat{U}_{r}$, where $\hat{U}_r$ is defined on the Hilbert space of the remaining degrees of freedom ${\cal H}_r = {\cal F}_a \otimes {\cal F}_b \otimes {\cal F}_{d_1} \otimes {\cal F}_{d2} \otimes \ldots$. Then, for any subsequent measurement on the neutrinos, the full information for this sub-ensembles is contained in the reduced neutrino density matrix
\begin{eqnarray}
\hat{\rho}_{\nu, \xi} = \frac{Tr_{{\cal H}_r}\left[ \hat{U}_{\nu}\otimes \hat{U}_{r})\sqrt{\hat{\Pi_{\xi}}} |\Psi\rangle \langle \Psi|  \sqrt{\hat{\Pi_{\xi}}}(\hat{U}^{\dagger}_{\nu} \otimes \hat{U}_r^{\dagger}) \right]}{\langle \Psi| \hat{\Pi}_{\xi}|\Psi\rangle} = \hat{U}_{\nu} \frac{Tr_{{\cal H}_r}\left[ \sqrt{\hat{\Pi_{\xi}}} |\Psi\rangle \langle \Psi|  \sqrt{\hat{\Pi_{\xi}}} \right]}{\langle \Psi| \hat{\Pi}_{\xi}|\Psi\rangle} U_{\nu}^{\dagger}.
\end{eqnarray}

The probabilities for the full ensemble will be obtained from the convex combination of $\hat{\rho}_{\nu, \xi}$ weighted by the probabilities $\langle \Psi|\hat{\Pi}_{\xi}|\Psi \rangle$ corresponding to each sub-ensemble. Hence, the probabilities for any measurements of the neutrino will be determined by the reduced density matrix
\begin{eqnarray}
\hat{\rho}_{\nu} = \hat{U}_{\nu} \hat{\rho}_{\nu 0} \hat{U}^{\dagger}_{\nu},
\end{eqnarray}
where
\begin{eqnarray}
\hat{\rho}_{\nu 0} = \sum_{\xi} Tr_{{\cal H}_r}\left(\sqrt{\hat{\Pi_{\xi}}} |\Psi\rangle \langle \Psi|  \sqrt{\hat{\Pi}_{\xi}}\right) \label{init}
\end{eqnarray}
is the initial reduced density matrix of the produced
neutrinos.

We showed therefore that, whatever the neutrino production may be, any subsequent measurement on the neutrinos  {\em depends only on the reduced density matrix of the neutrinos} that evolves under the neutrino self-dynamics. In other words, the details of the particle production are all encoded in the initial density matrix, the same way that the details of the detection are encoded in a POVM. Moreover, the existence of any macroscopically retrievable information from the production process, implies the existence of different macroscopically distinguishable sub-ensembles. Thus, $\hat{\rho}_{\nu}$ is necessarily a mixed state.

We also note the importance of having a probability assignment linear with respect to the initial state $\hat{\rho}_{\nu}$: without linearity it is impossible to obtain a consistent combination of different macroscopically distinguishable sub-ensembles. This means that any probability assignment  must be linear with respect to the initial state, otherwise it cannot provide the basis of sound probabilistic reasoning. This point is important in the critique of existing approaches to particle oscillations.

\end{appendix}


\begin{thebibliography}{}
\bibitem{inter} If $f(t)$ is not sharply peaked around a point
then the pointer variable is correlated to the time average of
$\hat{A}$ in the time interval $[0, t]$. See, Ref. \cite{AnSav07, BoHa}.
\bibitem{comment} Eq. (\ref{trrp}) presupposes that the momentum content of the mass eigenspaces, as encoded in the definition of $|\Psi^D(y) \rangle$, is the same in both the initial state and the detector. This is not necessarily the case: the precise form $\hat{\Pi}_L$ depends on the physics of the detector and its determination would require a careful modeling of the detection process.
\bibitem{negat} To see this, consider the elementary example of the probability current for the Schr\"odinger equation in one dimension $J(x, t) = - i [\psi^*(x,t) \partial_x \psi(x,t) - \psi(x,t) \partial_x \psi^*(x,t)]$. This is equal to the expectation value of the operator $\hat{J}(x) = \hat{p}|x\rangle\langle x| + |x\rangle\langle x| \hat{p}$, which is not positive-definite.
\bibitem{repr}
The restricted propagator $\hat{S}_t$ is obtained by the solution of the evolution equation $
i \dot{S}_t = \hat{Q} \hat{H} \hat{Q} \hat{S}_t$,
with initial condition $\hat{S}_{0} = \hat{Q}$ and its restriction onto ${\cal H}_-$ defines an one-parameter group of unitary transformations \cite{MiSu}.
\bibitem{indep} For the proof that the events corresponding to detection at different times $t$ are independent, so that integration is justified at the level of amplitudes, see Sec. 4.B in Ref. \cite{AnSav}.
\bibitem{del}The presence of the
factor of $2$ in  the normalization for $f_{\tau}$ is due to the fact that we normalize over positive values of $t$ and not in the full real axis.
The Dirac $\delta$-function refers to integration over the whole real axis, hence, the condition  $\lim_{\tau \rightarrow 0} f_{\tau}(s) =  \delta(s)$ would lead to probabilities normalized over the whole real axis.
To
restrict to the physically relevant range of $t > 0$, we would have
to employ smeared $\delta$-functions over the half-line that
are rather unwieldy. However, if the amplitudes Eq.
(\ref{ampl5a}) have support in values of $t >> \tau$, then  the
contribution of the terms for $t >0 , t' < 0$ in the integral is
negligible, and  $\int_{-\infty}^{\infty} dt  p_{\tau}( t) = 2 \int_0^{\infty} dt
p_{\tau}( t)$. In this regime it suffices to include a factor of $2$ in the normalization
of $f_{\tau}$.
\bibitem{comment3}
 Note that $\tau_{sup}$ formally vanishes if photons are included among the product particles. This case, however, is not part of the measurement scheme examined here. A POVM corresponding to photodetection should be used, rather than a POVM for the position of a massive particle.
\bibitem{microloc} The property Eq. (\ref{microl}) holds for sufficiently regular regions, i.e., regions that do not exhibit much structure at the scale of $\hbar$---see Ref. \cite{Omn4} for details.
\end{thebibliography}
\end{document}